\documentclass[12pt]{iopart}

\usepackage{iopams}
\usepackage{amssymb}
\expandafter\let\csname equation*\endcsname\relax

\expandafter\let\csname endequation*\endcsname\relax

\usepackage{mathtools,amsmath}
\usepackage{graphicx}
\usepackage{marginnote}
\usepackage{color}
\usepackage{siunitx}
\usepackage{changes}
\definechangesauthor[color=red]{Moha}
\usepackage[perpage]{footmisc}
\usepackage{xcolor}
\usepackage{hyperref}
\hypersetup{
	colorlinks   = true,
	citecolor    = darkblue,
	linkcolor    = darkred,
	urlcolor     = pink
}
\usepackage{bbm}
\usepackage{times,txfonts}
\usepackage{esdiff}

\definecolor{darkblue}{rgb}{0.,0.,0.5}
\definecolor{darkred}{rgb}{0.7,0.,0.}

\DeclareMathOperator{\sech}{sech}
\DeclareMathOperator{\csch}{csch}

\DeclareMathAlphabet{\pazocal}{OMS}{zplm}{m}{n}

\newcommand{\red}[1]{\textcolor{red}{#1}}
\newcommand{\ket}[1]{\left\vert#1\right\rangle}
\newcommand{\bra}[1]{\left\langle#1\right\vert}

\def\ketbra#1#2{{\vert#1\rangle\!\langle#2\vert}}

\newcommand{\average}[1]{\left<#1\right>}

\begin{document}

\topical[Thermometry in the quantum regime]{Thermometry in the quantum regime: \\ Recent theoretical progress}

\author{Mohammad Mehboudi $ ^1 $, Anna Sanpera $ ^{2,3} $ and Luis A. Correa $ ^{4,5,6} $}

\address{$ ^1 $ ICFO---Institut de Ciencies Fotoniques, The Barcelona Institute of Science and Technology, 08860 Castelldefels (Barcelona), Spain}
\address{$ ^2 $ ICREA, Pg. Llu\'is Companys 23, E-08010 Barcelona, Spain}
\address{$ ^3 $ F\'{\i}sica Te\`{o}rica: Informaci\'{o} i Fen\`{o}mens Quantics. Departament de F\'{\i}sica, Universitat Aut\`{o}noma de Barcelona, 08193 Bellaterra, Spain}
\address{$ ^4 $ CEMPS, Physics and Astronomy, University of Exeter, Exeter, EX4 4QL,
	United Kingdom}
\address{$ ^5 $ School of Mathematical Sciences and CQNE, University of Nottingham, University Park Campus, Nottingham NG7 2RD, United Kingdom}
\address{$ ^6 $ Kavli Institute for Theoretical Physics, University of California, Santa Barbara, CA 93106, USA}

\begin{abstract}
Controlling and measuring the temperature in different devices and platforms that operate in the quantum regime is, without any doubt, essential for any potential application. In this review, we report the most recent theoretical developments dealing with accurate estimation of very low temperatures in quantum systems. Together with the emerging experimental techniques and developments of measurement protocols, the theory of \textit{quantum thermometry} will decisively impinge and shape the forthcoming quantum technologies. While current quantum thermometric methods differ greatly depending on the experimental platform, the achievable precision, and the temperature range of interest, the theory of quantum thermometry is built under a unifying framework at the crossroads of quantum metrology, open quantum systems, and quantum many-body physics. At a fundamental level, theoretical quantum thermometry is concerned with finding the ultimate bounds and scaling laws that limit the precision of temperature estimation for systems in and out of thermal equilibrium. At a more practical level, it provides tools to formulate precise, yet feasible, thermometric protocols for relevant experimental architectures. Last but not least, the theory of quantum thermometry examines genuine quantum features, like entanglement and coherence, for their exploitation in enhanced-resolution thermometry. 
\end{abstract}


\maketitle

{\hypersetup{linkcolor=black}
\tableofcontents
}

\pagebreak{}

\section{Introduction}\label{sec1}

The spectacular level of precision needed to prepare, manipulate and detect systems of very small size that operate in the quantum regime is opening new domains in physics. In order to monitor and control such quantum systems, measurement schemes of unprecedented accuracy are required. For example, atomic gases can presently be cooled down to the lowest temperatures in the Universe (see \cite{leanhardt2003cooling,PhysRevLett.106.195301, Bloch2005} and references therein)---even below the nanokelvin regime---which demands measurement protocols capable of high precision at very low temperatures. In addition, quantum systems \textit{themselves} can be exploited to attain an astonishing resolution in the estimation of parameters and fundamental constants of Nature~\cite{RevModPhys.89.035002,Giovannetti2011}. For instance, the most precise atomic clock could measure the estimated age of the Universe with an error of less than $100\,\si{\milli\second}$ \cite{marti2018imaging,mann2018core,PhysRevLett.116.063001}. A thorough analysis of the ultimate bounds on measurement precision is thus in order. This is the central question addressed in this review---specifically, the fundamental limits on the accuracy of \textit{temperature measurements}.

Temperature is an intuitive notion deeply rooted in our daily life and yet it is subtle and surprisingly difficult to formalise \cite{Puglisi2017}. The existence of temperature can be taken as the first step in an axiomatic construction of thermodynamics \cite{sommerfeld2012lectures}, which was empirically developed to explain the interconversion between heat and work \cite{carnot1890fire}. Alternatively, one can start by introducing the concept of \textit{entropy} $ S $ and take temperature as the variation of the energy of a system with respect to $ S $ \cite{callen2006thermodynamics}. The link established by Boltzmann between entropy and molecular dynamics through the celebrated H-theorem, provides further insights: from a statistical viewpoint, temperature determines how the particles in a system arrange their energies in order to maximise the disorder and thus, attain \textit{thermal equilibrium}. For a system in equilibrium, the energy of its particles might fluctuate but not its temperature, which is a constant; or rather, a \textit{name tag}.

The measurability of temperature follows from the so-called zeroth law of thermodynamics \cite{sommerfeld2012lectures}. Quoting Planck \cite{planck1927vorlesungen}:
\begin{quote}
	{\it If a body $ A $ be in thermal equilibrium with two other bodies, $ B $ and $ C $, then $ B $ and $ C $ are in thermal equilibrium with one another.}
\end{quote}
Therefore, any two systems in equilibrium with some fixed reference can be tagged with the \textit{same} empirical temperature, and would remain unchanged if put in thermal contact with each other. When it comes to the seemingly arbitrary task of attaching numerical values to those temperature tags, thermodynamics comes to rescue once again. The fact that the efficiency of a (reversible) Carnot heat engine is solely determined by the quotient of the external temperatures \cite{carnot1890fire} provides a protocol to \textit{measure} temperature ratios, thus removing ambiguities in the definition of an absolute thermometric scale. In other words, the second law of thermodynamics, which is intimately related to the reversibility of a Carnot engine, enables thermometry \cite{planck1927vorlesungen}.

So far, we have simply skipped through the textbook account of temperature. However, as soon as we move away from the standard scenario of large Hamiltonian systems---possibly in weak thermal contact with infinite heat baths---serious difficulties start to arise. Quantum thermodynamics \cite{binder2018thermodynamics} is precisely concerned with the challenge of consistently redefining the usual thermodynamic variables (i.e., `heat', `work', or `temperature') \cite{esposito2010entropy,RevModPhys.83.771,PhysRevLett.118.070601}, or even reformulating the laws of thermodynamics \cite{brandao2015second,uzdin2018global,kolavr2012quantum,levy2012quantum,freitas2017fundamental,masanes2017general,bera2017thermodynamics} to make them applicable to systems which are not macroscopic, but fully \textit{quantum}. These questions are far from settled and continue to motivate a vibrant activity \cite{e15062100,1310.0683v1,gelbwaser2015thermodynamics,vinjanampathy2016quantum,goold2016role}.

Throughout this review, we always remain `on the safe side'---in most situations discussed here, the subject of study are large Hamiltonian quantum systems with (effectively) infinite heat capacity, and whose energy follows a Boltzmann--Gibbs distribution with some well-defined temperature $ T $. We therefore avoid dealing with more subtle points on the very foundations of quantum statistical mechanics, such as `equilibration' and `thermalisation' on average of closed (many-body) quantum systems in unitarily evolving pure states. For a recent review on the topic see, e.g., reference \cite{gogolin2016equilibration}. Also, importantly, we will make no mention as to how long did the equilibration process take for our systems (see, e.g., Ch.~18 of \cite{binder2018thermodynamics} for an up-to-date overview).
 
Temperature can thus be estimated via global measurements, provided that the energy spectrum of the system is known \cite{stace2010quantum}. If this were not the case or if we simply seek to minimise the disturbance on the system, we could couple an individual quantum `probe' or `thermometer' with well-known Hamiltonian through a very weak dissipative interaction \cite{PhysRevLett.114.220405}. It is usually assumed that, as a result of this weak-coupling constraint and the comparatively large `thermal mass' of the system, the probe eventually thermalises at the system temperature $ T $ \textit{without disturbing it}. Probe and system would thus end up being uncorrelated, and measuring a suitable temperature-dependent property of the probe alone would allow to build the desired estimate of the system's temperature. Essentially, this is no different from checking our body temperature with a liquid-in-glass thermometer. In some cases (especially in sections \ref{sec3} and \ref{sec5}), we will leave the system entirely out of the picture and focus on how to extract information about its temperature from the thermalised probe. We must, nevertheless, bear in mind that there \textit{is} always a system in the background.

As we shall see, the formalism reviewed here---hereafter, simply \textit{quantum thermometry} (see also Ch.~21 in reference \cite{binder2018thermodynamics})---can go well beyond this `classical-like' weak-coupling limit and account in full generality for large probe--system (quantum) correlations \cite{correa2016low,de2016local,miller2018mean}. In return, however, one must pay a toll in terms of detailed knowledge of the dissipative interactions between them. In particular, quantum thermometry allows us to make statements about the precision of ultra-cold temperature measurements, i.e., deep in the quantum regime \cite{hovhannisyan2017probing,hofer2017fundamental}. These are particularly challenging and also relevant to enable upcoming quantum technological applications.

In the discussion so far, we have left \textit{dynamics} entirely out of the picture. However, it is possible to build temperature estimates from information retrieved during the transient evolution of a probe as it approaches its steady state \cite{PhysRevLett.114.220405,jevtic2015single,jarzyna2015quantum,de2017estimating}. Interestingly, the accuracy of these temperature readings may be even larger than that of steady-state measurements although, once again, a careful modelling of the dissipation becomes essential to exploit this advantage. 

In order to place bounds on the precision of a given temperature-sensing protocol, quantum thermometry resorts to the theory of (local) quantum parameter estimation \cite{PhysRevLett.72.3439,barndorff2000fisher}. Crucially, in addition to setting precision bounds, one can also single out which temperature-dependent quantity is the most sensitive to thermal fluctuations and thus, produces the most accurate temperature estimates. Often, however, the corresponding observable is not measurable in practice, due to technological limitations. In such cases, quantum thermometry can assist in identifying accessible (sub-optimal) alternatives which approximate the fundamental precision limits closely \cite{correa2016low,1367-2630-19-10-103003,1367-2630-17-5-055020,cavina2018bridging}.

Ultimately, this review aims to provide a primer to a newly-appeared \textit{unifying theoretical framework} capable of encompassing the (very diverse) techniques used in nanoscale low-temperature sensing. We illustrate how this framework can allow us to identify highly sensitive measurement protocols and make general statements about the desirable features of precise temperature probes. As more refined models for experimentally relevant setups are considered \cite{lampo2018open}, quantum thermometry might inform the design of new thermometric techniques that could outperform the state of the art in low-temperature sensing \cite{mehboudi2018using}. 

We have organised this review in three main self-contained blocks, which cover the situations discussed in the previous paragraphs. Namely, in section \ref{sec3}, we focus on temperature estimation from measurements on stationary quantum probes in thermal equilibrium. We discuss the structure of the energy spectrum of an optimal thermometer as well as various versatile sub-optimal albeit feasible alternatives. 

In section \ref{sec4} we move on to consider local quantum thermometry, i.e., temperature estimates based on incomplete information obtained from measurements on a small fraction of a large equilibrium many-body quantum system. This situation can be regarded as a special case of a probe--system setting with strong dissipative interactions. In particular, we comment on the impact of the strong interactions on the sensitivity of local thermometry and the attempts to harness them to achieve more precise low-temperature readings. We pay particular attention to the scaling of thermometric precision as the temperature drops to zero. 

In section \ref{sec5}, we continue to consider quantum thermometry on non-equilibrium probes but, this time, not due to strong dissipative interactions but rather, to incomplete equilibration. Specifically, we discuss finite-time thermometry with probes undergoing Markovian and non-Markovian dynamics, interferometric setups, and dynamically controlled temperature probes. 

The basic tools needed from quantum estimation theory are briefly reviewed in the introductory section \ref{sec2}, where we also fix notation. Finally, in section \ref{sec6}, we conclude this review by summarising the theoretical developments on quantum thermometry. The interested reader willing to get only a quick overview of the main results discussed here may directly jump to section \ref{sec6}. There, we also briefly comment on experimental advances and highlight some interesting open problems and future perspectives. We must stress, however, that a thorough discussion on the current state of the art in experimental quantum thermometry, although necessary to provide the big picture of the field, is entirely beyond the scope of the present review. Our focus is \textit{exclusively theoretical} and the interested reader is deferred to the existing experimental literature on nanoscale thermometry at ultra-low temperatures. Once again, section \ref{sec6} might be a good starting point.

\section[Preliminaries]{Preliminaries} \label{sec2}

\subsection{The basic problem of thermometry}

We start by briefly formulating the basic problem of temperature estimation. Let the Hamiltonian of the system of interest be $ \hat{H} = \sum_{k}~\epsilon_k~\ketbra{\epsilon_k}{\epsilon_k} $, where $ \epsilon_k $ and $ \ket{\epsilon_k} $ are, respectively, energy eigenvalues and eigenvectors. In all cases considered here, the system is described by the Gibbs (equilibrium) density operator 
\begin{align}
\label{eq-Gibbs-Distribution}
{\hat \varrho}(T) = \sum_k \frac{e^{-\epsilon_k/k_B T}}{{\mathcal Z}}~\ketbra{\epsilon_k}{\epsilon_k},
\end{align}
with $ \mathcal{Z} \coloneqq \sum_k {\rm exp}(-\epsilon_k/k_B T) $ being the partition function, and $ k_B $ the Boltzmann constant. For convenience, we henceforth set $ k_B = \hbar = 1 $. The temperature $ T $ is then merely a parameter which determines the probability of occupying different energy states. In principle, its value reflects the constraint on the global energy of an enlarged closed {\it system-environment} unit \cite{pathria1996statistical,2010heat}. However, as already stated, we simply take equation \eqref{eq-Gibbs-Distribution} as given, and leave the process that drove the system to equilibrium out of the discussion.

From \eqref{eq-Gibbs-Distribution} it follows that $ T $ may be estimated by studying the statistics of energy measurements \cite{PhysRevE.83.011109}. In fact, as we shall see in section \ref{sec-Thermometry-Precision}, this turns out to be the most accurate protocol for temperature estimation on a state like \eqref{eq-Gibbs-Distribution}. Of course, any observable other than energy can be measured instead, provided that it also has a temperature-dependent statistics. Exploring feasible sub-optimal measurements becomes practically relevant whenever measuring energy becomes costly \cite{1367-2630-17-5-055020}. In order to propose accurate thermometric protocols in different physical scenarios, it is necessary to quantify and compare the uncertainty of temperature estimates built from various measurement choices \footnote{Although the estimate is affected by both systematic and statistical errors, we focus here only on the minimisation of the latter.}. In particular, benchmarking against the ultimate precision bound allows to quantify the `optimality' of any thermometric scheme. As we shall see in section \ref{sec:Metro} below, quantum metrology provides the solution to both problems.

In most cases, only a small part of the equilibrium system can be locally measured. Temperature must be then inferred from the information contained in the marginal of $ \hat{\varrho}(T) $ on the `accessible subspace'. Importantly, this marginal is not generally of the simple form \eqref{eq-Gibbs-Distribution} due to the internal interactions within the system. Therefore, before we use the toolbox of quantum metrology for making statements about the accuracy of temperature estimates, we must solve another more difficult problem---finding a suitable description for the state of the temperature probe. As we show in sections \ref{sec4} and \ref{sec5}, an open system approach \cite{BRE02} can be particularly useful in this respect.  

\subsection{Ultimate precision bounds for parameter estimation} \label{sec:Metro}

Consider a system whose state depends on some unknown parameter $ \xi $ that we wish to infer; we denote the corresponding density operator by $\hat{\varrho}(\xi) $. The only way to acquire information about $ \xi $ is to perform measurements on the system. In the most general case, these can be described by a `positive operator-valued measure' (POVM) measurement (see e.g., references \cite{Helstrom1969,preskilllecture,holevo2011probabilistic,nielsen2010quantum,kraus1983states,Wilde:2013:QIT:2505455}). More specifically, we focus on discrete sets of positive Hermitian operators $ \hat{\mathbf{\Pi}} = \{\hat{\Pi}_m\} $, referred-to as `POVM elements'. The corresponding measurement outcomes occur according to the probability distribution $ p_m(\xi) = \mathrm{tr}\{\hat{\Pi}_m~\hat{\varrho}(\xi)\} $, that depends on the unknown parameter. To ensure normalisation, the POVM elements must thus satisfy $ \sum_m \hat{\Pi}_m = \mathbbm{1} $.

Repeating the measurement a large number of times, say $ \mathcal{N} $, on identically prepared copies of the state, $ \hat{\varrho}(\xi) $, provides us with a set of outcomes $ \mathbf{x} $. From these, one may eventually build an \textit{estimator} $ \xi_{\text{est}}(\mathbf{x}) $ for $ \xi $. The accuracy of the resulting estimate is negatively affected by the uncertainty in the outcomes of the POVM---quantitatively, its `error bars' can be gauged by the mean squared error
\begin{equation} 
\delta^2(\hat{\mathbf{\Pi}};~\xi_{\text{est}}) = \average{\big(\xi_{\text{est}}(\mathbf{x})-\average{\xi_{\text{est}}(\mathbf{x})}_\mathbf{x}\big)^2}_\mathbf{x},
\end{equation}
where the argument $\hat{\mathbf{\Pi}}$ indicates that $ \delta $ depends on the specific POVM considered. On the right hand side, the averages are taken over the set of all measurement outcomes, i.e., $ \average{f(\mathbf{x})}_\mathbf{x} = \sum_{\mathbf{x}} p(\mathbf{x} | \xi) f(\mathbf{x})$, with $p(\mathbf{x}|\xi)$ being the probability of obtaining the data set $ \mathbf{x} $, conditioned to the value of the parameter being exactly $ \xi $. In this review, our main focus is on \textit{unbiased estimators}, i.e., those for which $ \average{\xi_{\text{est}}(\mathbf{x})}_\mathbf{x} = \xi$. The mean squared error thus coincides with the variance $ [\Delta\xi_{\text{est}}(\hat{\boldsymbol{\Pi}})]^2 $.

An interesting question to ask is whether there exists a limit on how small the estimation error can be for a given number of measurements $ \mathcal{N} $. It is well known in statistics that the Cram\'er--Rao bound (CRB) proves the existence of such a limit, and further quantifies it for \textit{any} unbiased estimator \cite{frieden1998physics,gershenfeld1999nature,cramer2016mathematical}. The bound reads \begin{equation}
{\Delta\xi_{\text{est}}}(\hat{\boldsymbol{\Pi}})\geq\frac{1}{\sqrt{\mathcal{N}\mathcal{F}_c(\hat{\mathbf{\Pi}},\xi)}},
\label{eq-CRB}
\end{equation}
and can be saturated by choosing the maximum likelihood estimator\footnote{In general, the CRB is tight only in the asymptotic limit of $ \mathcal{N}\rightarrow \infty $. However, when $ p(\mathbf{x} | \xi) $ belongs to the so-called exponential family, the CRB can be saturated for any $ \mathcal{N} $; in particular, also \textit{in a single shot} \cite{kolodynski2014precision} ($ \mathcal{N} = 1 $). See also section \ref{sec-Thermometry-Precision}.}.
Whenever $ \Delta\xi_{\text{est}}(\hat{\boldsymbol{\Pi}}) \sim 1/\sqrt{\mathcal{N}} $ we speak of shot-noise-limited (or `classical') estimation. This scaling is also referred-to as the standard quantum limit and relates to the central limit theorem. On the contrary, a faster decrease in the mean squared error would be the hallmark of \textit{quantum-enhanced} estimation~\cite{giovannetti2004quantum,Giovannetti2011,PhysRevLett.94.020502,PhysRevLett.72.3439,paris,toth2014quantum}. Such enhancement is enabled by quantum entanglement \cite{Toth2012,PhysRevA.85.022321,strobel2014fisher} and ranges from $ \Delta\xi_{\text{est}}(\hat{\boldsymbol{\Pi}}) \sim 1/\sqrt{\mathcal{N}^\alpha} $ with $ 1 < \alpha < 2 $ (`sub-Heisenberg' scaling) to `Heisenberg' scaling ($ \alpha = 2 $). When a parameter is encoded making use of nonlinear interactions among the constituents of an $ \mathcal{N} $-partite probe the scaling of the bound with the number of particles $ \mathcal{N} $ can reach $ \alpha > 2 $ \cite{PhysRevLett.98.090401,Napolitano:12,PhysRevLett.101.040403,RevModPhys.90.035005}.

The quantity $ \mathcal{F}_c(\hat{\mathbf{\Pi}},\xi) $ introduced in equation \eqref{eq-CRB}, stands for the so-called \textit{Fisher information} (FI), and captures the response of the probability distribution $ p(\mathbf{x}|\xi) $ to small changes in the parameter $ \xi $. It is explicitly defined as 
\begin{equation}
\mathcal{F}_c(\hat{\mathbf{\Pi}},\xi)\coloneqq\average{\big(\partial_{\xi}\log p(\mathbf{x}|\xi)\big)^2}_{\mathbf{x}}=\sum_{\mathbf{x}}\frac{\big(\partial_{\xi}p(\mathbf{x}|\xi)\big)^2}{p(\mathbf{x}|\xi)}.
\label{eq5:Fisher} 
\end{equation}

If we chose to perform projective measurements onto the eigenstates of some observable $ \hat{O} $ to estimate $ \xi $, we could resort to the error propagation formula which quantifies the ensuing uncertainty as $ \Delta\hat{O}/\vert\partial_{\xi}\langle\hat O\rangle\vert $ (see e.g., references \cite{Helstrom1969,holevo2011probabilistic,toth2014quantum}). Of course, the CRB would also apply to this case, and therefore
\begin{align}
\frac{\Delta\hat{O}}{\vert\partial_{\xi}\langle\hat O\rangle\vert} \geq \frac{1}{\sqrt{\mathcal{F}_c(\hat{O},\xi)}},
\label{eq-error-propagation}
\end{align}
where $ \langle\circ\rangle = \mbox{tr}\{\circ\,\hat{\varrho}(\xi)\} $.

Identifying the measurement which saturates the CRB is very useful. The corresponding uncertainty sets the so-called \textit{quantum} Cram\'er--Rao bound (QCRB)
\begin{equation}
\Delta\xi_{\text{est}}(\hat{\boldsymbol{\Pi}})\geq\frac{1}{\sqrt{\mathcal{N}\mathcal{F}_c(\hat{\boldsymbol{\Pi}},\xi)}}\geq\frac{1}{\sqrt{\mathcal{N}\mathcal{F}(\xi)}},
\label{eq-QCRB}
\end{equation}
where we have introduced the quantum Fisher information (QFI) $ \mathcal{F}(\xi) $ associated with the parameter $ \xi $. This is nothing but an optimisation of the Fisher information over all possible measurements $ \mathcal{F}(\xi) = \max_{\hat{\mathbf{\Pi}}} \mathcal{F}_c(\hat{\mathbf{\Pi}},\xi) $. Alternatively, it can be written as
\begin{equation}
\mathcal{F}(\xi)\coloneqq\mathrm{tr}\lbrace\hat{\varrho}(\xi)~\hat{\Lambda}_{\xi}^2\rbrace,
\label{eq-QFI}
\end{equation}
where the self-adjoint operator $ \hat{\Lambda}_{\xi} $ (satisfying $ \langle\hat{\Lambda}_\xi\rangle = 0 $) is the \textit{symmetric logarithmic derivative} (SLD). In turn, the SLD is connected to the density matrix and its first derivative through
\begin{equation}
\hat{\Lambda}_{\xi}~\hat{\varrho}(\xi)+\hat{\varrho}(\xi)~\hat{\Lambda}_{\xi}\coloneqq 2\partial_{\xi^{\prime}}\hat{\varrho}(\xi^{\prime})\big|_{\xi^{\prime}=\xi}.
\label{eq-SLD}
\end{equation}

Interestingly, in light of equation \eqref{eq-SLD}, the SLD can also be used to rewrite \eqref{eq-error-propagation} as
\begin{equation}
\frac{\Delta\hat{O}}{\vert\partial_{\xi}\langle\hat O\rangle\vert} = \frac{\Delta\hat{O}}{{\rm cov}(\hat{O},\hat{\Lambda}_\xi)}\geq \frac{1}{\Delta\hat{\Lambda}_\xi} = \frac{1}{\sqrt{{\mathcal F}(\xi)}},
\label{eq2:SLD-meaning}
\end{equation}
where the covariance is defined as $ {\rm cov}(\circ , \bullet) \coloneqq \average{\circ~\bullet + \bullet~\circ}/2 - \average{\circ}\average{\bullet} $. The first equality follows directly from \eqref{eq-SLD} and the fact that the expectation value of the SLD vanishes. One can then apply the Cauchy-Schwartz inequality and the definition of the QFI from equation \eqref{eq-QFI} in order to arrive to the last equality on the right. 

Looking at equation \eqref{eq2:SLD-meaning}, it becomes clear that the ultimate precision bound is attained by choosing $ \hat{O} = \hat{\Lambda}_\xi $. That is, \textit{the most accurate estimate of $ \xi $ can be built from projective measurements onto the eigenbasis of the SLD} \cite{paris, toth2014quantum, PhysRevLett.72.3439}. Unfortunately, such measurements are often experimentally unfeasible (e.g., highly non-local collective measurements). It is thus very important to consider more practical sub-optimal measurements and benchmark their performance against the QFI.

Finally, let us point out that the QFI is additive under tensor products, i.e., given a fully uncorrelated $ M $-partite system in the state $ \hat{\varrho}(\xi) = \bigotimes_i \hat{\varrho}_i $, its QFI is~\footnote{The corresponding basis vectors must be independent of the parameter $ \xi $.}
\begin{align}\label{eq-QFI-Additive}
\mathcal{F}(\xi,\hat{\varrho}) = \sum_i \mathcal{F}(\xi,\hat{\varrho}_i).
\end{align}
The additivity property proves helpful when dealing with thermometry in those many-body systems that can be mapped into non interacting quasi-fermions. This is the case because the thermal state of such models can always be expressed as a product of uncorrelated marginals (cf. section \ref{sec3:criticality}). In what follows, we explicitly write the corresponding state as an argument of the QFI wherever confusion might arise. 

\subsection{Geometric interpretation of the quantum Fisher information}\label{sec:geometric_QFI}

One can gain further insights for the interpretation of the QFI \eqref{eq-QFI} as a quantifier of the responsiveness of a state $ \hat{\varrho}(\xi) $ to a variation of the relevant parameter $ \xi $, by considering its connection to the Uhlmann fidelity (hereafter, simply fidelity). Given any two (mixed) quantum states $ \hat{\varrho}_1 $ and $ \hat{\varrho}_2$, their fidelity is defined as (see, e.g., reference \cite{UHLMANN1976273,doi:10.1080/09500349414552171}) 
\begin{align}\label{eq-Fidelity}
\mathbb{F}(\hat{\varrho}_1, \hat{\varrho}_2) \coloneqq \left({\rm tr}\,\sqrt{\sqrt{\hat{\varrho}_1}\hat{\varrho}_2\sqrt{\hat{\varrho}_1}}\right)^2.
\end{align} 
$ \mathbb{F}(\hat{\varrho}_1,\hat{\varrho}_2) $ is symmetric under the exchange of its arguments and bounded between $ 0 $ and $ 1 $. In particular, $ \mathbb{F}(\hat{\varrho}_1,\hat{\varrho}_2) = 1 $ iff $ \hat{\varrho}_1 = \hat{\varrho}_2 $. Therefore, we can use fidelity in order to introduce a measure of \textit{distance} between quantum states---the so-called Bures distance
\begin{align}
d_B^2(\hat{\varrho}_1, \hat{\varrho}_2) \coloneqq 2\left(1-\sqrt{\mathbb{F}(\hat{\varrho}_1,\hat{\varrho}_2)}\right).
\label{eq-Bures}
\end{align}

Letting the density matrices in equation \eqref{eq-Bures} be parametrised by $ \xi $, so that $ \hat{\varrho}_1 = \hat{\varrho}(\xi) $ and $ \hat{\varrho}_2 = \hat{\varrho}(\xi+\delta) $, and taking the limit $\delta \to 0$, allows to cast the QFI as \cite{sommers2003bures}  
\begin{align}\label{eq-Fid-QFI}
{\mathcal F}\left(\xi\right) & = 4 \lim_{\delta \to 0} \frac{d^2_B\left(\hat{\varrho}(\xi),\hat{\varrho}(\xi+\delta)\right)}{\delta^2}\nonumber\\
&= 8~\lim_{\delta \to 0}\frac{1 - \sqrt{\mathbb{F}\left(\hat{\varrho}(\xi),\hat{\varrho}(\xi + \delta)\right)}}{\delta^2}\nonumber\\
&= -2 \lim_{\delta\to 0}\partial_\delta^2\,\mathbb{F}\left(\hat{\varrho}(\xi),\hat{\varrho}(\xi + \delta)\right)\nonumber\\
&\coloneqq -2\,\chi_{\mathbb{F}}[\hat{\varrho}(\xi)],
\end{align}
where, in the last line, we have introduced the \textit{fidelity susceptibility} \footnote{
	Note that the fidelity is often defined as $ \mathbb{F}(\hat{\varrho}_1, \hat{\varrho}_2) \coloneqq {\rm tr}\,\sqrt{\sqrt{\hat{\varrho}_1}\hat{\varrho}_2\sqrt{\hat{\varrho}_1}} $. In that case, the fidelity susceptibility would gain a prefactor of $ 1/2 $ and the QFI would become ${\mathcal F}(\xi,\hat{\varrho}) = -4\, \chi_{\mathbb{F}}[\hat{\varrho}(\xi)]$.     
}. Since the fidelity has an extremum at $ \delta = 0 $, its first derivative should vanish, so that $ \chi_{\mathbb{F}}[\hat{\varrho}(\xi)] $ becomes the leading-order response in the fidelity to fluctuations with respect to $ \delta $---hence the name `fidelity susceptibility'. $ \chi_{\mathbb{F}}[\hat{\varrho}(\xi)] $ has been extensively used in the study of quantum many-body systems, as it helps to identify quantum phase transitions (see, e.g., references \cite{Gu2010,zanardi2007ground,PhysRevB.75.014439,PhysRevLett.98.110601,PhysRevLett.99.100603,PhysRevE.76.061108}).

\section{Quantum thermometry in thermal equilibrium}\label{sec3}

Now that we have introduced the basic tools needed from quantum estimation theory, we can tackle the basic problem of estimating the temperature of a system in thermal equilibrium. We start by showing that energy measurements are optimal for temperature estimation and that the corresponding thermal responsiveness is proportional to the heat capacity---or, equivalently, to the energy variance---of the system \cite{PhysRevE.83.011109,stace2010quantum,PhysRevLett.114.220405} (cf. section \ref{sec-Thermometry-Precision}). We then discuss, in section \ref{sec-best-individual}, which are the desirable properties that endow a quantum system with the maximum possible thermal sensitivity at any given temperature \cite{PhysRevLett.114.220405,reeb2015tight}. Interestingly, we shall also see how systems whose energy spectrum is not optimised for temperature sensing can also be useful and versatile thermometers in many situations \cite{PhysRevA.88.033607,2058-9565-3-2-025002}. 

We shall then discuss the role of internal couplings in the global thermal responsiveness of multipartite interacting probes by resorting to a simple bipartite example (see section \ref{sec-Coupled-ho-Thermal}) \cite{1367-2630-19-10-103003}. Next, in section \red{\ref{sec3:criticality}}, we exploit the paradigmatic XY model to illustrate the interplay between quantum criticality and near-ground-state thermometry in many-body systems \cite{1367-2630-17-5-055020}. In particular, we are concerned with the size and temperature scaling of the QFI at low temperatures and in the thermodynamic limit.

We also illustrate how the thermometric performance of a system can be dramatically affected by imposing conservation laws in, e.g., particle number or total angular momentum \cite{1367-2630-19-10-103003,guo2015ring} (cf. section \ref{sec-GGS}). Finally, in section \ref{sec-inf-dist}, we briefly comment on the interplay between thermometric precision and the back-action on the measured system \cite{seveso2018trade,PhysRevA.98.012115}.

\subsection{The ultimate thermometric bounds} \label{sec-Thermometry-Precision}

In this section we consider an $ N $-dimensional system with Hamiltonian
\begin{equation}
\hat{H} = \sum\nolimits_{k=1}^N \epsilon_k\ketbra{\epsilon_k}{\epsilon_k}
\end{equation}
in the equilibrium state \eqref{eq-Gibbs-Distribution} at temperature $ T $. Recall that this could also be a fully equilibrated temperature probe in weak thermal contact with the system. For that reason, we use the terms `probe' and `system' interchangeably.

To begin with, we search for the observable $ \hat{O} $ with the largest sensitivity to temperature fluctuations. As discussed in section \ref{sec:Metro}, performing a large number $ \mathcal{N} $ of projective measurements onto the eigenbasis of $ \hat{O} $ provide us with a data set that can be used to produce an estimate $ T_{\text{est}} $. The uncertainty of such estimate, as measured by the error propagation formula \eqref{eq-error-propagation}, contains the key quantity
\begin{equation}
\partial_{T}\langle\hat O\rangle = \partial_T\,{\rm tr}\{{\hat O}~\hat{\varrho}(T)\} \coloneqq \chi(\hat{O},T)
\label{eq3:temperature_susceptibility}
\end{equation}
that we shall refer to as the (static) `temperature susceptibility' of $ \hat{O} $ in $ \hat{\varrho}(T) $. In particular, the temperature susceptibility of the Hamiltonian $ \hat{H} $ is the `heat capacity' of the system, which we denote by $ C_T $. It can be easily verified that
\begin{equation}
C_T = \partial_T\langle\hat{H}\rangle = (\Delta\hat{H}/T)^2.
\end{equation}
In fact, the temperature susceptibility of \textit{any} observable $ \hat{O} $ is proportional to its correlation with the Hamiltonian $ \hat{H} $; that is,
\begin{align}
\chi({\hat O},T) = \frac{1}{T^2}\,\mathrm{tr}\big\lbrace\hat{O}\,\big(\hat{H}-\langle\hat{H}\rangle\big)\,\mathcal{Z}^{-1}e^{-\hat{H}/T}\big\rbrace = \frac{1}{T^2}\,\mathrm{cov}\big({\hat O},\hat{H}-\langle\hat{H}\rangle\big).
\label{eq3:suscept_cov}
\end{align}
Comparing equations \eqref{eq3:suscept_cov} and \eqref{eq2:SLD-meaning} we thus see that the SLD is, in this case, 
\begin{equation}\label{eq-SLD-Temperature}
\hat{\Lambda}_T = \frac{1}{T^2}(\hat{H}-\langle\hat{H}\rangle), 
\end{equation}
and as expected, \textit{energy measurements are the most informative ones} when it comes to temperature estimation with a thermalised probe. The corresponding (minimum) statistical uncertainty is thus given by
\begin{align}
\sqrt{\mathcal{N}}\Delta T \geq \frac{1}{\sqrt{\mathcal{F}(T)}} = \frac{1}{\sqrt{\mbox{tr}\{\hat{\Lambda}_T^2\hat{\varrho}(T)\}}} = \frac{T^2}{\sqrt{\mbox{tr}\{(\hat{H}-\langle \hat{H} \rangle)^2 \hat{\varrho}(T)\}}}=\frac{T^2}{\Delta\hat{H}},
\label{eq-QFIT-Variance}
\end{align}
which, in the single-shot case $ {\cal N} = 1 $ takes the form 
\begin{equation}\label{eq:T-Energy-uncertainty}
\Delta\beta\Delta{\hat H}\geq 1,
\end{equation}
with the inverse temperature $ \beta = 1/T $. Hence, the signal-to-noise ratio for any equilibrium probe is \cite{PhysRevE.83.011109}
\begin{align}\label{eq-Temperature-Uncertainty}
(T / \Delta T)^2 \leq \mathcal{N} C_T.
\end{align}
Equation \eqref{eq-Temperature-Uncertainty} tells us that the thermal sensitivity of an equilibrium probe is fundamentally limited by its heat capacity. It is important to note that, due to the Gibbs form of the state $ \hat{\varrho}(T) $, the probability distribution resulting from projective measurements in the energy basis belongs to the \textit{exponential family} and, therefore, it allows for the saturation of the CRB at finite ${\cal N}$ \cite{kolodynski2014precision} (even at $ \mathcal{N} = 1 $).

Alternatively, one may also obtain equation \eqref{eq-QFIT-Variance} from \eqref{eq-Fid-QFI}, thus bridging a gap between thermodynamics and the geometry of quantum states \cite{PhysRevLett.99.100603, PhysRevLett.114.220405}. Indeed, the fidelity between two thermal states at temperatures $ T_1 $ and $ T_2 $ writes as \cite{PhysRevA.75.032109}
\begin{align}
\mathbb{F}(\hat{\varrho}(T_1),\hat{\varrho}(T_2)) = \frac{{\cal Z}^2\big(\frac{2 T_1 T_2}{T_1 + T_2}\big)}{{\cal Z}(T_1) {\cal Z}(T_2)},
\end{align}
where the temperature has been explicitly included as an argument in the partition functions $ \mathcal{Z}(T_i) \coloneqq \mbox{tr}\{ e^{-\hat{H}/T_i} \} $. Using equation \eqref{eq-Fid-QFI} for $ T_2 = T_1 + \delta $, and expanding the fidelity to second order in $ \delta $, eventually brings us back to \eqref{eq-QFIT-Variance}.

\subsection{Optimal and sub-optimal quantum thermometers}\label{sec-best-individual}

\begin{figure}
	\includegraphics[width=1\linewidth]{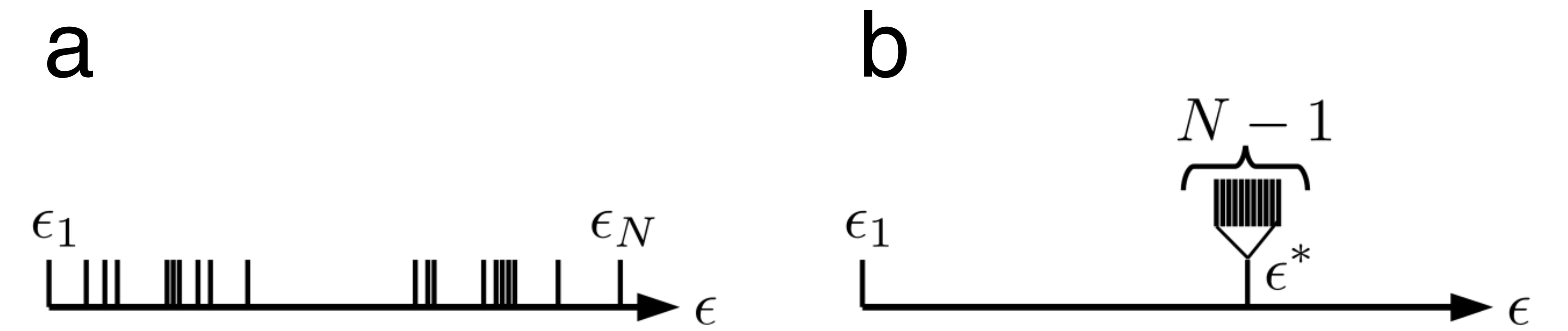}
	\caption{{\bf (a)} Diagrammatic representation of energy spectrum of a generic $ N $-dimensional system. {\bf (b)} Optimal energy-level structure maximising the thermal sensitivity for the same dimension. This configuration corresponds to an effective two-level system with an $ (N-1) $-fold degenerate excited state. The optimal energy gap $ \epsilon^* $ is a function of both the temperature $ T $, and the dimension $ N $ of the probe.}
	\label{fig3-1}
\end{figure}

From the relation $ \mathcal{F}(T) = (\Delta\hat{H})^2/T^4 = C_T/T^2 $ it becomes clear that the energy-level structure of $ \hat{H} $ plays a central role in limiting thermometric precision \cite{PhysRevLett.114.220405,paris2015achieving,1367-2630-19-10-103003,2058-9565-3-2-025002,plodzien2018few}. Hence, we now wish to find the optimal spectrum for estimating a given temperature $ T $. To that end, we can simply impose the $ N $ simultaneous conditions 
\begin{equation}
\frac{\partial}{\partial\epsilon_k}\big(\Delta\hat{H}\big)^2 = 0, \qquad k \in \{1,\cdots, N\},
\end{equation}
which herald the maximisation of the energy variance and thus, of the QFI. As shown in figure \ref{fig3-1}(b), the maximum is attained in an effective two-level configuration with an $ (N-1) $-fold degenerate excited state. The ratio of the corresponding energy gap to the temperature $ x^*\coloneqq\epsilon^*/T = (\epsilon_{k>1}-\epsilon_1)/T $ only depends on the dimension $ N $, and its value is the solution to the following transcendental equation \cite{PhysRevLett.114.220405}: 
\begin{align}\label{eq-optimal-energy}
e^{x^*} = (N-1)\,\frac{x^* + 2}{x^* - 2}.
\end{align}
Consequently, the quantum Fisher information for such optimised probe reads
\begin{align}
\mathcal{F}^*(T,N) = \frac{x^{*\,2}\,e^{x^*}}{T^2}\frac{N-1}{(N-1+e^{x^*})^2},
\label{Energy-Var-quNit}
\end{align}
and the relative error $ \Delta T / T $ is lower-bounded by
\begin{equation}
\left(\frac{\Delta T}{T}\right)^2\geq\frac{(N-1+e^{x^*})^2}{(N-1)\,x^{*\,2}\,e^{x^*}},
\label{eq3:relative-error-dimension}
\end{equation}
which, again, depends only on the dimensionality of the probe. In particular, taking the limit $ N\rightarrow\infty $ yields a best-case relative error $(\Delta T/T)^2  \sim 4/\log{N} $ \cite{plodzien2018few}.

\begin{figure}
	\includegraphics[width=0.495\linewidth]{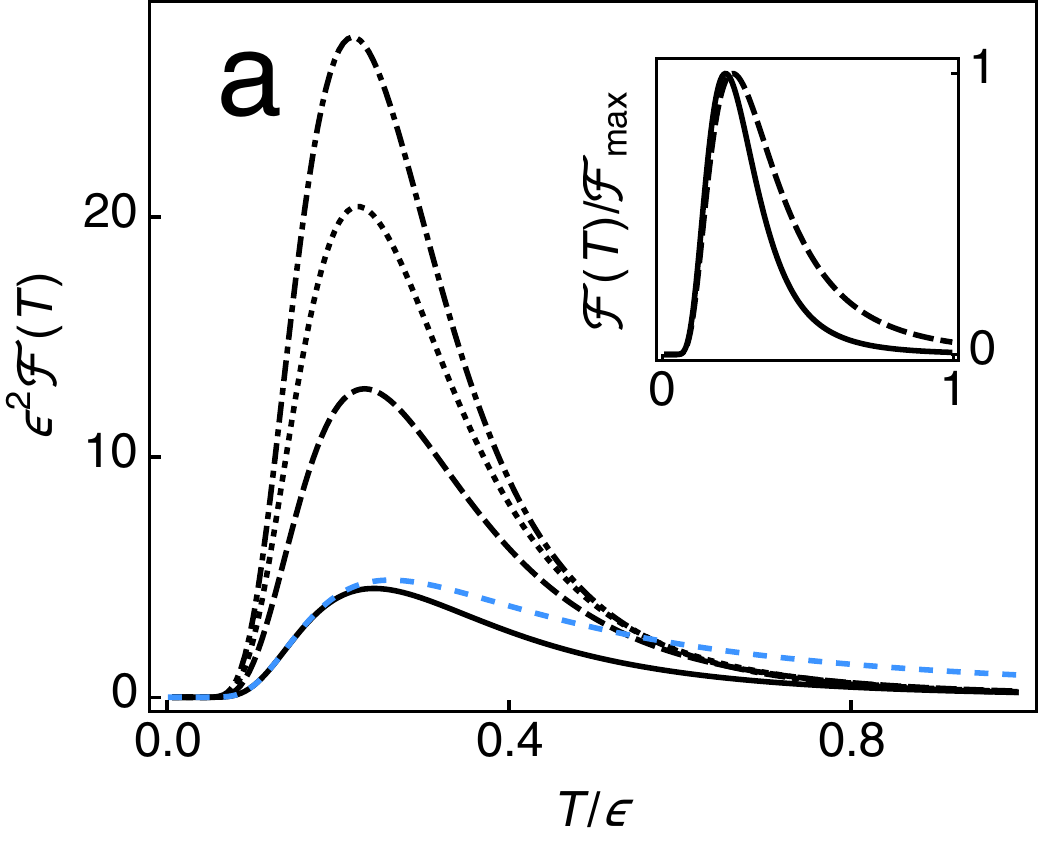}
	\includegraphics[width=0.505\linewidth]{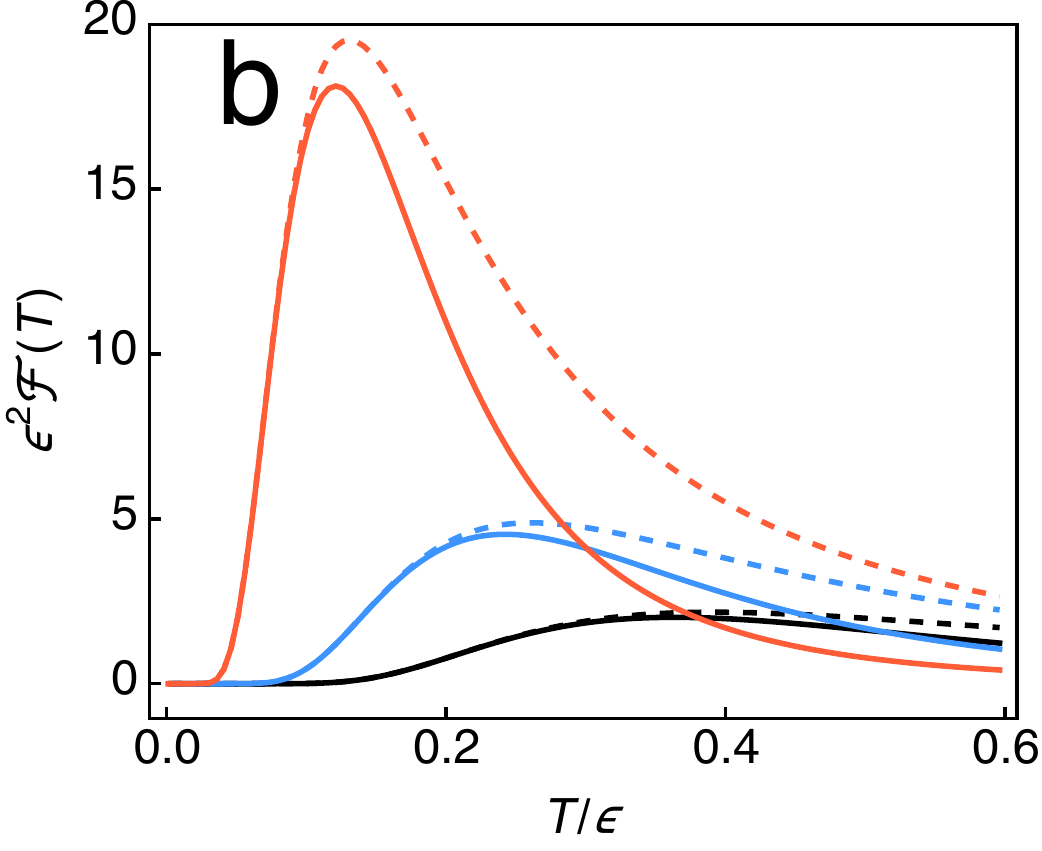}
	\caption{\textbf{(a)} QFI $ \mathcal{F}(T,N) $ of thermalised two-level probes as a function of the temperature $ T $, for a fixed energy gap $ \epsilon $ and $ (N-1) $-fold excited-state degeneracy. The cases $ N = 2 $ (solid black), $ N = 4 $ (dashed black), $ N = 6 $ (dotted black), and $ N = 8 $ (dot-dashed black) are plotted. The QFI of a harmonic oscillator with frequency $ \epsilon $ (dashed blue) has been superimposed for comparison. In the inset, the normalised QFI of a qubit (solid) is compared to that of an effective two-level probe with $ N = 8 $ (dashed). As it can be seen, a two-level system performs efficiently over a wider range of temperatures. \textbf{(b)} $ \mathcal{F}(T, N = 2) $ versus $ T $ for $ \epsilon = 1.5 $ (solid black), $ \epsilon = 1 $ (solid blue), and $ \epsilon = 0.5 $ (solid orange). The dashed black, blue, and orange curves stand for the QFI of harmonic oscillators at frequencies $ \epsilon = 1.5 $, $ \epsilon = 1 $, and $ \epsilon = 0.5 $, respectively. We can acknowledge that the oscillators outperform their two-level counterparts and remain accurate over a wider range of temperatures. Note as well that, for small temperatures, qubits and harmonic probes showcase exactly the same thermometric performance. 
	}
	\label{fig3-2}
\end{figure}

In figure \ref{fig3-2}(a), we plot the QFI from equation \eqref{Energy-Var-quNit} as a function of $ T $ for various $ N $ and \textit{fixed energy gap} $ \epsilon $; that is, not at the temperature-dependent $ \epsilon_* = x^*\, T $. To avoid confusion, we use the notation $ \mathcal{F}(T,N) $ (without asterisk). As it can be seen $ \mathcal{F}(T,N) $ drops rapidly in the low temperature limit, regardless of $ N $. This is not surprising since, at low-enough temperatures ($ T/\epsilon \ll 1 $), a probe with finite gap $ \epsilon $ collapses into its ground state $ \hat{\varrho}(T)\sim \ketbra{\epsilon_1}{\epsilon_1} $. Small variations of $ T $ would then be insufficient to pump population to the excited-state manifold and the probe would thus become insensitive to temperature changes. A similar argument applies to situations in which the temperature is so large that $ \hat{\varrho}(T) \simeq \hat{\varrho}(T+\delta) \simeq \mathbbm{1}/N $. 

If we let the gap depend on temperature so that $ \epsilon(T)/T $ saturates to a constant as $ T \rightarrow 0 $, the smallest statistical uncertainty scales as $ \Delta T \sim T $ in the low temperature limit. On the contrary, if $ \epsilon $ is bounded from below the estimation suffers from a large uncertainty, diverging \textit{exponentially} as $ T \rightarrow 0 $---no better than $ \Delta T \sim (T^2/\epsilon_{\min})\,e^{\epsilon_{\min}/2T} $---as we shall see in section \ref{sec:exponential_ineff_gapped} below (see also \cite{paris2015achieving}). Figure \ref{fig3-2}(a) also shows that $ \mathcal{F}(T,N) $ increases monotonically with $ N $ at any $ T $. Nevertheless, as illustrated in the inset, a larger degeneracy translates into a narrower temperature range over which the sensitivity remains close to its optimal value. 

The highly degenerate optimal spectra described above can be hard to craft in practice. Alternatively, one could try to realise more practical thermometers with a far less exotic energy spectrum. In particular, let us consider the thermal responsiveness of a quantum harmonic oscillator with frequency $ \epsilon $. At thermal equilibrium, its average energy is given by $ \langle \hat{H}_{\text{ho}}\rangle = (1-e^{-\epsilon/T})^{-1}\,e^{-\epsilon/T}\,\epsilon $. Deriving with respect to $ T $ leads to the energy variance, $ \Delta\hat{H}_{\text{ho}} = (1-e^{-\epsilon/T})^{-1}\,e^{-\epsilon/2T}\,\epsilon $. Therefore, the corresponding QFI can be written as
\begin{align}
\mathcal{F}_{\text{ho}}(T) = \frac{1}{T^4}\frac{\epsilon^2\,e^{-\epsilon/T}}{(1-e^{-\epsilon/T})^2}.
\label{eq-ho-temp-sensitivity}
\end{align}
In figure \ref{fig3-2}, $ \mathcal{F}_{\text{ho}}(T) $ appears (coloured dashed lines) superimposed to $ \mathcal{F}(T,N) $. On the one hand, we see that the harmonic thermometer performs efficiently in a wider temperature range than \textit{any} finite-dimensional probe. On the other hand, although harmonic oscillators are superior to two-level thermometers at any temperature, the two types of probe converge towards the \textit{same} sensitivity as $ T \rightarrow 0 $. This is to be expected, since, at low-enough temperatures, a thermal oscillator can be reliably truncated to its first two energy levels. More rigorously, comparing equations \eqref{Energy-Var-quNit} and \eqref{eq-ho-temp-sensitivity} suggests that $\mathcal{F}_{\text{ho}}(T)\sim\mathcal{F}(T,2) \sim (\epsilon^2/T^4)\,e^{-\epsilon/T}$. In fact, this holds also for any $ N $-dimensional probe with equispaced spectrum \cite{2058-9565-3-2-025002}. 

As proposed in reference \cite{plodzien2018few}, another interesting type of probe consists of a mixture of two interacting species of fermions confined in a 1D harmonic potential. Indeed, these fermionic models have been realised experimentally with a high degree of control \cite{serwane2011deterministic,wenz2013few,PhysRevLett.108.075303,PhysRevLett.111.175302}, and their energy spectrum can be precisely tuned through the interaction strength \cite{PhysRevA.88.033607,PhysRevLett.111.045302,PhysRevA.87.060502}. In particular, the many-body ground state of such system becomes quasi-degenerate in the large interaction limit which, in principle, could endow them with high precision at low temperatures. Adjusting the interaction strength to optimise the thermal sensitivity for decreasing $ T $ does approximately yield a scaling $ \mathcal{F}(T) \sim T^2 $ and thus, $ \Delta T \sim T $ for $ T\rightarrow 0 $. The numerical value of $ \mathcal{F}(T) $, however, lies below the ultimate bound established by the optimal `qubit-like' degenerate probes described in \ref{sec-best-individual}. In addition, counting the number of particles in the lowest orbitals, or detecting particles above the Fermi level, are feasible sub-optimal measurements displaying the same qualitative features as the global optimal ones \cite{plodzien2018few}. 

The thermometers considered so far perform optimally at a single temperature. On the contrary, one could be interested in designing an equilibrium thermometer with a ``multi-peaked'' QFI, i.e., capable of measuring multiple temperatures efficiently \cite{2058-9565-3-2-025002}. Energy spectra featuring several highly degenerate energy states would achieve this effect, although the required degeneracies can become extremely large. As we shall see in section \ref{sec:dynamical_control}, a more practical solution to achieve such multi-peaked sensitivity profiles is to consider \textit{dynamically controlled} probes \cite{mukherjee2017high}.

\subsection{The role of interactions in multipartite systems} \label{sec-Coupled-ho-Thermal}

We just discussed that the internal interactions in a two-component mixture of fermions can be harnessed to improve the thermometric precision at low temperatures. We now take a closer look at the role of interactions in multipartite probes. We consider the simple case of two coupled harmonic oscillators in a global equilibrium state. Let the Hamiltonian be  
\begin{align}
\hat{H} = \omega\,\hat{a}^\dagger_1 \hat{a}_1 + (\omega + \Delta)\,\hat{a}^\dagger_2 \hat{a}_2-J\left(\hat{a}^\dagger_1 \hat{a}_2 + \hat{a}_1 \hat{a}^\dagger_2\right)+\frac{U}{2}\left(\hat{a}^{\dagger 2}_1 \hat{a}^{2}_1 + \hat{a}^{\dagger 2}_2 \hat{a}^{2}_2\right),
\label{eq-H-DoubleWell}
\end{align}
where $ \hat{a}_i $ is the annihilation operator for oscillator $ i \in \{1,2\} $. The first two terms of equation \eqref{eq-H-DoubleWell} represent the free Hamiltonian of each mode, while the third one takes into account the coupling between them. Finally, the last term introduces a non-linear self-interaction. For now, we limit ourselves to the case $ U = 0 $, in which the problem is exactly solvable. We shall return to this Hamiltonian in section \ref{sec-GGS} below, where we stress the thermometric consequences of fixing the total number of excitations.

\begin{figure*}[t]
	\begin{center}
		\includegraphics[width=0.49\textwidth]{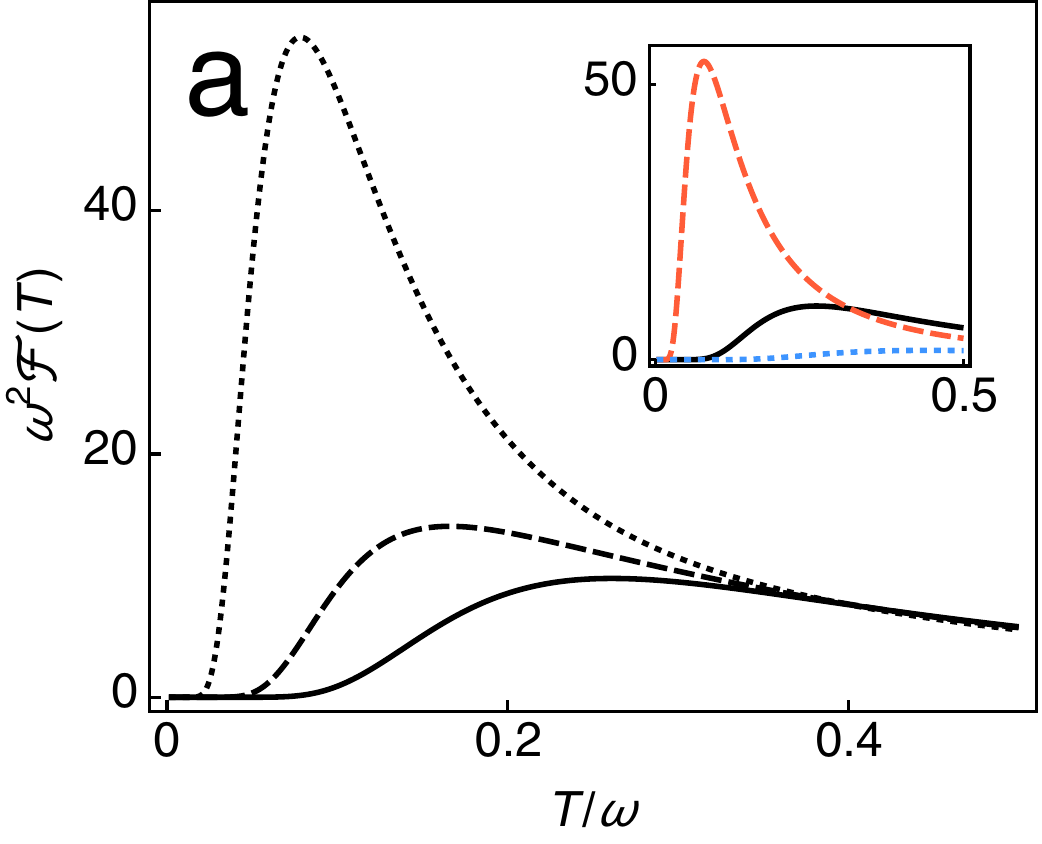}
		\includegraphics[width=0.49\textwidth]{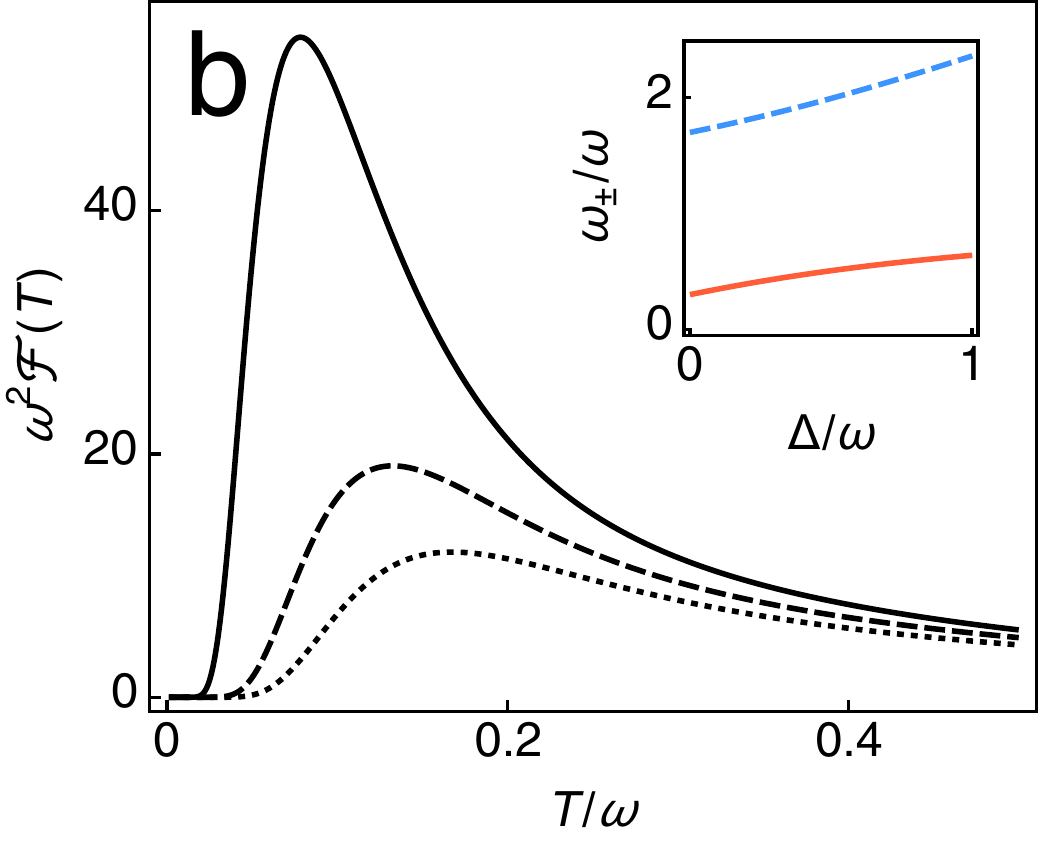}
	\end{center}
	\caption[fig]{\textbf{(a)} $\mathcal{F}(T)$ of the bipartite probe in equation \eqref{eq-H-DoubleWell} versus temperature for different interaction strengths; namely $ J = 0 $ (solid), $ J = 0.4 $ (dashed), and $ J = 0.7 $ (dotted). Here, $ U = \Delta = 0 $. In the inset, the sensitivity of two uncoupled harmonic oscillators at frequency $ \omega $ (solid black) is compared with that of the individual normal modes of two coupled oscillators at frequency $ \omega $ and interaction $ J = 0.7 $, i.e., $ \mathcal{F}(T,\omega_-) $ (dashed red) and $ \mathcal{F}(T,\omega_+) $ (dotted blue). \textbf{(b)} Same as in (a) for $ J = 0.7 $, $ U = 0 $, and different detuning values; specifically, $ \Delta = 0 $ (solid), $ \Delta = 0.5 $ (dashed), and $ \Delta = 1 $ (dotted). In the inset, the two normal-mode frequencies of equation \eqref{eq3:normal-mode} are plotted as a function of $ \Delta $.}
	\label{fig3-3}
\end{figure*}

Owing to the coupling $ J $, the two modes are correlated when at thermal equilibrium. As a result, their global QFI is \textit{not} additive. Nevertheless, by using a `mode-mixing' transformation one can bring such Hamiltonian into the non-interacting form $ \hat{H} = \sum_{l\in\{+,-\}} \omega_l ~ \hat{b}_l^{\dagger}\hat{b}_l$, so that
\begin{equation}
\mathcal{F}(T) = \mathcal{F}_{\text{ho}}(T,\omega_+) + \mathcal{F}_{\text{ho}}(T,\omega_-),    
\end{equation}
and $ \mathcal{F}_{\text{ho}}(T,\omega_\pm) $ corresponds to equation \eqref{eq-ho-temp-sensitivity} evaluated at the normal-mode frequencies  
\begin{equation}
\omega_{\pm} = 1 + \frac12
\left(\Delta \pm \sqrt{\Delta^2 + 4 J^2}\right).
\label{eq3:normal-mode}
\end{equation}

In figure \ref{fig3-3} we plot the impact of tuning the internal interaction $ J $ and the detuning $ \Delta $ on $ \mathcal{F}(T) $. As it can be seen, increasing the interaction strength $ J $ significantly enhances the thermal sensitivity; particularly at low temperatures. This can be easily explained from equation \eqref{eq3:normal-mode}---at larger interaction $ J $, $ \omega_- $ decreases, thus boosting the contribution of $ \mathcal{F}(T,\omega_-) $ to the total QFI. On the other hand, $ \omega_+ $ increases with $ J $, which reduces $ \mathcal{F}(T,\omega_+) $. As should be expected, at low temperatures the sensitivity of the probe is indeed dominated by the contribution of the fundamental mode \cite{1367-2630-19-10-103003}, which explains the net enhancement [see inset of figure \ref{fig3-3}(a)]. On the contrary, increasing the detuning between the modes at fixed coupling strength causes \textit{both} $ \omega_+ $ and $ \omega_- $ to grow, which worsens the overall thermal sensitivity at any temperature [see figure \ref{fig3-3}(b)]. A very similar argument is used in section \ref{sec:strong_coupling_thermometry} to explain the dissipation-driven enhancement of the low-temperature thermal sensitivity of a Brownian thermometer. 

Even if we have just looked at a simple example, the general take-home message is that internal interactions can be levered to substantially increase thermometric sensitivity. For completeness, let us also mention that ``switching on'' the non-linear term proportional to $ U $ in equation \eqref{eq-H-DoubleWell} does not have a significant impact in the qualitative features of the QFI, as revealed by numerical inspection. While $ \mathcal{F}(T) $ generally decreases by increasing the self interaction $ U $, the effects are minor. This is specially so at low temperatures, due to the fact that the first and second lowest energy levels---the most relevant ones for near-ground-state thermometry---are insensitive to $ U $ \cite{1367-2630-19-10-103003}.

\begin{figure}
	\centering
	\includegraphics[width=0.49\textwidth]{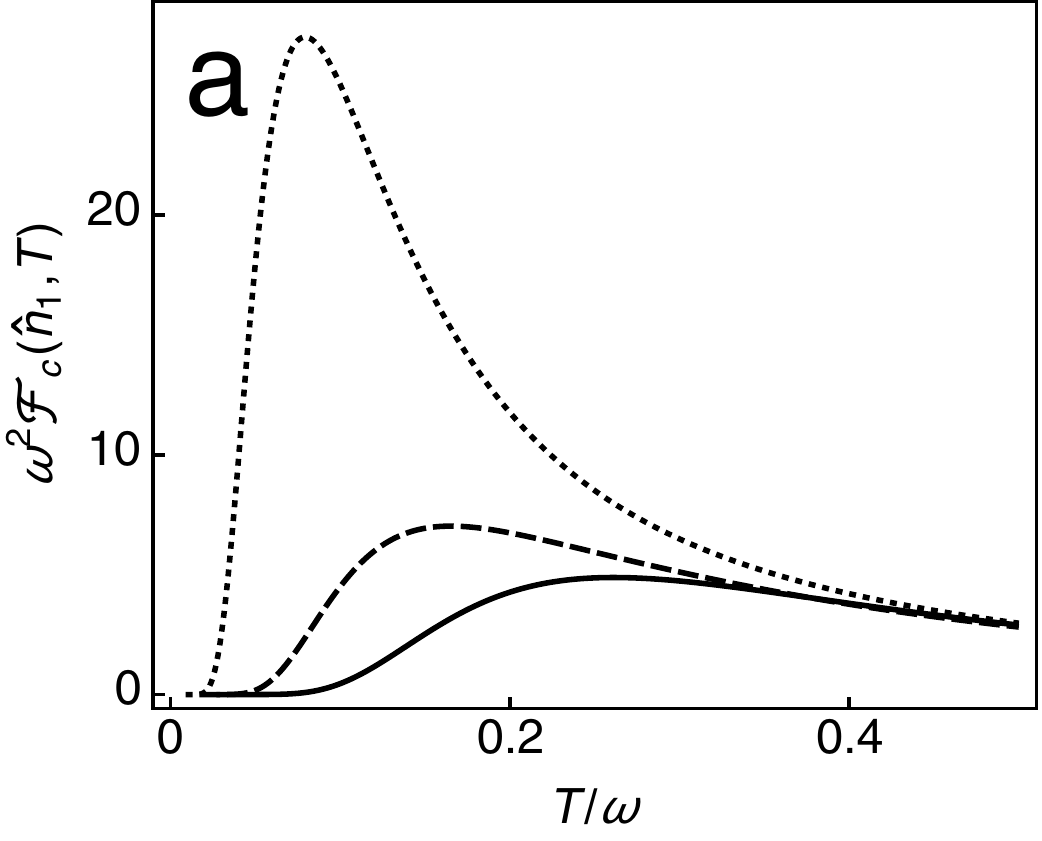}
	\includegraphics[width=0.49\textwidth]{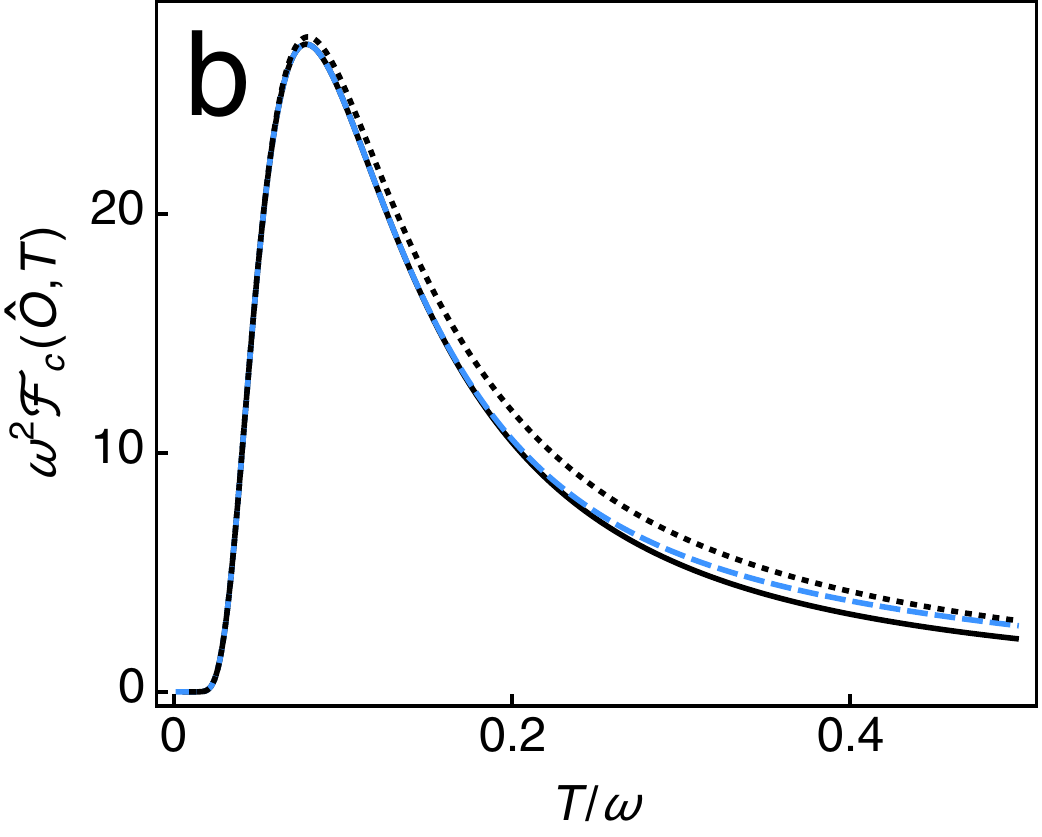}
	\caption[fig]{\textbf{(a)} Fisher information $ \mathcal{F}_c(\hat{n}_1,T) $ as a function of $ T $ for the local measurement $ \hat{n}_1\otimes\mathbbm{1}_2 $. As in figure \ref{fig3-3}(a), $ \Delta = U = 0 $, and the coupling strengths are $ J = 0 $ (solid), $ J = 0.4 $ (dashed), and $ J = 0.7 $ (dotted). \textbf{(b)} Fisher information versus $ T $ at fixed coupling $ J = 0.7 $. The sensitivity per oscillator $ \frac12\mathcal{F}_c(\hat{n}_1 + \hat{n}_2,T) $ (solid black) is compared with $ \mathcal{F}_c(\hat{n}_1,T) $ (dotted black), while the QFI per oscillator $ \frac12\mathcal{F}(T) $ (dashed blue) has been superimposed for comparison.}
	\label{fig3-4}
\end{figure}

In general, building an estimate with uncertainty $ \sim 1/\sqrt{\mathcal{F}(T)} $ would require to perform global measurements on both parties. But, which is the maximum precision of temperature estimates from \textit{local} measurements on a single party? To answer this, in figure \ref{fig3-4}(a) we plot the Fisher information of the local occupation number $ \hat{n}_1\otimes\mathbbm{1}_2 $, where $ \hat{n}_i \coloneqq \hat{a}_i^\dagger\hat{a}_i $. By comparing figures \ref{fig3-4}(a) and \ref{fig3-3}(a), we first note that the qualitative behaviour of local and global sensitivities is, indeed, identical. Remarkably, in figure \ref{fig3-4}(b) we see that $ \mathcal{F}_c(\hat{n}_1\otimes\mathbbm{1}_2 , T) $ can be \textit{larger} than $ \frac12\mathcal{F}_c(\hat{n}_1 + \hat{n}_2, T) $. Note that this is simply a consequence of the Fisher information not being additive when the probes interact, i.e., $ \mathcal{F}_c(\hat{n}_1+\hat{n}_2,T) \neq \mathcal{F}_c(\hat{n}_1\otimes\mathbbm{1}_2,T) + \mathcal{F}_c(\mathbbm{1}_1\otimes\hat{n}_2,T) $. Nonetheless, one obviously always finds that $ \mathcal{F}_c(\hat{n}_1 + \hat{n}_2, T ) \leq \mathcal{F}(T) $.  

\subsection{Criticality and thermometry on cold quantum gases} \label{sec3:criticality}

Strongly correlated quantum many-body systems can be very sensitive platforms for temperature measurements. Indeed, it is well known that phase transitions can be harnessed for metrology \cite{PhysRevA.78.042105,PhysRevA.76.062318,PhysRevA.75.032109}. To intuitively understand why, let us go back to equation \eqref{eq-Fid-QFI}: When being driven through a phase transition the state of a critical system changes dramatically, which translates into a very low fidelity between the states on both sides of the critical point. This increased distinguishability results in a high sensitivity to fluctuations in the driving parameter [e.g., $ \xi $ in equation \eqref{eq-Fid-QFI}] and, eventually, in a diverging QFI. Concretely, many-body systems undergoing classical phase transitions---generated by temperature driving---display a diverging heat capacity at the critical point, which can lead to extremely precise thermometry \cite{PhysRevLett.99.100603, PhysRevE.51.1006, PhysRevA.20.1608}. 

Our focus here is, however, on \textit{quantum} phase transitions---occurring as a function of non-thermal parameters, such as an external magnetic field---in many-body systems at very low, but finite, temperature.  It is very natural to ask whether there is an interplay between quantum criticality and the ultimate thermometric precision bounds. At finite temperatures, critical points expand into smoother crossovers and yet, by being driven through such regions, a many-body system can experience drastic changes. Therefore, it is to be expected that at low enough temperatures, improved metrological bounds become possible due to a combination of quantum and thermal correlations \cite{PhysRevA.78.042106,PhysRevA.90.022111,1367-2630-17-5-055020,de2016local,guo2015ring,PhysRevA.94.042121,1367-2630-17-5-055020}. 

\begin{figure}
	\centering
	\includegraphics[width=0.8\linewidth]{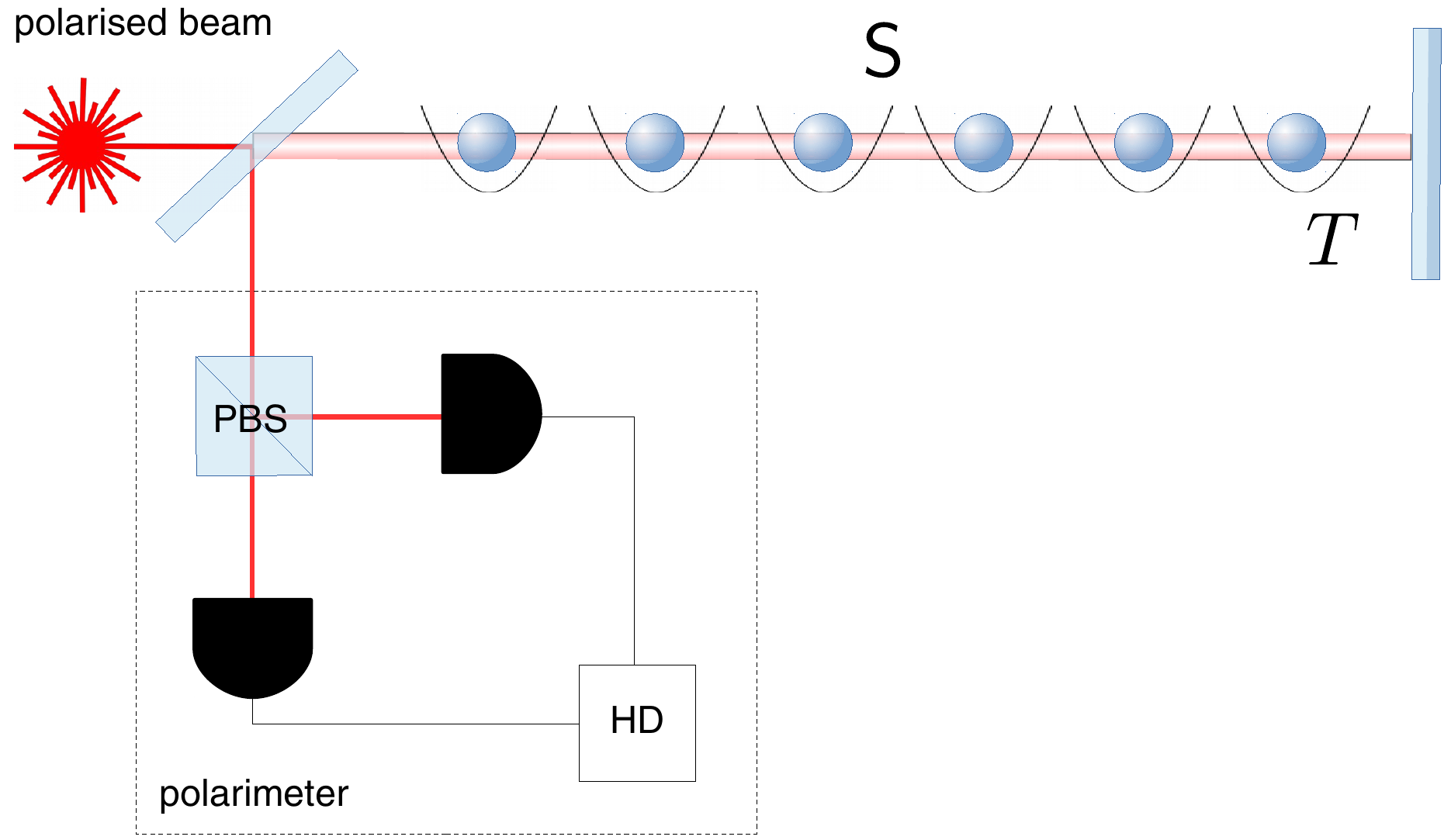}
	\caption{Schematic representation of the setup for measuring the collective angular momenta of an ultra-cold lattice gas using Faraday interferometry. A strongly polarized laser beam is directed onto the many-body system of interest and its reflection on a mirror forms a standing wave. Homodyne detection on the scattered light reveals its polarization, which encodes information about the components of the global angular momenta of the system. In turn, this allows for non-demolition temperature estimation; in some cases, with nearly minimal statistical uncertainty.}
	\label{fig-HD}
\end{figure}

Unfortunately, strongly correlated quantum many-body systems are also very fragile and difficult to prepare and, specially, to measure \cite{RevModPhys.86.153,Bloch2005,PhysRevLett.107.255301,schneider2012experimental,barreiro2011open,schmied2011quantum,diehl2008quantum,campbell2006imaging,greiner2002quantum}. In fact, many of the available measurement schemes are \textit{destructive} \cite{Bloch2005,schneider2012experimental,stewart2008using,leanhardt2003cooling,gati2006noise,PhysRevLett.98.240402,PhysRevA.85.023623}. However, proposals for \textit{non-demolition} measurements in many-body systems simulated in ultra-cold lattice gases have been put forward in recent years \cite{Eckert2007a,PhysRevX.4.021045,PhysRevLett.80.3487,colangelo2013quantum,PhysRevLett.105.093602,sewell2013certified}. For instance, exploiting the atom-light interface in Faraday interferometry has been suggested as an efficient non-destructive probing technique (see figure \ref{fig-HD}). Essentially, the correlations among the sites of the atomic lattice gas can be imprinted on the state of a probing beam of light (see e.g., references \cite{Eckert2007a,RevModPhys.90.035005,RevModPhys.82.1041}); subsequent measurements on the light can reveal information about the many-body system without disturbing it significantly. Due to the temperature-dependence of these correlations, the statistics of light would, in principle, also enable minimally invasive thermometry. In the neighbourhood of a quantum phase crossover, this could be also particularly precise \cite{1367-2630-17-5-055020}.

\subsubsection{Thermometry on the XY model.}\label{sec-XY}

\begin{figure}
	\centering
	\includegraphics[width=0.65\linewidth]{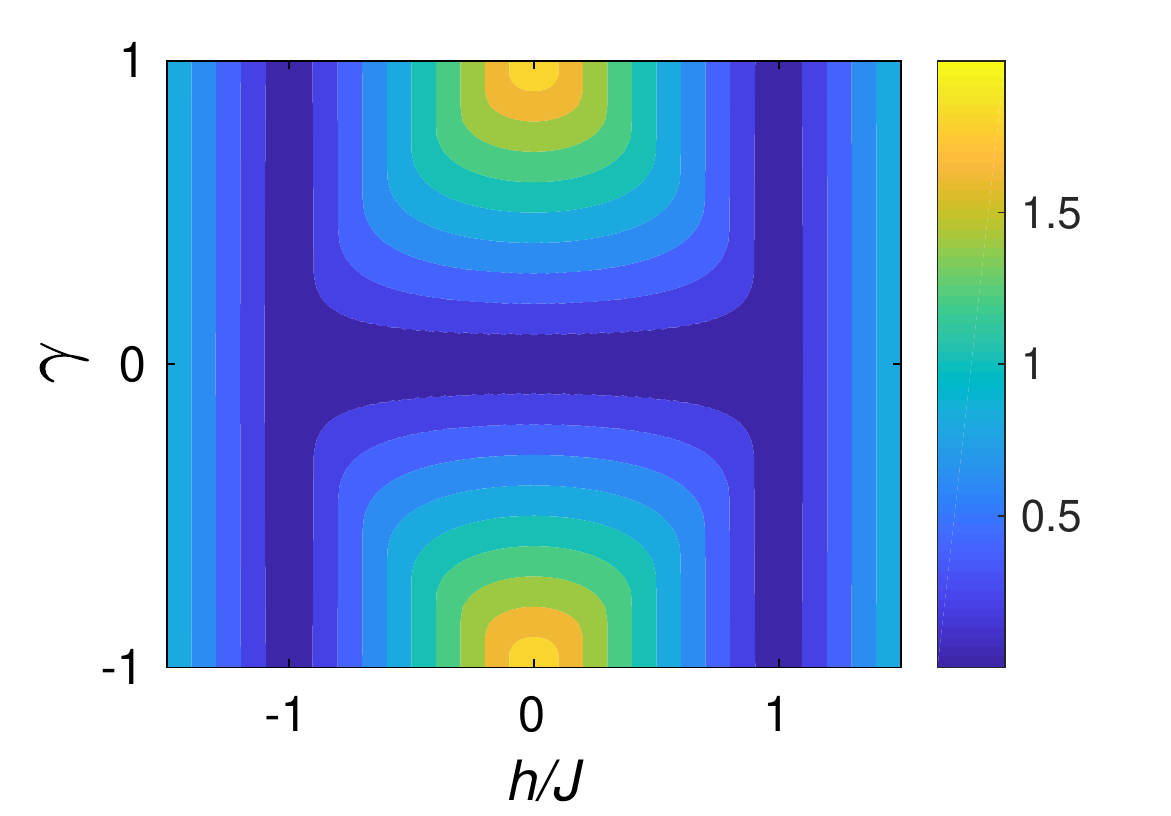}
	\caption{Gap $ \Delta E $ of the XY model versus the transverse field $ h $ and the anisotropy $ \gamma $. The system remains gapped except in the critical regions, which includes the second-order phase transitions at $ h/J = \pm 1 $ and the critical phase at $ \gamma = 0 $.}
	\label{fig-XY-Phase}
\end{figure}

To better understand the interplay between quantum and thermal fluctuations in many-body systems, let us consider the paradigmatic XY model. It consists of a chain of $ N $ interacting spin-$ 1/2 $ particles in a transverse field \cite{lieb1961two}, and stands as one of the most extensively studied models \cite{its2005entanglement,PhysRevLett.90.227902,PhysRevA.2.1075,osterloh2002scaling,PhysRevB.70.064405}, since it is both analytically solvable \cite{lieb1961two,PhysRevA.3.786,PhysRevA.2.1075,sachdev2011quantum} and experimentally realisable \cite{PhysRevA.79.012305,islam2011onset,kim2010quantum,PhysRevB.82.060412,Simon2011}. The XY Hamiltonian reads
\begin{align}
\hat{H} = -J\sum_{l=1}^{N}\left[\frac{1+\gamma}{2}\,{\hat \sigma}^{x}_{l}\,{\hat \sigma}^{x}_{l+1}+\frac{1-\gamma}{2}\,{\hat \sigma}^{y}_{l}\,{\hat \sigma}^{y}_{l+1}\right]-h\sum_{l=1}^N{\hat \sigma}_{l}^{z}.
\label{eq-XYModel}
\end{align}
Here, ${\hat \sigma}_l^\alpha$ are the Pauli matrices ($ \alpha\in\{x,y,z\} $) at site $ l $ in the chain. The `anisotropy parameter' $ -1\leq \gamma \leq 1 $ allows to interpolate between the Ising model ($ \gamma = \pm 1 $) and the XX model ($ \gamma = 0 $) \cite{sachdev2011quantum}. The parameter $ h $ stands for the strength of an external transverse magnetic field. In turn, the nearest-neighbour coupling $ J $ is taken as positive (ferromagnet), although analogous results would hold for $ J < 0 $. Assuming periodic boundary conditions allows for an exact solution over the whole parameter space, since the Hamiltonian \eqref{eq-XYModel} can be then mapped onto a collection of non-interacting fermions. Namely, 
\begin{equation}
{\hat H} = \sum_{k} \epsilon_{k} {\hat \lambda}_{k}^{\dag}{\hat \lambda}_{k}.
\label{eq-XYJordan}
\end{equation}
This is achieved through a Jordan--Wigner transformation, followed by a Bogoliubov transformation \cite{lieb1961two, Mikeska}. ${\hat \lambda}_{k}$ ($ {\hat \lambda}^{\dagger}_k $) are annihilation (creation) operators with a fermionic algebra. Note that, in general, each of these $ \hat{\lambda}_k $ is a combination of the spin operators of all sites of the chain and hence, they are highly non-local. When it comes to the corresponding energies, these are
\begin{equation}
\epsilon_k = 2 J\sqrt{\left(\cos{k}- h/J\right)^2 + \left(\gamma\sin{k} \right)^2},
\label{positive-energy}
\end{equation}
with $ k = \frac{\pi}{N}(2j+1) $ and $ j \in \{-N/2,\dots,N/2-1\} $. Since all energies are positive, we can build excited states by merely filling the vacuum, or ground state, with free fermions. Let us focus on the \textit{energy gap} $ \Delta E = \min_{k} \,\epsilon_k $ of the system, which is particularly relevant for low-temperature thermometry \cite{hovhannisyan2017probing, paris2015achieving, de2016local}; we have plotted it in figure \ref{fig-XY-Phase}. As we can see, the gap closes only at critical lines of the model---i.e., at $ h/J = \pm 1 $---which corresponds to the Ising transitions where paramagnetic and ferromagnetic phases meet; and at $ \gamma = 0 $, and $ |h/J|\leq 1 $, which corresponds to the anisotropic transition in the XX model.

\begin{figure}[t!]
	\centering
	\includegraphics[width=\linewidth]{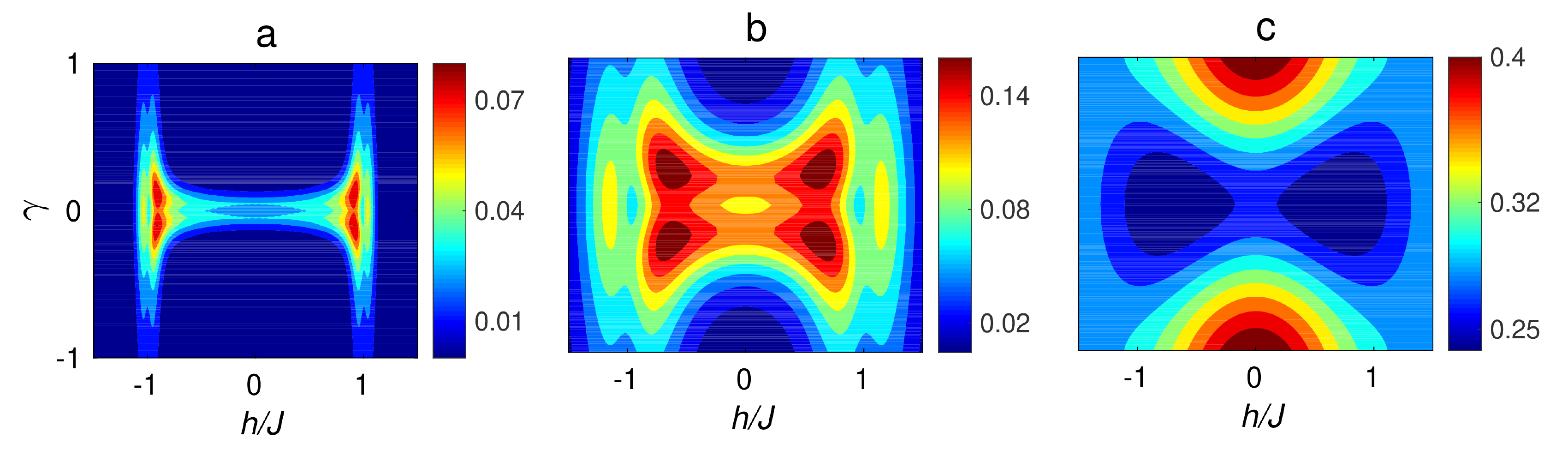}
	\caption{Maximum signal-to-noise ratio (SNR) $ (T/\Delta T)^2 \leq T^2 \mathcal{F}(T,\hat{\varrho}) $ versus the transverse field $ h $ and the anisotorpy $ \gamma $ in the XY model (compare with figure \ref{fig-XY-Phase}). \textbf{(a)} At low temperatures ($ T/J = 0.05 $), the SNR is negligible everywhere but in the vicinity of the critical regions. At higher temperatures, the overall responsiveness grows and the most sensitive regions move away from criticality. Specifically $ T/J = 0.2 $ for panel {\bf (b)} and $ T/J = 1 $ for panel {\bf (c)}. All plots are normalised to $ N $ (here, $ N=50 $).}
	\label{fig-XY-sensitivity}
\end{figure}

As already mentioned, a higher thermometric precision at low temperatures is to be expected in the vicinity of these critical regions due to the vanishing gap. Indeed, in light of equation \eqref{eq-XYJordan}, the thermal state $ \hat{\varrho}(T) $ can be written as
\begin{equation}
{\hat \varrho}(T) = \mathcal{Z}^{-1}  e^{-{\hat H}/T} = \bigotimes_k {\hat \varrho}_k(T),
\label{XYthermal}
\end{equation}
where $ {\hat \varrho}_k(T) $ denotes the thermal state of the $ k $-th free fermion; that is,
\begin{equation}
{\hat \varrho}_k(T) = \frac{\ket{0}_k\bra{0} + e^{-\epsilon_k / T}\ket{1}_k\bra{1}}{1+e^{-\epsilon_k/T}}.
\label{Thermal-Product-State}
\end{equation}
Here, $ \ket{0}_k $ ($ \ket{1}_k $) refers to the empty (occupied) state of the $ k $-th free fermion. Exploiting the additivity of the QFI we find that
\begin{align}
\mathcal{F}(T,{\hat \varrho})= \sum_{k} \mathcal{F}(T,{\hat \varrho}_k)
= \sum_{k} \left(\frac{\epsilon_k}{T^{2}}\right)^2  n_k (1-n_k),
\label{Eq:QFI_Allk}
\end{align}
with $ n_k = (1+e^{\epsilon_k/T})^{-1} $. Note that we have also dropped the argument $ T $ from the states to lighten notation. Using equation \eqref{Eq:QFI_Allk}, we plot the maximum signal-to-noise ratio $ N^{-1}(T /\Delta T)^2 \leq T^2 \mathcal{F}(T,{\hat \varrho}) $ in figure \ref{fig-XY-sensitivity}. As it can be seen, the strong suppression of thermal fluctuations at very low temperatures yields a vanishing signal-to-noise ratio everywhere in the space of parameters, except in the immediate neighbourhood of the critical regions---essentially, the system lies in its ground state unless the gap becomes small enough so that thermal fluctuations can cause measurable changes. On the contrary, at higher temperatures the baseline signal-to-noise ratio is also larger and the fact that the gap of the many-body system closes is not as relevant.

As we know from \ref{sec-Thermometry-Precision}, in order to saturate those precision bounds, we must perform projective measurement onto the many-body energy eigenstates. These generally require complex non-local manipulations besides being destructive. Alternatively, we can resort to the temperature-dependent statistics of components of the total angular momentum (i.e., $ {\hat J}_{\alpha} = \sum_{l = 1}^{N} {\hat \sigma}_{l}^{\alpha} $). Specifically, according to the error-propagation formula \eqref{eq-error-propagation} $ {\hat J}_x^{\, 2} $ can yield nearly optimal estimates at low $ T $ within the ferromagnetic region \cite{1367-2630-17-5-055020}. Furthermore, this is the type of observable that can be measured using the aforementioned non-demolition technique based on Faraday spectroscopy. 

\subsubsection{Scaling of the QFI with temperature and system size.}\label{sec3:size_temp_scaling} It is also interesting to study the functional dependence of the QFI with $ N $ and $ T $. To that end, we can follow reference \cite{PhysRevA.94.042121} and carefully take the large $ N $ and low-$ T $ limits in equation \eqref{Eq:QFI_Allk}. We assume $ \gamma = 0 $ and thus, focus on the XX model. This yields 
\begin{align}
\mathcal{F}(T,{\hat \varrho}) \simeq C \frac{N}{J T \sqrt{1-(h/J)^2}},
\label{eq:qfi_approximation}
\end{align}
where the fitting parameter $ C \approx 0.53 $ is independent of $ \beta $, $ J $, and the chain size $ N $. As a rule of thumb, we can expect the approximation to hold so long as $ T < J-h $. Closer to the critical point, i.e. when $ J-T < h < J $, the QFI behaves instead as
\begin{equation}
\mathcal{F}(T = J-h,{\hat \varrho})\approx C \frac{N}{\sqrt{2\,J\,T^3}}
\end{equation}

Therefore, the low-$ T $ QFI of the XX model in the thermodynamic limit grows \textit{extensively} in the system size (i.e., $ \mathcal{F}(T,\hat{\varrho}) \sim N $) and diverges as $ \mathcal{F}(T,\hat{\varrho}) \sim 1/T $ as $ T \rightarrow 0 $. This implies $ (T/\Delta T)^2 \sim T $. Closer to the ferromagnetic-paramagnetic crossover, however, the scaling is \textit{enhanced} to $ (T/\Delta T)^2 \sim \sqrt{T} $. In any case, note that the low-temperature behaviour is \textit{power-law-like rather than exponential}, due to the ``gaplessness'' of the model (cf. figure \ref{fig-XY-Phase}). As we shall see in detail in section \ref{sec:exponential_ineff_gapped} below, this polynomial scaling also appears when estimating low temperatures from local measurements on other gapless systems \cite{hovhannisyan2017probing}. For completeness, let us mention that, at low $ T $ and large $ N $, the temperature QFI in the Ising model behaves as
\begin{align}
\mathcal{F}(T) \simeq C^{\prime} \frac{ N }{JT} \qquad(C^{\prime}\simeq 1.05).
\label{eq:qfi_approximation_Ising_Critical}
\end{align}

We remark that the exact same size and temperature scaling of \eqref{eq:qfi_approximation} and \eqref{eq:qfi_approximation_Ising_Critical} has been observed in the QFI of a tight-binding fermionic system in equilibrium with periodic boundaries \cite{hofer2017fundamental}. Likewise, polynomial temperature scalings in the signal-to-noise appear very generally in ideal ultra-cold quantum gases, as shown in reference \cite{PhysRevA.88.063609}. There, the problem of the \textit{simultaneous} estimation of temperature and chemical potential $ \mu $ of a bosonic or fermionic gas in a grand-canonical state $ \hat{\varrho}(T,\mu) \propto e^{-(\hat{H}-\mu\hat{N})/T} $ was studied (where $ \hat{N} = \sum_l \hat{a}_l^\dagger \hat{a}_l $ is the number operator). Both homogeneous and harmonically trapped gases were considered, even when undergoing Bose--Einstein condensation.

\subsection{The role of conserved quantities} \label{sec-GGS}

\begin{figure}[t]
	\begin{center}
		\includegraphics[width=0.49\columnwidth]{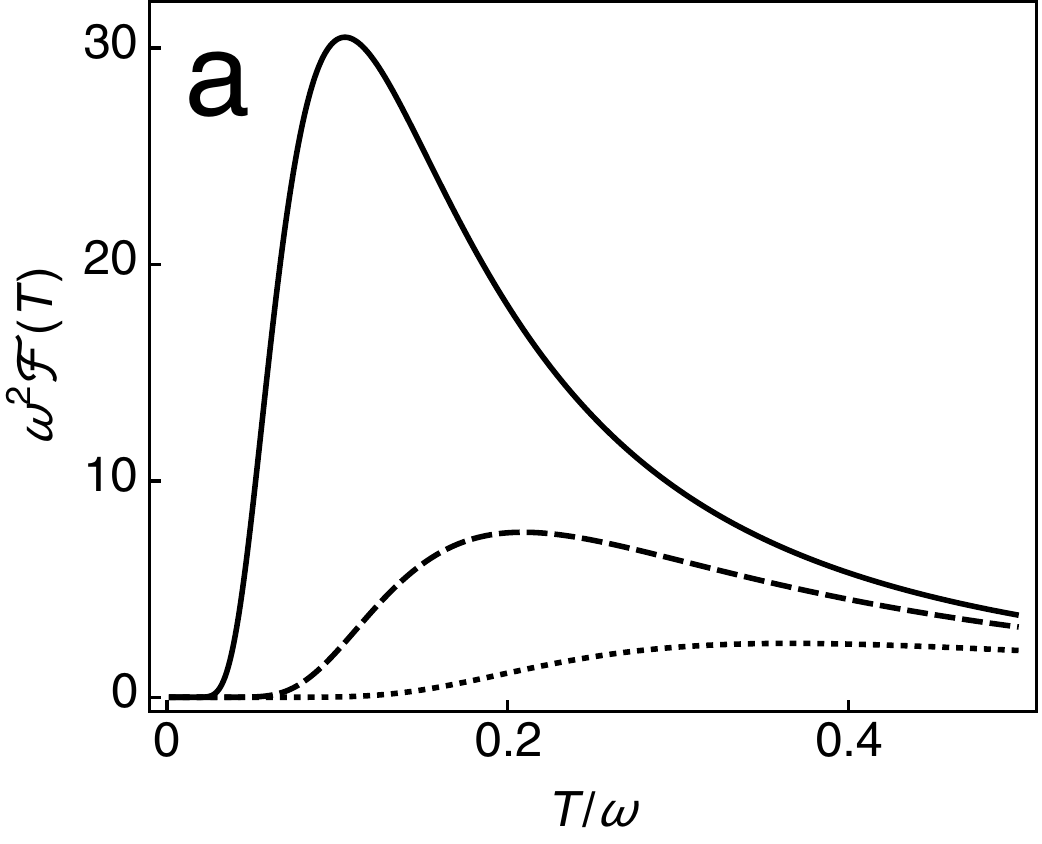}
		\includegraphics[width=0.49\columnwidth]{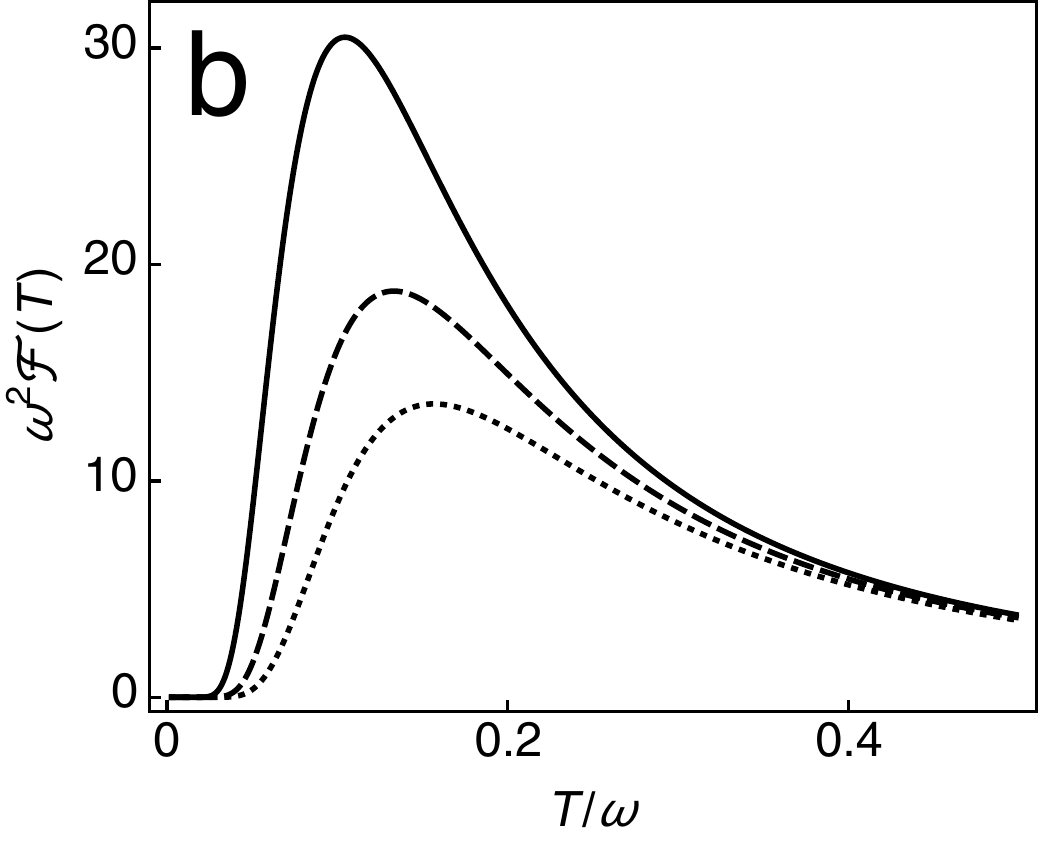}
	\end{center}
	\caption{QFI versus $ T $ for the double-well potential of equation \eqref{eq-H-DoubleWell} loaded with a cold atomic gas of $ N $ atoms for \textbf{(a)} $ U = 0 $ and tunnelling $ J = 0.2 $ (solid), $ J = 0.4 $ (dashed), and $ J = 0.7 $ (dotted). Note that for vanishing self-interaction, these do not depend on $ N $. In \textbf{(b)} the self-interaction is $ U = 5\times 10^{-3} $, $ J = 0.2 $, and $ N = 50 $ (dashed) and $ N = 100 $ (dotted). The curve for $ U = 0 $ and $ J = 0.2 $ (solid) has been super-imposed as a reference.}
	\label{fig1N}
\end{figure}

Imposing additional constraints on an equilibrium system can have drastic effects in the achievable precision of temperature estimates. We illustrate this here with two examples. As we shall see, conservation laws might be \textit{exploited} in practice to improve the precision of low-temperature sensing. 

Firstly, let us go back to the Hamiltonian in equation \eqref{eq-H-DoubleWell}; we shall now think of it as a two-mode Bose--Hubbard model, i.e., a double well loaded with a cold atomic gas, in which the local occupation numbers $ \langle \hat{a}_i^\dagger\hat{a}_i \rangle = n_i $ might change due to tunnelling (of strength $ J $) while the total number of particles $ N = n_1 + n_2 $ is fixed. Additionally, let us set $ \Delta = 0 $. To reflect particle conservation, the state of the system is now entirely contained within the $ N $-particle sector of the Hilbert space. As we shall see, this gives rise to qualitative changes in the behaviour of the QFI. 

Following reference \cite{1367-2630-19-10-103003}, we may introduce the Schwinger operators $ \hat{S}_x \coloneqq \frac{1}{2}\big( \hat{a}_1^\dagger \hat{a}_1 - \hat{a}_2^\dagger \hat{a}_2 \big) $, $ \hat{S}_y \coloneqq \frac{i}{2} \big( \hat{a}_1^\dagger \hat{a}_2 - \hat{a}_2^\dagger \hat{a}_1 \big) $, and $ \hat{S}_z \coloneqq \frac{1}{2}\big( \hat{a}_1^\dagger \hat{a}_2 + \hat{a}_2^\dagger \hat{a}_1 \big) $, which obey angular momentum commutation relations $ \big[\hat{S}_x,\hat{S}_y\big] = i\,\hat{S}_z $. Equation \eqref{eq-H-DoubleWell} can be thus conveniently rewritten as
\begin{equation}
\label{LMG}
\hat{H} = -2 J \hat{S}_z + U \hat{S}_x^2 + U \frac{\hat{N}^2}{4} + \omega \hat{N}. 
\end{equation}
Crucially, note that due to particle conservation, the last two terms in the right-hand side of \eqref{LMG} evaluate to a constant and can therefore be discarded. Using a Holstein--Primakoff transformation \cite{PhysRev.58.1098,PhysRevLett.93.237204} one can define collective bosonic operators $ \hat{a} $ and $ \hat{a}^\dagger $, such that
\begin{equation}
\hat S_x \simeq \frac{\sqrt{N}}{2}\left( \hat{a} + \hat{a}^\dagger \right) \qquad\text{and}\qquad \hat S_z \simeq \frac{N}{2} - \hat{a}^\dagger \hat{a}.
\label{eq3:primakoff}
\end{equation}
Importantly, equation \eqref{eq3:primakoff} holds only to lowest order in $ \hat{a}^\dagger\hat{a}/N $ and hence, $ N $ must be sufficiently large. A suitable rotation then brings the Hamiltonian into the simple form
\begin{equation}
\label{DWmapped}
\hat H = \Omega \left (\hat b^\dagger \hat b + 1/2 \right ) - J(N+1)
\end{equation}
with $ \Omega = \sqrt{2 J (2 J + N U)} $ \cite{1367-2630-19-10-103003}. The QFI for such system can be thus approximated using equation \eqref{eq-ho-temp-sensitivity}, and evaluates to 
\begin{equation}
\label{DWQFI}
\mathcal{F}(T) \simeq \mathcal{F}_{\text{ho}}(T,\Omega) = \frac{J\left(2J + N U \right)}{2T^4} \csch^2{\left( \frac{1}{T} \sqrt{J\left( J + \frac{N U}{2} \right)} \right)}.
\end{equation}

Focusing on the strict limit of $ U \rightarrow 0 $ (tunnelling-dominated regime), in which \eqref{DWQFI} is exact, we can see that $ \mathcal{F}(T) $ \textit{decreases} as the tunnelling rate increases [cf. figure \ref{fig1N}(a)]. This is in sharp contrast with the tunnelling-driven thermometric \textit{enhancement} described in section \ref{sec-Coupled-ho-Thermal} [compare figures \ref{fig3-3}(a) and \ref{fig1N}(a)], and is due to the fact that the effective frequency now grows with the tunnelling as $ \Omega \sim 2 J $. Allowing for small but finite self-interaction $ U $ further reveals that the QFI decreases as the system is scaled up in size [see figure \ref{fig1N}(b)].

As a second illustrative example, we discuss the sensitivity of a probe formed of $ N $ identical two-level atoms arranged in a ``ring structure''. These atoms are closely spaced and thermalise with the surrounding electromagnetic field in equilibrium \cite{guo2015ring}. The probe Hamiltonian is thus $ {\hat H} = \frac{\omega}{2}\sum_{l=1}^{N} {\hat \sigma}_l^z $, with the periodic boundary condition $ {\hat \sigma}_{N+l}^{z} = {\hat \sigma}_l^{z} $. As we shall see in section \ref{sec:coherence-entanglement}, the \textit{collective} dissipation of the atoms in such system can be modelled with a Markovian master equation \cite{higgins2014superabsorption}. In particular, the weak interactions with a common environment give rise to a coherent radiative coupling mechanism. This is captured by the Hamiltonian-like term
\begin{align}
\Delta{\hat H} = \Omega \sum_{l = 1}^{N} \left({\hat \sigma}_l^{+}~{\hat \sigma}_{l+1}^{-}+{\hat \sigma}_l^{-}~{\hat \sigma}_{l+1}^{+}\right),
\label{eq3:ring_interaction}
\end{align}
appearing in the corresponding quantum master equation (cf. section \ref{sec:markovian}). Here, $ {\hat \sigma}_{\pm}^{l} = \frac12({\hat \sigma}_x^l \pm i {\hat \sigma}_y^l) $ are the raising/lowering operators at site $ l $ and the effective coupling strength $ \Omega $ depends on the distance between neighbouring sites and the relative alignment between the atomic dipoles and the external field (for convenience $ -1/2 < \Omega/\omega < 1/2 $). 

Tackling the steady-state thermal sensitivity of this system via the `implicit interaction' approach entails replacing $ \hat{H} \mapsto \hat{H}_\text{eff} = \hat{H} + \Delta\hat{H} $ in the thermal state $ \hat{\varrho}(T) $. Due to the symmetries of the system we can diagonalise $ \hat{H}_\text{eff} = \sum_{S, S_z} \epsilon_{S,S_z}\,\ketbra{S,S_z}{S,S_z} $ in the simultaneous eigenstates of the total spin $ \hat{S}^2 $ and its $z$ component $ \hat{S}_z = \sum_{l=1}^N\hat{\sigma}_i^z $. Limiting ourselves to, e.g., the $ S = N/2 $ sector of the Hilbert space, leaves us with $ \hat{\varrho}'(T) \propto \sum_{S_z=-N/2}^{N/2}e^{-\epsilon_{N/2,S_z}/T}\ketbra{N/2,S_z}{N/2,S_z} $. The corresponding QFI can be seen to increase monotonically as $ \Omega/\omega \rightarrow -\frac12 $, becoming particularly large at low temperatures \cite{guo2015ring}. This is, again, in striking contrast with what we observed in figure \ref{fig-XY-sensitivity}(a). Indeed, notice that setting $ \omega = -2h $ and $ \Omega = -J $ turns $ \hat{H}_\text{eff} $ into the XX Hamiltonian of equation \eqref{eq-XYModel} ($ \gamma = 0 $). In particular, the range $ \vert \Omega/\omega \vert < \frac12 $ corresponds to the paramagnetic phase $ \vert h/J \vert > 1 $. There, the QFI of the unrestricted thermal state $ \hat{\varrho}(T) $ is vanishingly small at low temperatures, due to the finite gap of the system. In section \ref{sec:coherence-entanglement} below, we shall briefly revisit this model to comment on its dynamical features. 

\subsection{Information--disturbance trade-off}\label{sec-inf-dist}

To conclude this section, let us briefly discuss the issue of the back-action on the measured system. We have already pointed out that a good thermometric protocol should ideally be both precise and `minimally disturbing' (see discussion in section \ref{sec-XY}). However, one can intuitively expect very informative estimation protocols to be necessarily \textit{invasive} \cite{PhysRevA.96.012110}. By exploiting information--disturbance inequalities \cite{kretschmann2008information,maccone2006information,barnum2002information,PhysRevA.93.032134} this trade-off can be studied quantitatively. Although results differ depending on the disturbance measure chosen \cite{seveso2018trade}, let us here focus on `average information loss'. Defining the global unconditional QFI as
\begin{equation}
\mathcal{F}_\mathbf{\Pi}(T)\coloneqq\sum_{m}\,p_m(T)~\mathcal{F}(T,{\hat \varrho}_m),
\label{eq3:global_uncond_QFI}    
\end{equation}
the disturbance caused by $ \mathbf{\hat{\Pi}} $ may be thus gauged by the difference $ D(\mathbf{\hat{\Pi}},\hat{\varrho}(T))\coloneqq \mathcal{F}(T) - \mathcal{F}_\mathbf{\hat{\Pi}}(T) $. Recall from section \ref{sec2} that the notation $ \mathbf{\hat{\Pi}} $ stands for a POVM measurement that produces outcomes $ \hat{\varrho}_m $ with probability $ p_m(T) = \text{tr}\{\hat{\Pi}_m\,\hat{\varrho}\} $.

It is then natural to ask: \textit{What is the minimum disturbance $ D(\mathbf{\hat{\Pi}},\hat{\varrho}(T)) $ for a fixed amount of information $ \mathcal{F}_c(\mathbf{\hat{\Pi}},T) $ and which measurement achieves it?} Answering to the second part of the question, such measurements are referred-to as `efficient' and, for thermalised $ d $-dimensional systems, they belong to the so-called `semi-classical' type; i.e., those whose elements $ \hat{\Pi}_m $ commute with $ \hat{\varrho}(T) $ \cite{seveso2018trade}. The corresponding disturbance turns out to be equal to the extracted information, since $ \mathcal{F}_c(T,\mathbf{\hat{\Pi}}) \leq D(\mathbf{\hat{\Pi}},\hat{\varrho}(T)) $ \cite{PhysRevA.93.032134}.

In particular, the system under study could be bipartite $ \hat{\varrho}_{\mathsf{A B}}(T) $ and the allowed POVM measurements, local (e.g., $ \hat{\Pi}_m = \hat{\Pi}_m^{(\mathsf{A})}\otimes\mathbbm{1}^{(\mathsf{B})} $). In such restricted scenario, one could think of a two-step estimation protocol consisting of a projective measurement $ \mathbf{\hat \Pi}^{(\mathsf{A})}\otimes\boldsymbol{\mathbbm{1}}^{(\mathsf{B})} $ onto the eigenbasis of the SLD of the marginal $ \hat{\varrho}_\mathsf{A} = \text{tr}_\mathsf{B}\,\hat{\varrho}_{\mathsf{AB}}(T) $, followed by a second projection on the SLD of the residual post-measurement state on $ \mathsf{B} $, $ \hat{\varrho}_{\mathsf{B}\,\vert\,\hat{\Pi}_m^{(\mathsf{A})}} = \text{tr}_\mathsf{A}\,\hat{\varrho}_m(T) $. The corresponding (classical) Fisher information was dubbed ``LOCC QFI'' in \cite{PhysRevA.98.012115} which evaluates to
\begin{equation}
\mathcal{F}_{\mathsf{A}\rightarrow\mathsf{B}}(T) = \mathcal{F}(T,\hat{\varrho}_\mathsf{A}) + \mathcal{F}_{\mathbf{\Pi}^{(\mathsf{A})}\,\otimes\,\mathbbm{1}^{(\mathsf{B})}}(T).
\label{eq3:LOCC_QFI}    
\end{equation}
In the large-temperature limit, $ \mathcal{F}_{\mathsf{A}\rightarrow\mathsf{B}}(T) $ turns out to be related to the `quantum discord' \cite{PhysRevLett.88.017901,henderson2001classical} which is a quantifier of the disturbance caused by local measurements on a bipartite state, due to the sharing of non-classical correlations (for related measures of quantum correlations see, e.g., references \cite{RevModPhys.84.1655,adesso2016measures,farace2014discriminating,PhysRevA.92.042331}).

\section[Quantum thermometry out of equilibrium]{Quantum thermometry out of equilibrium}\label{sec4}

In this section, we focus on situations in which information about the temperature of a system may be acquired only by local measurements on a small accessible fraction. This scenario is particularly relevant when dealing with very large multi-particle composites in thermal equilibrium \cite{de2016local,correa2016low,de2017universal,correa2016low,de2017universal,miller2018mean}. We refer to the accessible fragment as the `probe' and, somewhat abusing of language, label the remaining inaccessible part as the `system'. Essentially, we are concerned with three aspects of such local thermometric schemes:
\begin{enumerate}
	\item How does the `thermodynamic uncertainty relation' $ \Delta \beta \Delta \hat{H} \geq 1 $, encountered in equilibrium thermometry, transform in presence of strong probe-sample correlations? 
	\item Do the internal interactions responsible for those correlations improve or undermine the accuracy of local thermometry? 
	\item How does the statistical uncertainty of local temperature estimates scale as $ T\rightarrow 0 $? 
\end{enumerate} 

Points (i) and (ii) are addressed in detail in sections \ref{sec:local_temperature_fluctuations} and \ref{sec:strong_coupling_thermometry}, respectively. 
In \ref{sec:exponential_ineff_gapped} we shall comment on recent results showing that low-$ T $ local thermometry is exponentially inefficient in gapped lattice systems while gapless spectra allow for a better power-law-like scaling 
\cite{hovhannisyan2017probing}. In the same section 
we also discuss low-temperature estimates from incomplete information. These can be thought-of as relying on \textit{global} POVM measurements with \textit{finitely many} possible outcomes, i.e., limited resolution \cite{hofer2017fundamental}. Finally, 
we discuss the practically relevant question of whether the relative error of local low-$ T $ thermometry can be kept from diverging as $ T\rightarrow 0 $ \cite{hovhannisyan2017probing,hofer2017fundamental}.

Finally, note that we do not intend to discuss here the `locality of temperature'; that is, whether or not a local temperature can be assigned to any given sub-lattice of a large many-body system, due to the exponential clustering of spatial correlations. For an updated review on this interesting topic see, e.g., Ch.~20 of \cite{binder2018thermodynamics}. Although results on the spatial distribution of correlations in thermal many-body systems are instrumental for some of the points presented below, we merely make statements about the `temperature-information' content of marginals of global thermal states of large systems.  

\subsection{Local temperature fluctuations.}\label{sec:local_temperature_fluctuations}

\begin{figure*}[t!]
	\centering
	\includegraphics[width=0.7\linewidth]{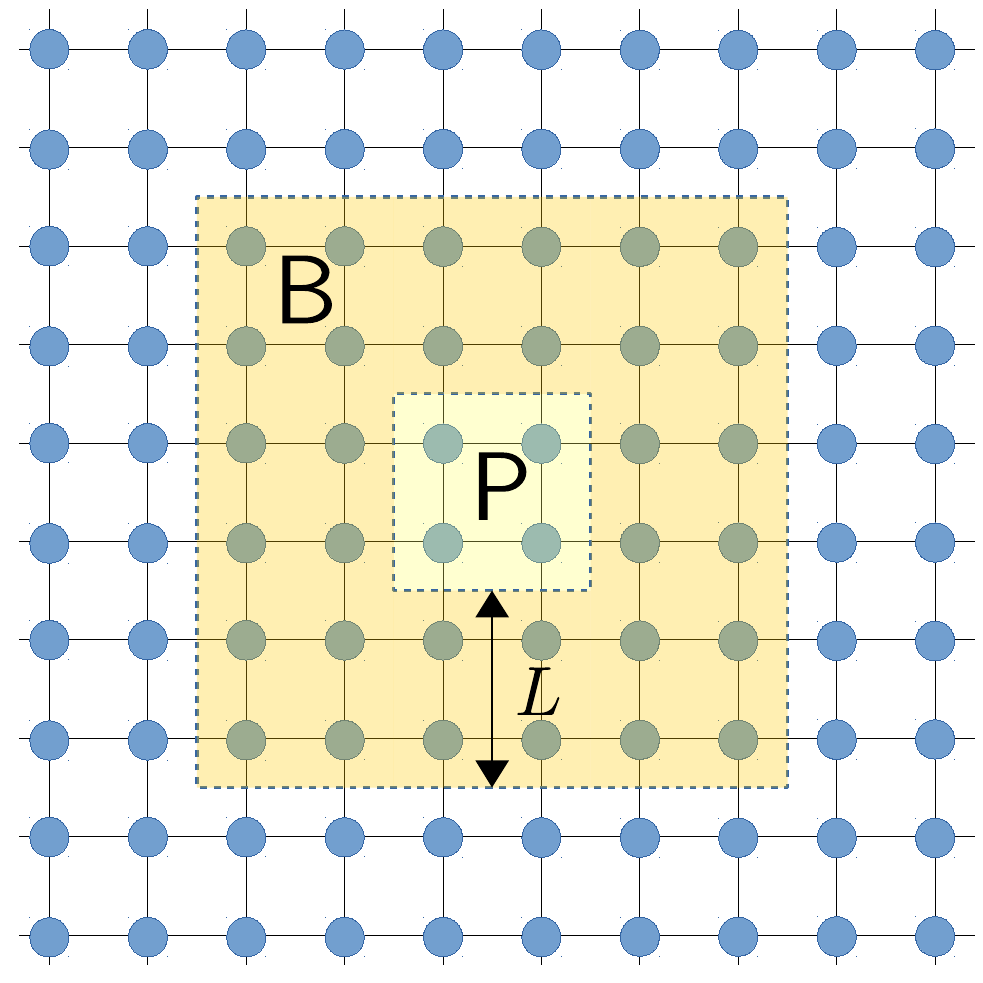}
	\caption{Diagrammatic representation of a translationally invariant short-range-interacting lattice system. The accessible sites within the central shaded yellow region make up the probe $ \mathsf{P} $, while the rest of the lattice acts as the system $ \mathsf{S} $. A boundary region $ \mathsf{B} $ of thickness $ L $ appears shaded in orange, surrounding the probe.}
	\label{fig4-1}
\end{figure*}

Consider a translationally invariant lattice in a global thermal equilibrium state at inverse temperature $ \beta $. The probe $ \mathsf{P} $ consists here of a small sub-lattice made up of a few sites (see figure \ref{fig4-1}), while the rest constitutes the system $ \mathsf{S} $. The reduced state of the probe reads $ \hat{\varrho}_\mathsf{P} = \mbox{tr}_\mathsf{S}\,\hat{\varrho}(\beta) $, where $ \hat{\varrho}(\beta) = \mathcal{Z}^{-1} e^{-\beta\hat H} $ and the total Hamiltonian can be split as $ \hat{H} = \hat{H}_{\mathsf{P}} + \hat{H}_{\mathsf{S}} + \hat{H}_{\mathsf{P}\leftrightarrow\mathsf{S}} $, i.e., as the sum of probe, system, and probe--system interaction terms. Recall that, in general, $ \hat{\varrho}_\mathsf{P} $ can be far from the local thermal state $ \mathcal{Z}_\mathsf{P}^{-1} e^{-\beta \hat{H}_\mathsf{P}} $, specially if the interactions are strong \cite{correa2016low}. As a result, the uncertainty relation 
\begin{equation}
\Delta\beta \Delta\hat{H}_\mathsf{P} \geq 1,
\label{eq4:uncertainty}\end{equation}
or alternatively, $ \Delta T \Delta\hat{H}_\mathsf{P} \geq T^2 $, does no longer hold. Nevertheless, it is legitimate to ask to which extent the \textit{local} quantum Fisher information $ \mathcal{F}(\beta,\hat{\varrho}_\mathsf{P}) $ might be well approximated by the variance of the local Hamiltonian of the probe $ (\Delta\hat{H}_\mathsf{P})^2 $. Notice that in this section, we explicitly include $ \hat{\varrho}_\mathsf{P} $ as an argument of the QFI to emphasise that this is calculated locally on the probe subsystem. 

The quantity $ \mathcal{F}(\beta,\hat{\varrho}_\mathsf{P}) $ was tagged `local quantum thermal susceptibility' in reference \cite{de2016local}. This was shown to display the same qualitative behaviour as the global energy variance around quantum phase \textit{crossovers} in locally interacting spin chains at finite temperatures \cite{PhysRevA.76.062318,1367-2630-17-5-055020,PhysRevA.94.042121}. Indeed one can rigorously prove that the relative error made when approximating $ \mathcal{F}(\beta,\hat{\varrho}_\mathsf{P}) $ by $ (\Delta\hat{H}_\mathsf{P})^2 $---i.e., $ \vert \sqrt{\mathcal{F}(\beta,\hat{\varrho}_\mathsf{P})} - \Delta\hat{H}_\mathsf{P} \vert / \Delta\hat{H}_\mathsf{P} $---decreases as the `volume-to-surface' ratio of $ \mathsf{P} $ grows \cite{de2017universal}. 

To prove this, the probe must be enveloped with a boundary layer $ \mathsf{B} $ of thickness $ L \gg \xi(\beta) $, where $ \xi(\beta) $ is the correlation length of the lattice (cf. figure \ref{fig4-1}). A large volume-to-surface ratio for $ \mathsf{P} $ thus translates into a probe containing many more sites than its corresponding boundary. For the statement to hold, one also needs to impose that $ T $ is above certain critical temperature, so that results on the exponential clustering of correlations on short-range-interacting locally finite lattice systems can be invoked \cite{kliesch2014locality}. Under these conditions one has
\begin{equation}
\mathcal{F}(\beta,\hat{\varrho}_\mathsf{P}) = (\Delta\hat{H}_\mathsf{P})^2 + \mathcal{O}(1) L^4 e^{-L/\xi(\beta)},
\end{equation}
and hence, equation (\ref{eq4:uncertainty}) would still approximately hold.

Another perspective on this problem is provided by the `potential-of-mean-force' approach to strongly coupled open quantum systems (see e.g., reference \cite{PhysRevE.79.061105}). Essentially, it consists in adopting the modified partition function
\begin{equation}
\mathcal{Z}_\mathsf{P}^* \coloneqq \frac{\mbox{tr}\,e^{-\beta(\hat{H}_\mathsf{P}+\hat{H}_\mathsf{S}+\hat{H}_{\mathsf{P}\leftrightarrow\mathsf{S}})}}{\mbox{tr}\,e^{-\beta\hat{H}_\mathsf{S}}}
\label{eq4:partition_hmf}
\end{equation} 
for the probe, so that its internal energy writes as \cite{seifert2016first}
\begin{equation}
-\partial_\beta\log{\mathcal{Z}_\mathsf{P}^*} = \langle \hat{H} \rangle - \langle \hat{H}_\mathsf{S} \rangle = \langle \hat{E}_\mathsf{P}^* \rangle,
\end{equation}
where
\begin{subequations} 
	\begin{align}
	\hat{E}_\mathsf{P}^* &\coloneqq  \partial_\beta\,\left[\beta\hat{H}_\mathsf{P}^*(\beta)\right],\\
	\hat{H}_\mathsf{P}^*(\beta) &\coloneqq -\beta^{-1}\log{\frac{\mbox{tr}_\mathsf{S}\exp{[{-\beta(\hat{H}_\mathsf{P}+\hat{H}_\mathsf{S}+\hat{H}_{\mathsf{P}\leftrightarrow\mathsf{S}})]}}}{\mbox{tr}\,\exp{(-\beta\hat{H}_\mathsf{S}})}},
	\end{align}
\end{subequations}
and thus, $ \hat{\varrho}_\mathsf{P} = {\mathcal{Z}_\mathsf{P}^*}^{-1}e^{-\beta\hat{H}_\mathsf{P}^*} $. The variance of the modified energy operator $ \hat{E}_\mathsf{P}^* $ (or any other observable), can be expressed as the sum of two contributions \cite{PhysRevLett.91.180403} 
\begin{equation}
\big(\Delta\hat{E}_{\mathsf{P}}^*\big)^2 = \mathcal{Q}_\alpha(\hat{\varrho}_\mathsf{P},\hat{E}_\mathsf{P}^*) + \mathcal{C}_\alpha(\hat{\varrho}_\mathsf{P},\hat{E}_\mathsf{P}^*),
\end{equation}
representing its `quantum' and `classical' shares, respectively. In particular, 
\begin{equation}
\mathcal{Q}_\alpha(\hat{\rho},\hat{A}) \coloneqq -\frac12\mbox{tr}\,\{[\hat{A},\hat{\rho}^\alpha][\hat{A},\hat{\rho}^{1-\alpha}]\}
\end{equation}
is the so-called Wigner-Yanase skew information \cite{wigner1963information,PhysRevLett.91.180403}, where $ \alpha\in(0,1) $. Hence, the decomposition of $ (\Delta \hat{A})^2 $ into a quantum and a classical part is not unique. The connection between $ (\Delta\hat{E}_{\mathsf{P}}^*)^2 $ and $ \mathcal{F}(\beta,\hat{\varrho}_\mathsf{P}) $ can be established by noticing that \cite{miller2018mean}
\begin{equation}
\mathcal{C}(\hat{\varrho}_\mathsf{P},\hat{E}_\mathsf{P}^*)\coloneqq \int_0^1 d\alpha~\mathcal{C}_\alpha(\hat{\varrho}_\mathsf{P},\hat{E}_\mathsf{P}^*) \geq \mathcal{F}(\beta,\hat{\varrho}_\mathsf{P})
\label{eq4:classicalskew_fisher}
\end{equation}
and hence, from the Cram\'{e}r-Rao bound, $ (\Delta\beta)^2 \geq \mathcal{C}(\hat{\varrho}_\mathsf{P},\hat{E}_\mathsf{P}^*)^{-1} = [(\Delta\hat{E}_\mathsf{P}^*)^2-\mathcal{Q}(\hat{\varrho}_\mathsf{P},\hat{E}_\mathsf{P}^*)]^{-1} $, where $ \mathcal{Q}(\cdot,\cdot) $ is defined analogously to $ \mathcal{C}(\cdot,\cdot) $ in equation \eqref{eq4:classicalskew_fisher}. As a result one has
\begin{equation}
\Delta\beta\geq\frac{1}{\sqrt{(\Delta\hat{E}_\mathsf{P}^*)^2-\mathcal{Q}(\hat{\varrho}_\mathsf{P},\hat{E}_\mathsf{P}^*)}}\geq\frac{1}{\Delta\hat{E}_\mathsf{P}^*},
\label{eq4:uncertainty_hmf}
\end{equation}
which generalises equation (\ref{eq4:uncertainty}). Roughly speaking, replacing $ \hat{H}_\mathsf{P} $ by $ \hat{E}^*_\mathsf{P} $ accounts for the strong probe--system coupling. Additionally, \eqref{eq4:uncertainty_hmf} features a (quantum) correction term due to the \textit{non-commutativity} between $ \hat{\varrho}_\mathsf{P} $ and $ \hat{E}_\mathsf{P}^* $ \cite{miller2018mean}. What is not clear from equation \eqref{eq4:uncertainty_hmf} is whether or not $ \Delta \beta $ might be reduced by strengthening the probe--system interaction. As we shall see next, this is indeed the case in suitable parameter ranges \cite{correa2016low}.

\subsection{Probe--system interactions as a thermometric resource.}\label{sec:strong_coupling_thermometry} 

Rather than arguing on generic many-body lattice systems, we adopt a concrete model to \textit{quantitatively} assess the impact of strong interactions on the achievable thermometric precision across different temperature ranges. Specifically, consider a translationally invariant chain of harmonic oscillators in thermal equilibrium, with short-ranged two-body interactions---a toy model for a 1D solid. We focus on estimating its temperature by means of local measurements performed on one of its nodes. Importantly, we can always choose the frequency of the nodes so that the energy spectrum of the chain is \textit{gapless}. 

\begin{figure*}[t!]
	\centering
	\includegraphics[width=0.60\linewidth]{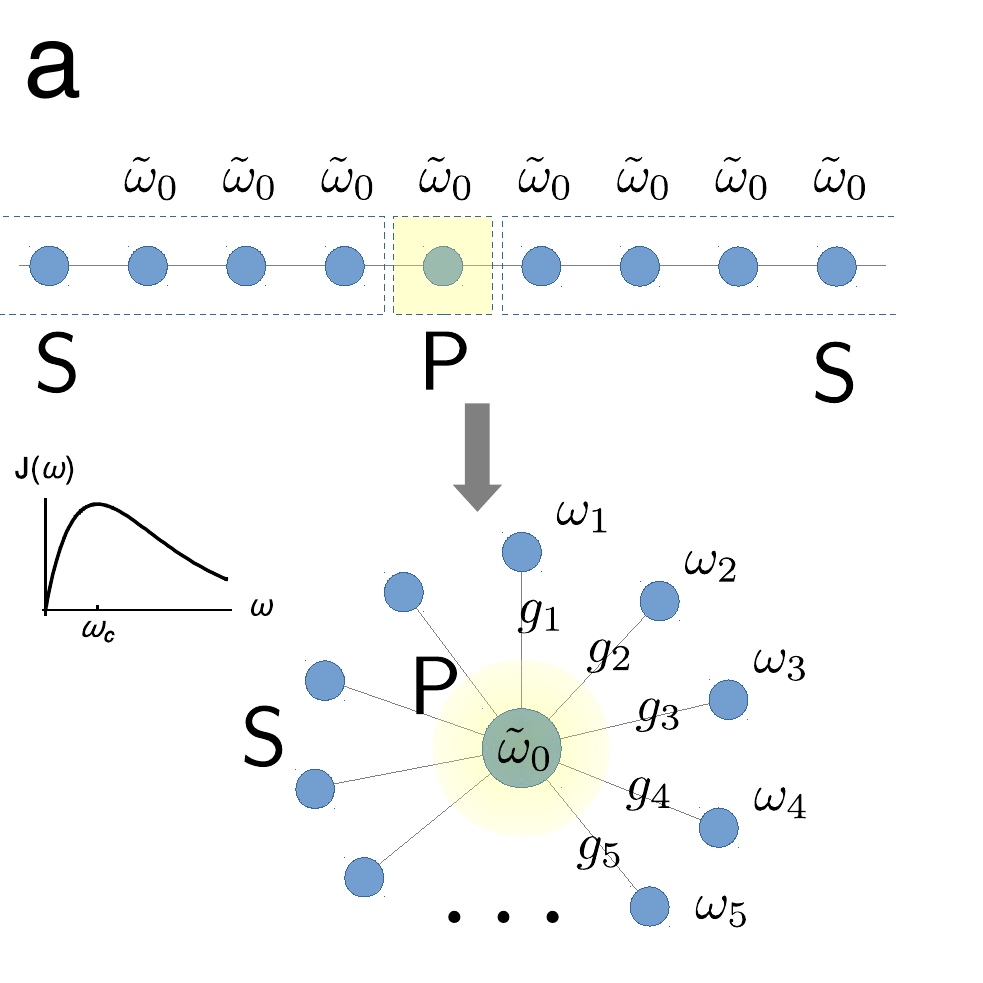}
	\includegraphics[width=0.35\linewidth]{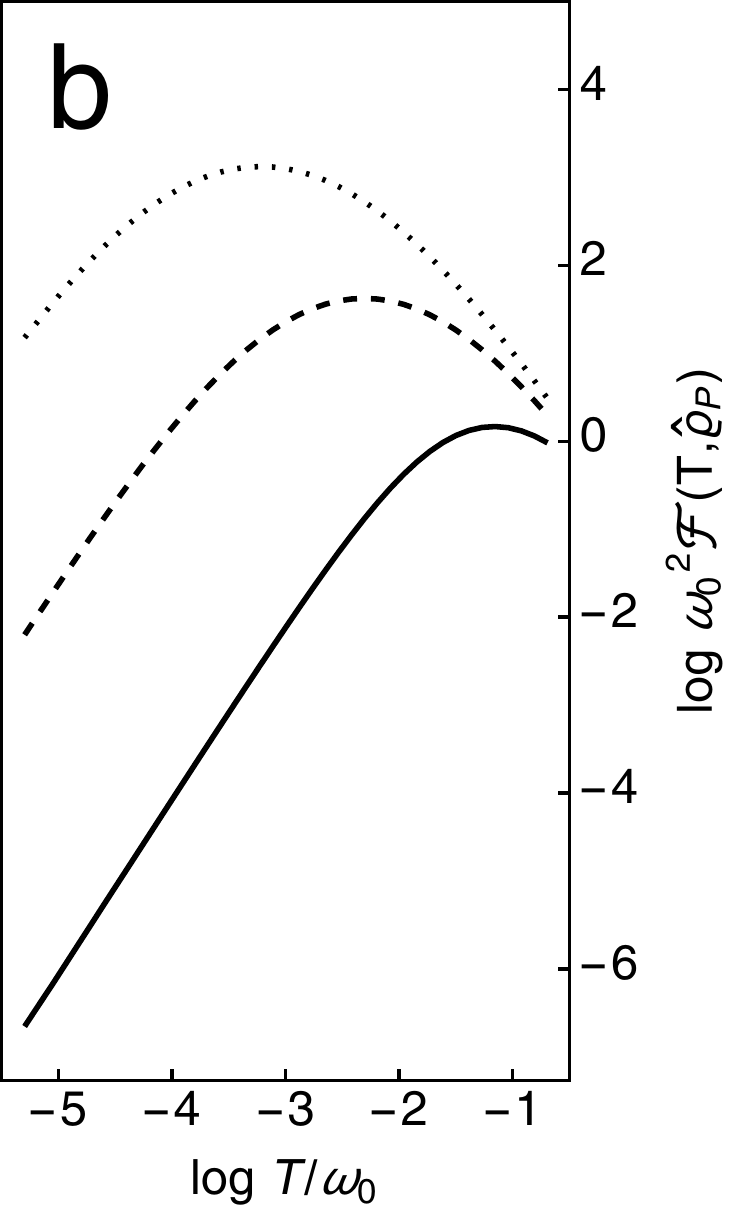}
	\caption{\textbf{(a)} Schematic representation of the equivalence between the problem of local thermometry on one node of a translationally-invariant gapless chain of harmonic oscillators with two body interactions (not necessarily restricted to nearest-neighbours), and thermometry of a bosonic bath from measurements on a Brownian probe undergoing Ohmic dissipation. \textbf{(b)} Quantum Fisher information of a Brownian thermometer coupled to an equilibrium reservoir through an Ohmic-algebraic spectral density (see main text), as a function of temperature in a log-log scale. The various curves correspond to different dissipation strengths: $ \gamma/\omega_0 = 1 $ (solid), $ \gamma/\omega_0 = 5 $ (dashed), and $ \gamma/\omega_0 = 15 $ (dotted). The bare frequency of the probe is $ \omega_0 = 1 $ and the high-frequency cutoff is set to $ \omega_c = 100 $.}
	\label{fig4-2}
\end{figure*}

As shown in reference \cite{hovhannisyan2017probing}, in the thermodynamic limit, this is formally equivalent to interrogating a Brownian particle coupled to an equilibrium bath through an Ohmic interaction scheme, once they have reached a steady state [see figure~\ref{fig4-2}(a) and section~\ref{sec:exponential_ineff_gapped} below]. Owing to this analogy, we can borrow powerful techniques from the theory of open quantum systems to compute the stationary state of the probe $ \hat{\varrho}_\mathsf{P} $ \textit{exactly}. Importantly, this connection between local temperature estimation and open systems is not only technically convenient but can also prove insightful when discussing the ultimate limitations on low-$ T $ thermometry \cite{hovhannisyan2017probing}.

Coming back to our problem, the reduced state of the single node acting as probe equals the (non-equilibrium) steady state of a harmonic oscillator linearly coupled to a bosonic bath prepared at the original temperature. The model would be specified by
\begin{subequations}
	\begin{align}
	\hat{H}_\mathsf{P} &= \frac12(\tilde{\omega}_0^2 \hat{X}^2 + \hat{P}^2), \\
	\hat{H}_\mathsf{S} &= \frac12\sum\nolimits_{\mu}\big(m_\mu\omega_\mu^2\hat{x}_\mu^2 + \hat{p}_\mu^2/m_\mu\big), \\
	\hat{H}_{\mathsf{P}\leftrightarrow\mathsf{S}} &= \hat{X}\sum\nolimits_{\mu}g_\mu \hat{x}_\mu.
	\end{align}
	\label{eq4:CL_Hamiltonian}
\end{subequations}

Here, we have defined $ \tilde{\omega}_0^2 \coloneqq \omega_0^2 + \omega_R^2 $, where $ \omega_0 $ is the \textit{bare} frequency of the Brownian particle serving as temperature probe, and $ \omega_R^2 = \sum_{\mu}\frac{g_\mu^2}{m_\mu\omega_\mu^2} $ is a renormalisation term ensuring that the probe--system energy spectrum is bounded from below, regardless of the interaction strength \cite{weiss2008quantum}. Let the initial condition be $\hat{\varrho}(0) = \hat{\varrho}_\mathsf{P}(0)\otimes\big(\mathcal{Z}_\mathsf{S}^{-1}e^{-\beta\hat{H}_\mathsf{S}}\big) $, where the probe can be initialised in \textit{any} state. When it comes to the coupling constants $ g_\mu $, these are conveniently specified by the spectral density
\begin{equation}
J(\omega)\coloneqq\pi\sum\nolimits_\mu\frac{g_\mu^2}{2 m_\mu\omega_\mu}\delta(\omega-\omega_\mu),
\label{eq4:spectral_density}
\end{equation}
which can be given a smooth functional form. Equations \eqref{eq4:CL_Hamiltonian} correspond to the paradigmatic Caldeira--Leggett model for quantum Brownian motion \cite{caldeira1985physica}. Starting from the Heisenberg equations for all degrees of freedom, one can obtain the following \textit{exact} equation for the position of the probe \cite{weiss2008quantum,AoP326_1207,PhysRevA.31.471}
\begin{equation}
\frac{d^2\hat{X}(t)}{dt^2} + \tilde{\omega}_0^2 \hat{X}(t) - \int_{-\infty}^{\infty}ds\,\chi(t-s)\hat{X}(s) = \hat{F}(t).
\label{eq4:QLE}
\end{equation}  
This is the quantum Langevin equation, originally derived by Ford, Kac and Mazur \cite{ford1965statistical,PhysRevA.37.4419}. It captures the fact that the Brownian probe is subjected to an external stochastic quantum force $ \hat{F}(t) $, due to interaction with the system. The exact expression for ${\hat F}(t)$ is cumbersome, and can be found in, e.g., \cite{correa2016low}. In addition, the dissipation kernel $ \chi(t) $ in the integral term implies that the probe keeps a certain memory of its past dynamics \footnote{To preserve causality in equation~\eqref{eq4:QLE}, $ \chi(t)\coloneqq\frac{2}{\pi}\Theta(t)\int_0^\infty d\omega\,J(\omega)\sin{\omega t} $, i.e., it includes a Heaviside step function in its definition.}. 

As already mentioned, in our case $ J(\omega) $ is Ohmic, i.e., it grows linearly with $ \omega $ at low frequencies and decays quickly to zero for $ \omega $ above a certain cutoff $ \omega_c $. Specifically, the choice $ J(\omega) = \gamma \omega\omega_c^2/(\omega^2 + \omega_c^2) $---that is, Ohmic spectral density with algebraic cutoff, or Ohmic-algebraic, for short---allows for a relatively simple analytical steady-state solution \cite{correa2016low}. Analytic results may also be found for, e.g., generalised (super-)Ohmic spectral densities with exponential cutoff (i.e., $ J(\omega) = \frac{\pi}{2}\gamma \omega^s \omega_c^{1-s} e^{-\omega/\omega_c} $ with $ s \geq 1 $) \cite{PhysRevA.88.042303,correa2016low}.  

Since the Hamiltonian \eqref{eq4:CL_Hamiltonian} is quadratic in positions and momenta, the steady state of the probe is necessarily Gaussian \cite{grabert1984quantum} and hence, fully determined by the stationary second-order moments $ \sigma_{11}\coloneqq\langle\hat{X}^2\rangle $, $ \sigma_{22}\coloneqq\langle\hat{P}^2\rangle $, and $ \sigma_{12}\coloneqq\langle\hat{X}\hat{P}\rangle $ (the first moments vanish). In turn, these can be explicitly evaluated from equation \eqref{eq4:QLE} \textit{without making any approximations} beyond our assumption of an initially uncorrelated probe--system state. Equation \eqref{eq4:QLE} thus enables a totally unrestricted analysis of the thermal sensitivity of the probe, including arbitrarily low temperatures and strong probe--system interactions. The resulting QFI of the probe can be obtained by using techniques from Gaussian metrology~\cite{monras2013phase,PhysRevA.89.032128,PhysRevA.98.012114}, and reads
\begin{subequations}
	\begin{equation}
	\mathcal{F}(T,\hat{\varrho}_\mathsf{P}) = 2\,C_1^2\,\sigma_{11}^2 + 2\,C_2^2\,\sigma_{22}^2 - C_1\,C_2,
	\label{eq4:QFI_Moha}
	\end{equation}
	where
	\begin{equation}
	C_1 \coloneqq \frac{2\sigma_{22}^2\,\partial_T\sigma_{11} + \frac12\partial_T\sigma_{22}}{4\sigma_{11}^2\sigma_{22}^2-\frac14},~~~\mbox{and}~~~
	C_2 \coloneqq \frac{\frac12\partial_T\sigma_{11} + 2\sigma_{11}^2\partial_T\sigma_{22}}{4\sigma_{11}^2\sigma_{22}^2-\frac14}.
	\label{eq4:QFI_Moha_coeffs}
	\end{equation}
\end{subequations}
In figure \ref{fig4-2}(b) we depict $ \mathcal{F}(T,\hat{\varrho}_\mathsf{P}) $ versus $T/\omega_0$ for various probe--system interaction strengths. As it can be seen, increasing the interactions can result in a significant enhancement of $ \mathcal{F}(T,\hat{\varrho}_\mathsf{P}) $. In fact, in the limit $ T/\omega_0 \ll 1 $ the dissipation-driven thermometric advantage grows \textit{monotonically} with the dissipation strength \cite{correa2016low}. 

In order to turn the growth of the QFI into a practical improvement on thermometry, one needs to come up with an experimentally feasible measurement scheme whose thermal sensitivity approaches $ \mathcal{F}(T,\hat{\varrho}_\mathsf{P}) $. As it turns out, the variance of the position quadrature $ \langle\hat{X}^2\rangle $ is an ``experimentally friendly'' \cite{PhysRevA.54.R25,PhysRevA.53.R1966,bastin2006measure} quasi-optimal temperature estimator in this regime and, in particular, it largely outperforms the precision achieved via local energy measurements \cite{correa2016low}. One can thus conclude that the strength of the internal interactions in a many-body system---or, equivalently, the dissipative probe--system coupling---can be harnessed as a practical resource for low-temperature local thermometry.

To better understand the range of temperatures for which these dissipation-enabled effects might actually be useful, one needs to focus on a specific experimental realisation. For instance, the atoms of a very dilute probe gas overlaid on a Bose--Einstein condensate would, in certain limits, be exactly described by the Caldeira--Leggett Hamiltonian \eqref{eq4:CL_Hamiltonian} \cite{mehboudi2018using,lampo2018open}. For realistic parameters, a \textit{weak coupling} between the impurities and the condensed system turns out to yield better precision, even for temperatures as low as few hundreds of \si{\pico\kelvin}. This is because the ratio $ k_B T/\hbar \omega_0 $ would not typically qualify as `cold' in this specific setup (indeed, $ k_B T/\hbar \omega_0 \simeq 0.4 $ for the parameters in \cite{mehboudi2018using}). Proposals like this are particularly relevant since that the most common thermometric techniques for condensates (at temperatures $\sim 100 \si{\nano\kelvin} $) are based on the `time of flight' absorption method. This can achieve noise-to-signal ratios down to $ 1\% $ \cite{leanhardt2003cooling,gati2006noise,gati2006primary,el2013thermodynamic}, although it is \textit{destructive}, as it collapses the atomic gas. Furthermore, it features very low precision in the sub-$\si{\nano\kelvin} $ regime \cite{leanhardt2003cooling,olf2015thermometry,spiegelhalder2009collisional}. 

Finally, just like in the simple bipartite case discussed in section \ref{sec-Coupled-ho-Thermal}, the underlying physical mechanism responsible for this dissipation-driven enhancement can be understood by looking at the normal modes of the global probe--system composite. For oscillators arranged in the ``star-like'' configuration of figure \ref{fig4-2}(a), it can be shown that the normal-mode frequencies below $ \tilde{\omega}_0 $ decrease monotonically with $ \gamma $; the corresponding collective degrees of freedom thus become more sensitive to small thermal fluctuations. As could be expected in light of the discussion of section \ref{sec-Coupled-ho-Thermal}, this increased low-temperature sensitivity also extends to the central Brownian thermometer \cite{correa2016low} [cf. equation \eqref{eq-ho-temp-sensitivity}]. Interestingly, note as well that figure \ref{fig4-2}(b) shows that the low temperature scaling of the QFI is \textit{power-law-like}. Below, we look into the various possible low-temperature scalings of the thermal sensitivity. 

\subsection{Local thermometry on gapped and gapless systems.}\label{sec:exponential_ineff_gapped}

Let us start by considering a finite-dimensional non-degenerate system in thermal equilibrium. When the temperature is sufficiently low, only the ground and first excited states (with energies $ \epsilon_0 = 0 $ and $ \epsilon_1 = \Delta $, respectively) are significantly populated so that, for all practical purposes, we would be dealing with a two-level atom in $ \hat{\varrho}(T) = \mathcal{Z}^{-1}(e^{-\Delta/T}\ket{1}\bra{1} + \ket{0}\bra{0}) $. The corresponding QFI reads [cf. equation \eqref{Energy-Var-quNit}]
\begin{equation}
\mathcal{F}(T,\hat{\varrho}) = \frac{\Delta^2\sech^2{\frac{\Delta}{2T}}}{4T^4} = \frac{\Delta^2}{T^4}\,e^{-\Delta/T} + \mathcal{O}\big(e^{-2\Delta/T}\big). 
\label{eq4:exponential_inefficiency}
\end{equation}
Therefore, thermometry on a finite-dimensional system becomes \textit{exponentially inefficient} at low temperatures. 

The same adverse scaling would bear upon local thermometry on any part of a finite gapped equilibrium system, due to the monotonicity of the QFI under partial tracing---that is, $ \mathcal{F}(T,\hat{\varrho})\geq\mathcal{F}(T,\hat{\varrho}_\mathsf{P}) $ \cite{petz2002covariance}. However, when the system size is infinitely large, one has to be cautious: Indeed, the extensivity of the heat capacity of locally-interacting many-body systems entails a \textit{diverging} global QFI \cite{hovhannisyan2017probing}. As a result, the upper bound $ \mathcal{F}(T,\hat{\varrho}) $ is too loose to infer the low-$ T $ scaling of $ \mathcal{F}(T,\hat{\varrho}_\mathsf{P}) $. Nevertheless, it can be proven that the exponential inefficiency of local thermometry showcased by equation \eqref{eq4:exponential_inefficiency} does extend to the thermodynamic limit, whenever the many-body system under consideration is short-range-interacting, gapped, and translationally invariant \cite{hovhannisyan2017probing}.

The aforementioned exponential inefficiency has been reported even in a dynamical scenario, in reference \cite{brunelli2011qubit}. There, the temperature of a cold mechanical resonator was estimated by connecting it to a two-level probe through a Jaynes--Cummings Hamiltonian \cite{jaynes1963comparison} and the effects of environmental noise were introduced via a phenomenological model (cf. section \ref{sec:micromechanical}). In this scenario, the QFI can be seen to decay exponentially with decreasing temperature, even when optimised over the initial preparation and the evolution time.

A better low-temperature scaling of the local thermal sensitivity may only be achieved if the many-body system in question is \textit{gapless} \cite{hovhannisyan2017probing}. This is indeed the case for the exact results shown in figure \ref{fig4-2}(b), where the thermal sensitivity decays \textit{polynomially} at low temperatures; namely, as $ \mathcal{F}(T\rightarrow 0,\hat{\varrho}_\mathsf{P}) \sim T^2 $. In fact, this specific scaling is encountered generically in any Brownian thermometer with a non-vanishing bare frequency under Ohmic dissipation, regardless of the details of its spectral density \cite{hovhannisyan2017probing}. As already mentioned in section \ref{sec3:size_temp_scaling}, the exact same scaling has also been observed for local thermometry on a tight-binding model of interacting fermions \cite{hofer2017fundamental}. Interestingly, low-temperature sensitivity also appeared when measuring temperature \textit{globally} in the (gapless) ferromagnetic phase of the XX model (cf. section \ref{sec-XY}). 

\begin{figure*}[t!]
	\centering
	\includegraphics[width=0.60\linewidth]{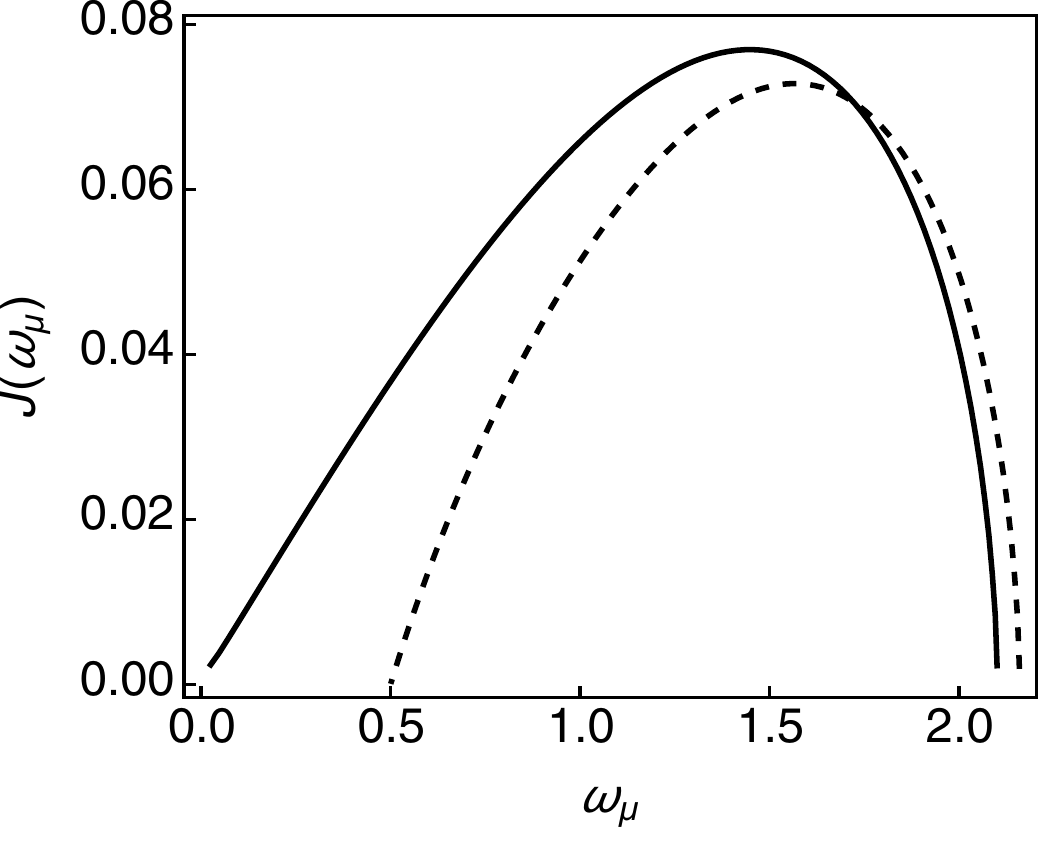}
	\caption{Effective spectral density for local thermometry on a single site of a 100-node equilibrium chain of harmonic oscillators with periodic boundary conditions versus the normal-mode frequencies of the system (see text). The inter-node couplings are $ G_{\vert i-j \vert} = \vert i-j \vert^{-5/2} $. For the solid (dashed) curve the frequency $ \tilde{\omega}_0 $ of the nodes is $ \tilde{\omega}_0 \simeq 1.3169 $ ($ \tilde{\omega}_0 \simeq 1.4086 $), so that the gap is $ \Delta = 0 $ ($ \Delta = 0.5 $).}
	\label{fig4-3}
\end{figure*}

In order gain additional insights as to why low-$ T $ local thermometry becomes so efficient on gapless systems it is best to go back to the simple (translationally invariant) harmonic chain depicted in figure \ref{fig4-2}(a) and its open-system analogue. Let its Hamiltonian be
\begin{equation}
\hat{H} = \frac12\big(\boldsymbol{\hat{X}}^\mathsf{T}\mathsf{V}\boldsymbol{\hat{X}} + \boldsymbol{\hat{P}}^\mathsf{T}\boldsymbol{\hat{P}}\big),
\label{eq4:TIHC_Hamiltonian}
\end{equation} 
where $ \boldsymbol{\hat{X}} \coloneqq ( \hat{X}_1,\cdots,\hat{X}_{2N + 1} )^\mathsf{T} $ and $ \boldsymbol{\hat{P}} \coloneqq ( \hat{P}_1,\cdots,\hat{P}_{2N + 1})^\mathsf{T} $ are vectors containing the $ 2N + 1 $ positions and momenta of the nodes of the chain (all masses are set to one for simplicity). The interaction matrix $ \mathsf{V} $ is real and symmetric and, in particular, all of its diagonal elements equal $ \tilde{\omega}_0^2 $. Very generally, let us choose interactions that decay with the `distance' between nodes as $ [\mathsf{V}]_{i \neq j} \propto \vert i-j \vert^{-\alpha} \coloneqq G_k $, with $ \alpha > 1 $ and $ k = \vert i-j \vert $. Note that the two-body interactions are thus not limited to nearest neighbours. Imposing periodic boundary conditions as per translational invariance, requires that $ G_k = G_{2N+1-k} $ for $ 1\leq k \leq 2N $. 

The lowest eigenvalue of $ \mathsf{V} $ is given by
\begin{equation}
\Delta^2 = \tilde{\omega}_0^2 + 2 \sum_{k=1}^N G_k \cos{\frac{2\pi k N}{2N+1}}.
\label{eq4:lowest-eigenvalue-circulant}
\end{equation}
In order to guarantee that the energy spectrum is bounded from below, the frequency of the individual nodes must thus satisfy $ \tilde{\omega}_0^2 \geq -2 \sum_{k=1}^N G_k \cos{\frac{2\pi k N}{2N+1}} $. In particular, when the equality is saturated, $ \Delta $ vanishes and the chain becomes \textit{gapless}.

Identifying, e.g., node $ \# 1 $ as the probe, and all the others as the system, $ \hat{H} $ becomes
\begin{multline}
\hat{H} = \hat{H}_\mathsf{P} + \hat{H}_\mathsf{S} + \hat{H}_{\mathsf{P}\leftrightarrow\mathsf{S}}  \\
= \frac12\big( \tilde{\omega}_0^2 \hat{X}_1^2 + \hat{P}_1^2 \big) + \frac12\Big( \sum_{i > 1} \big(\tilde{\omega}_0^2 \hat{X}_i^2 + \hat{P}_i^2 \big) + \sum_{\substack{i,j > 1 \\ i \neq j}} [\mathsf{V}]_{ij} \hat{X_i}\hat{X_j} \Big) + \hat{X}_1\sum_{i>1} [\mathsf{V}]_{1i} \hat{X}_i.
\label{eq4:TICH_Hamiltonian_2}
\end{multline}
Then, we can bring the system into a diagonal form through an orthogonal transformation. Denoting its $ 2N $ normal-mode coordinates as $ \hat{x}_{\mu>1} = \sum_{\nu>1}[\mathsf{O}]_{\mu\nu}\hat{X}_\nu $ (where $ \mathsf{O} $ is the corresponding orthogonal transformation), allows to rewrite the probe--system interaction term as $ \hat{H}_{\mathsf{P}\leftrightarrow\mathsf{S}} = \hat{X}_1\sum_{\mu>1} g_\mu \hat{x}_\mu $, where
\begin{equation}
g_\mu = \sum_{\nu > 1} [\mathsf{O}]_{\mu\nu}[\mathsf{V}]_{1\nu}.
\label{eq4:probe-sample-couplings}
\end{equation}
We can thus obtain the effective density $ J(\omega)\coloneqq\pi\sum\nolimits_\mu\frac{g_\mu^2}{2 m_\mu\omega_\mu}\delta(\omega-\omega_\mu) $ describing the dissipative interactions between probe and system in the gapped ($ \Delta > 0 $) and gapless ($ \Delta = 0 $) cases. 

As we can see in figure \ref{fig4-3}, when ``gaplessness'' is enforced on a finite chain by suitably choosing $ \tilde{\omega}_0 $, the residual spectral density is of the Ohmic type. On the contrary, when the chain features a finite gap, a non-zero minimal frequency emerges in the residual spectrum of the system---that is, the probe can no longer couple to arbitrarily low-frequency system modes, which are the most informative ones at low $ T $. This explains the switching between polynomial and exponential scaling of the performance of low-temperature thermometry in gapped many-body systems \cite{hovhannisyan2017probing}. Thus, we see that adopting an open system viewpoint does provide useful insights into local quantum thermometry, as advanced in section~\ref{sec:strong_coupling_thermometry}.  

\subsubsection{Low-$ T $ thermometry limited by measurement resolution.}\label{sec:measurement_resolution}

One can approach the problem of low-temperature thermometry from a different angle \cite{hofer2017fundamental}; namely, $ T $ can be estimated from a global POVM measurement for which the finite number of possible outcomes $ M $ is interpreted as a limitation on \textit{resolution}. The corresponding Fisher information may then be expressed in the familiar form 
\begin{subequations}
	\begin{equation}
	\mathcal{F}_c(\boldsymbol{\hat\Pi};T) = \frac{1}{T^4}\left[\left(\sum\nolimits_{m} p_m E_m^2\right) - \left(\sum\nolimits_{m} p_m E_m\right)^2\right],
	\label{eq4:CFI_POVM}
	\end{equation}
	by defining the POVM-dependent ``spectrum''
	\begin{equation}
	E_m \coloneqq {p_m}^{-1}\,\mbox{tr}\,\left\lbrace\hat{\Pi}_m \hat{H}\,\mathcal{Z}^{-1}e^{-\beta\hat{H}}\right\rbrace.
	\label{eq4:POVM_spectrum}
	\end{equation}
\end{subequations}
Therefore, the problem of temperature estimation from partial information can be mapped into thermometry on a non-equilibrium finite-dimensional probe, diagonal in the eigenbasis of a fictitious Hamiltonian. The ``energies'' $ E_m $ are generally temperature-dependent and so are the gaps between any level and the ``ground state'', i.e., $ \Delta_m(T) = E_m(T)-E_0(T) $. Whenever $ \Delta_m(0) = 0 $ and $ \Delta_m(\delta T) > 0 $ (for arbitrary $ \delta T $ and one or more $ m $), a power-law-like scaling of $ \mathcal{F}_c(\boldsymbol{\hat\Pi};T) $ becomes possible in the low-temperature limit \cite{hofer2017fundamental}. This is similar to the scenario depicted in reference \cite{paris2015achieving}, which proposed to reduce the gap of a multi-level probe as $ T\rightarrow 0 $ in order to keep $ \Delta T $ from diverging exponentially.

\subsubsection{Relative error of low-temperature estimates.}\label{sec:low-T_relative_error}

\begin{figure*}[t!]
	\centering
	\includegraphics[width=0.60\linewidth]{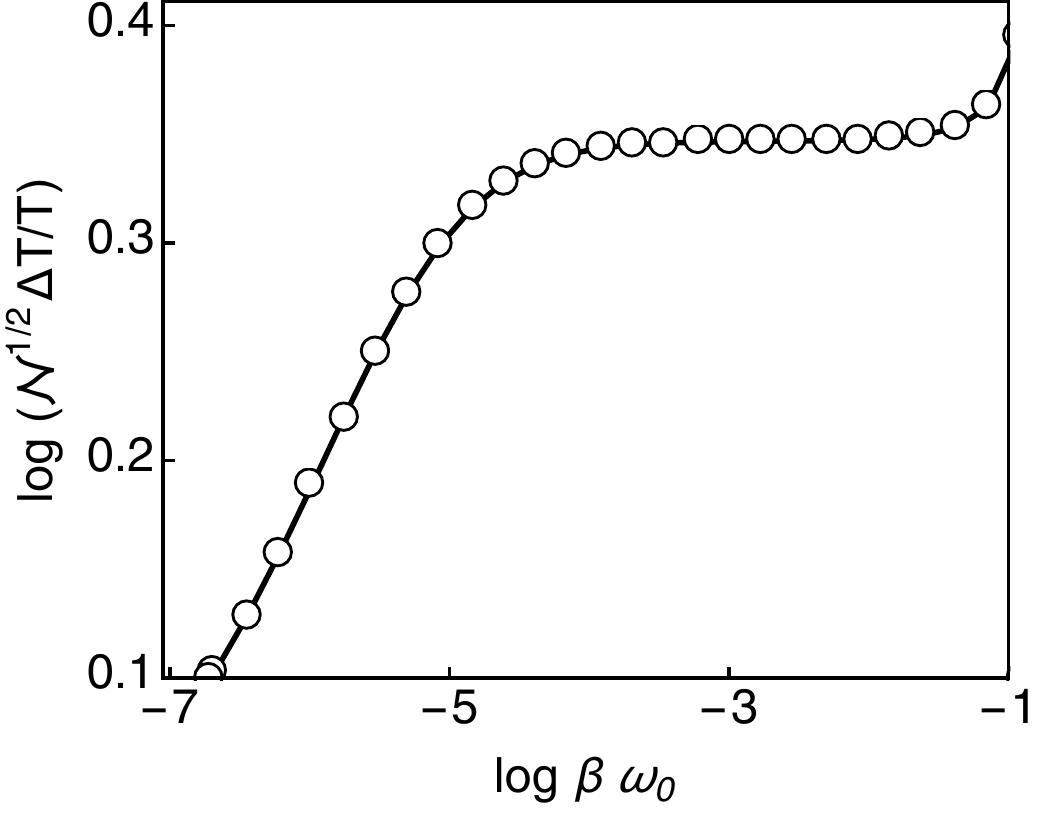}
	\caption{Best-case relative error $ \sqrt{\mathcal{N}}(\Delta T/T) $ for thermometry on an Ohmic system with a Brownian thermometer, versus the inverse temperature in a log-log scale. The bare frequency of the probe is $ \omega_0 = 10^{-3} $ and the spectral density is of the Ohmic-exponential type, i.e., $ J(\omega) = \frac{\pi}{2}\,\gamma\,\omega\,e^{-\omega/\omega_c} $, with $ \gamma = 0.1 $ and $ \omega_c = 100 $ ($ \hbar = k_B = 1 $). Note that even though $ \omega_0 \ll 1 $, the Brownian particle is still tightly confined, as its effective frequency within the Caldeira-Legget Hamiltonian \eqref{eq4:CL_Hamiltonian} is $ \tilde{\omega}_0^2 = \omega_0^2 + \gamma\omega_c \simeq 10 $.}
	\label{fig4-4}
\end{figure*}

In practice, the low-$ T $ scaling of the relative error (multiplied by the length of the data set for convenience)
\begin{equation}
\sqrt{\mathcal{N}}~\frac{\Delta T}{T} \geq \frac{1}{T\sqrt{\mathcal{F}(T,\hat{\varrho}_\mathsf{P})}}
\end{equation}
may be more relevant than that of $ \Delta T $ alone. For instance, even under the benign scaling $ \mathcal{F}(T,\hat{\varrho}_\mathsf{P}) \sim T^2 $, the relative error still diverges as $ 1/T^2 $ when $ T\rightarrow 0 $. 

As argued in reference \cite{hofer2017fundamental} one can make use of the relation
\begin{equation}
T^2\mathcal{F}(T,\hat{\varrho}) = C_T
\label{eq4:QFI_heat_capacity}
\end{equation}
for the global thermal sensitivity, and invoke the third law of thermodynamics to require $ \lim_{T\rightarrow 0} C_T = 0 $ and thus $ (\Delta T/T) \rightarrow \infty $, which would also hold for local estimates. Hence, measuring ultra-cold temperatures precisely seems to be an impossible task. 

Luckily, one might bypass this impediment in practice by choosing, e.g., a Brownian probe with `sufficiently low' bare frequency $ \omega_0 $. Note that, even in the strict limit $ \omega_0 = 0 $, the effective trapping frequency $ \tilde{\omega}_0 $ of such Brownian thermometer would still be finite due to the explicit renormalisation in equation \eqref{eq4:CL_Hamiltonian}. Assuming an Ohmic spectrum, it can be rigorously proven from the exact steady-state solution of this model that
\begin{equation}
\mathcal{F}(T\rightarrow 0,\hat{\varrho}_\mathsf{P}) \sim 1/T^2, \qquad\qquad (\omega_0\rightarrow 0)
\label{eq4:inverse_scaling_power}
\end{equation}
so that the relative error $ \sqrt{\mathcal{N}}(\Delta T/T) $ \textit{converges to a constant} in the low-$ T $ regime \cite{hovhannisyan2017probing}. A small but finite $ \omega_0 $ would allow for a constant relative error down to arbitrarily low temperatures, so long as $ \omega_0/T $ remains small. As $ \omega_0/T $ grows, however, the usual power-law-like divergence should be expected to take over (see figure \ref{fig4-4}). We shall come back to this topic in section \ref{sec5} below, where we comment on a recent proposal to use periodically driven thermometers which also enables precise thermometry at arbitrarily low temperatures \cite{mukherjee2017high}. 

\section[Dynamical quantum thermometry]{Dynamical quantum thermometry}\label{sec5}

We are now concerned with \textit{dynamical} scenarios, in which the (non-equilibrium) probe is interrogated at some finite time, without waiting for full relaxation. This section is thus a ``ragbag'' of diverse---albeit practically relevant---thermometric setups. 

In section \ref{sec:micromechanical} we study {\it very small} system-thermometers that give rise to `non-thermalising' coherent dynamics. In this case, minimising the error bars also demands to choose the optimal interrogation time, and the most sensitive preparation for the probe.
In contrast, section \ref{sec:markovian} deals with scenarios with thermalising dynamics. However, they require to interrogate the thermometer after some relatively short time, rather than allowing for full thermalization.
Interferometric thermometry, and the usefulness of entanglement to achieve {\it super-extensive} scaling is addressed in section \ref{sec:interferometric}. 
In section \ref{sec:dynamical_control} we address \textit{active} strategies for temperature estimation, in which an external control is applied on the probe and/or the system. This includes (i) estimating temperature from non-equilibrium work distributions, which are experimentally measurable; (ii) periodic modulation of the energy spectrum of the probe, that can endow it with enhanced sensitivity over a wide range of temperatures; and (iii) thermometers based on quantum heat pumps---realisable on circuit QED platforms.

\subsection{Thermometry under non-thermalising dynamics}\label{sec:micromechanical}

The observation of quantum effects in near-ground-state micro-mechanical setups has attracted a lot of attention in recent years (see, e.g., references \cite{rocheleau2010preparation,o2010quantum,clarke2018growing}). These systems can be modelled as single harmonic oscillators with frequency $ \Omega $ and Hamiltonian $ \hat{H}_\mathsf{S} = \Omega\,\hat{a}^\dagger\hat{a} $, in a thermal state $ \hat{\varrho}_\mathsf{S}(\beta) = \mathcal{Z}_\mathsf{S}^{-1}e^{-\beta\hat{H}_\mathsf{S}} $ at some very low temperature. 

One way to determine the temperature of the resonator is to couple it to a two-level probe (e.g., a superconducting qubit) \cite{o2010quantum,brunelli2011qubit,brunelli2012qubit}, effectively described by $ \hat{H}_\mathsf{P} = \frac{\omega}{2}\hat{\sigma}_z $, where $ \hat{\sigma}_{\alpha\in\{x,y,z\}} $ is a Pauli matrix in the Hilbert space of the probe. Close to resonance (i.e., $ \Omega \simeq \omega $) and provided that the probe--system coupling $ \lambda $ is not too strong, their mutual interaction can be approximated by the Jaynes--Cummings Hamiltonian \cite{jaynes1963comparison}
\begin{equation}
\hat{H}_{\mathsf{P}\leftrightarrow\mathsf{S}} = \lambda\left( \hat{\sigma}_-\hat{a}^\dagger + \hat{\sigma}_+\hat{a} \right),
\label{eq5:jaynes_cummings}    
\end{equation}
where $ \hat{\sigma}_\pm = (\sigma_x \pm i \sigma_y)/2$ are the ladder operators acting on the qubit. Disregarding the dissipative interactions with the environment, the reduced dynamics of the qubit $ \hat{\varrho}_\mathsf{P}(t) $ is periodic and can be solved exactly \cite{brunelli2011qubit} by assuming a factorised initial state [$ \hat{\rho}(0) = \hat{\varrho}_\mathsf{P}(0)\otimes\hat{\tau}_\mathsf{S}(\beta) $], and exploiting the fact that the global Hamiltonian $ \hat{H} = \hat{H}_\mathsf{P} + \hat{H}_\mathsf{S} + \hat{H}_{\mathsf{P}\leftrightarrow\mathsf{S}} $ conserves the total number of excitations. From the exact solution, it is easy to compute the FI for measurements on the energy basis of $ \hat{\varrho}_\mathsf{P}(t) $, i.e., $ \mathcal{F}_c(\mathbbm{1}_\mathsf{S}\otimes\hat{\sigma}_z , t) $, where the argument $ T $ has been unambiguously dropped. 

As already mentioned, it is necessary to optimise the Fisher information over the initial preparation $ \hat{\varrho}_\mathsf{P}(0) $ and the free evolution time $ t $ \cite{brunelli2011qubit}. As it turns out, preparing the qubit in the ground state $ \hat{\varrho}(0) = \ketbra{g}{g} $ and measuring its energy at $ t = \pi/2\lambda $ maximises the FI, which also coincides with the maximum possible thermal sensitivity given by the QFI (see also references \cite{PhysRevLett.114.220405,jevtic2015single,de2017estimating}). Other forms of probe system coupling yield similar results \cite{brunelli2012qubit}.

When the dissipative coupling of the resonator to the phonons of its substrate is such that it remains thermalised at all times, i.e., $ \hat{\varrho}_\mathsf{P}(t)\otimes\hat{\varrho}_\mathsf{S}(\beta) $, the standard Rabi model
\begin{subequations}
	\begin{align}
	\hat{H}_\mathsf{P} &= \frac{\epsilon}{2}\hat{\sigma}_z + \frac{\Delta}{2}\hat{\sigma}_x, \\
	\hat{H}_{\mathsf{P}\leftrightarrow\mathsf{S}} &= \gamma\,\hat{\sigma}_z\otimes(\hat{a}^\dagger + \hat{a}),
	\label{eq5:rabi_model}    
	\end{align}
\end{subequations}
can be solved by moving into the so-called `polaron' frame (i.e., by applying a suitable unitary transformation \cite{wagner1986unitary}). As a result of the coupling to the resonator, the qubit would undergo Rabi oscillations at a temperature-dependent ac-Stark-shifted frequency $ \tilde{\omega}(T) $ (see \cite{higgins2012quantum} for an explicit expression), instead of the Rabi frequency $ \omega = \sqrt{\epsilon^2 + \Delta^2} $. Therefore, measuring $ \tilde{\omega}(T) $ allows for precise quantum thermometry on micro-mechanical resonators \cite{higgins2012quantum}. 

This effect is reminiscent of the thermometry experiments based on temperature dependence of the zero-field splitting of the ground state of nitrogen-vacancy (NV) colour centres in diamond \cite{maletinsky2012robust,kucsko2013nanometre,schirhagl2014nitrogen}. Remarkably, in reference \cite{kucsko2013nanometre}, this technique was successfully exploited to estimate the temperature gradients \textit{inside living cells} containing nano-diamond sensors. Furthemore, the size-scaling of the accuracy of such temperature estimates with nano-diamonds was theoretically studied in \cite{alicki2014size}.

\subsection{Thermometry under thermalising dynamics}\label{sec:markovian}

The weak thermalising interactions between a quantum probe and its environment can be effectively described by a Markovian master equation in the standard Gorini--Kossakowski--Lindblad--Sudarshan (GKLS) form \cite{gorini1976completely,lindblad1976generators}. That is,
\begin{equation}
\frac{d\hat{\varrho}_\mathsf{P}}{dt} \coloneqq \mathcal{L}\hat{\varrho}_\mathsf{P} = -i [\hat{H}_\mathsf{P} + \Delta\hat{H}, \hat{\varrho}_\mathsf{P}] + \sum\nolimits_\omega \Gamma_{\omega}\big(\hat{A}_\omega\hat{\varrho}_\mathsf{P}\hat{A}_\omega^\dagger-\frac12\hat{A}_\omega^\dagger\hat{A}_\omega\hat{\varrho}_\mathsf{P}-\frac12\hat{\varrho}_\mathsf{P}\hat{A}_\omega^\dagger\hat{A}_\omega\big).
\label{eq5:lindblad}
\end{equation}
This may be derived from first principles, essentially by imposing that the thermalisation time is much longer than any other time scale in the problem \cite{BRE02}, which justifies the Born-Markov and secular approximations. Equation \eqref{eq5:lindblad} is ubiquitous in quantum optics, quantum information, or quantum thermodynamics, and comes with the advantage of guaranteeing a completely positive and trace-preserving dynamics, which converges monotonically to the thermal state $ \hat{\varrho}(T) = \mathcal{Z}_{\mathsf{P}}^{-1} e^{-\hat{H}_\mathsf{P}/T} $ \cite{spohn1978entropy}. 

Specifically, the form $ \hat{H}_{\mathsf{P}\leftrightarrow\mathsf{S}} = \hat{P}\otimes\hat{S} $ is assumed for the probe--system interaction, so that the probe operator $ \hat{P} $ decomposes as $ \hat{P} = \sum_\omega \hat{A}_\omega $. In equation \eqref{eq5:lindblad}, the sums run over the frequencies of the `open decay channels' and the operators $ \hat{A}_\omega $ satisfy $ [\hat{H}_\mathsf{P},\hat{A}_\omega] = -\omega \hat{A}_\omega $. The term $ \Delta \hat{H} $ accounts for the (usually neglected) Lamb and Stark shifts, and $ \Gamma_\omega $ denotes the `decay rate' of the open channel at frequency $ \omega $ \cite{BRE02}. In turn, this depends on the spectral density $ J(\omega) $ [cf. equation \eqref{eq4:spectral_density}].

\begin{figure*}[t!]
	\centering
	\includegraphics[width=0.49\linewidth]{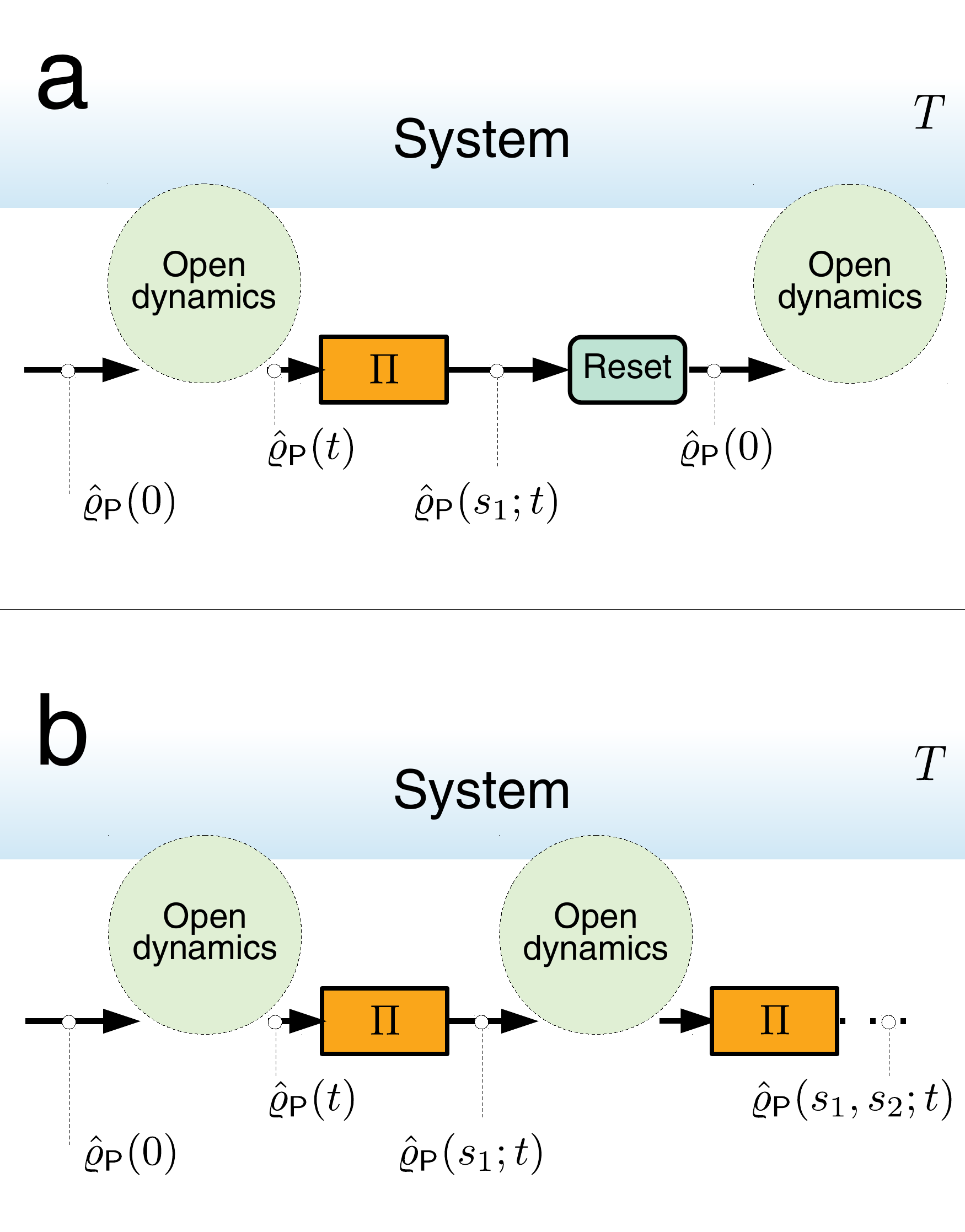}
	\includegraphics[width=0.49\linewidth]{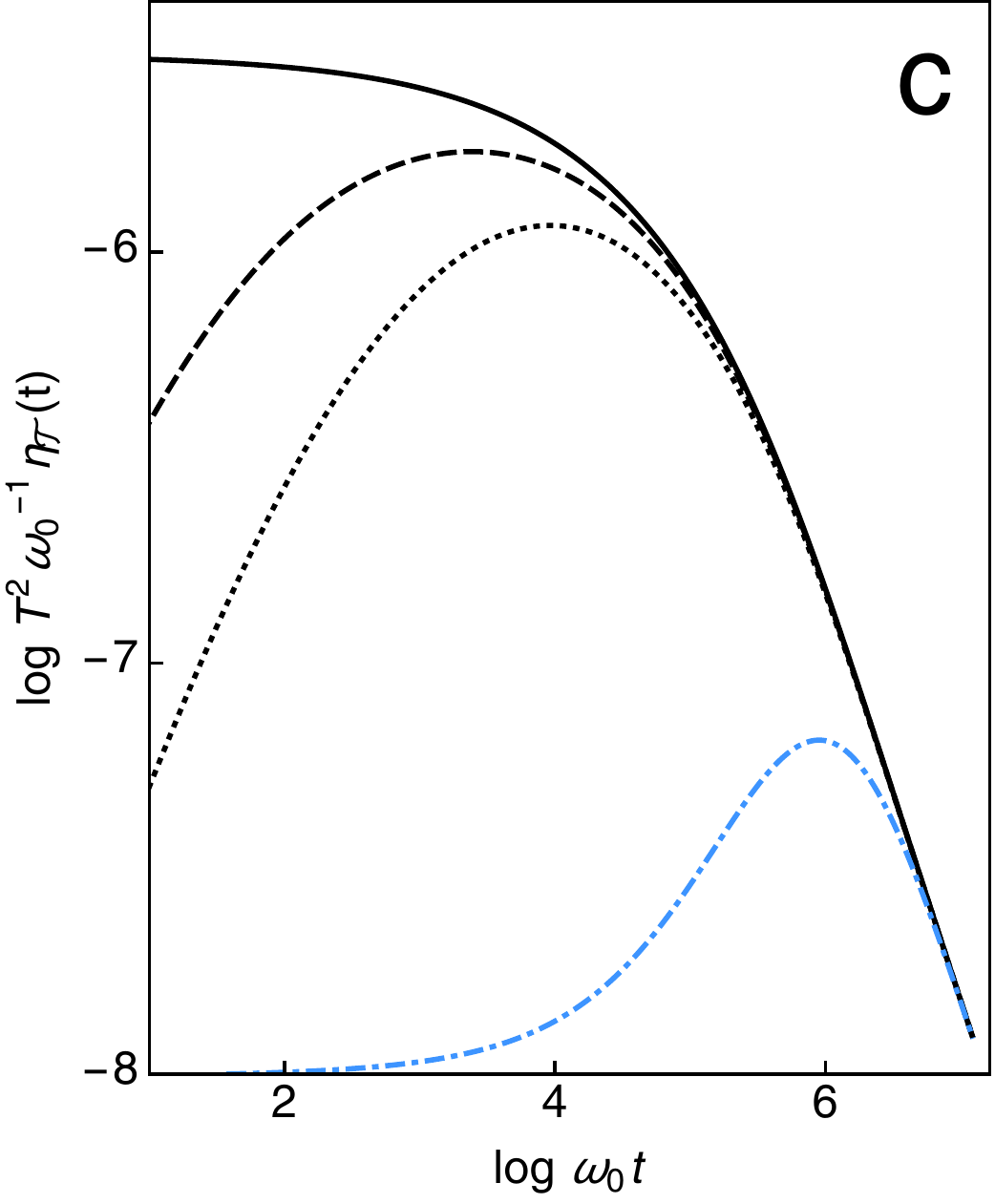}
	\caption{\textbf{(a)} Sequential setting in which the probe is initialised in the state $ \hat{\varrho}_\mathsf{P}(0) $ and evolved under a thermalising dissipative dynamics for time $ t $. The POVM $ \hat{\boldsymbol{\Pi}} $ is then performed, which yields the outcome $ s_1 $. Then, the probe is reset to its initial state $ \hat{\varrho}_\mathsf{P}(0) $ to iterate the process. \textbf{(b)} The same sequential setting as in (a), the only difference being that the probe is \textit{not} reset after every measurment. \textbf{(c)} Figure of merit $ \eta_{\mathcal{T}}(t) = \mathcal{F}(t)/t $ as a function of time for a two-level probe at fixed temperature $ T = 1 $, in a log-log scale (see text). The frequency of the probe has been set to the solution of $ e^{\omega_0/T} = (\omega_0/T+2)/(\omega_0/T-2) $ so as to maximise (asymptotically) the QFI at that temperature \cite{PhysRevLett.114.220405}. We compare the time-evolution of various preparations; namely, the ground state (solid), a thermal state at temperature $ T_\mathsf{P} = 0.8 $ (dashed), a thermal state at temperature $ T_\mathsf{P} = 1 $ (dotted), and a maximally coherent state (dot-dashed).}
	\label{fig5-1}
\end{figure*}

Let us assume that, for practical reasons, we need to base our temperature estimate on measurements performed during the transient dynamics to equilibrium. For instance, there might be constraints that severely limit the total available estimation time $ \mathcal{T} $. Therefore, we must partition it into an optimal number $ \mathcal{N} = \mathcal{T}/t $ of steps comprised of preparation, evolution, and measurement. Let us suppose as well that the probe is reset to its initial state after every measurement [see figure~\ref{fig5-1}(a)]. In this case, the Cram\'er--Rao bound would be
\begin{equation}
\Delta T \geq \frac{1}{\sqrt{\mathcal{N}\mathcal{F}(t)}} = \frac{1}{\sqrt{(\mathcal{T}/t)\,\mathcal{F}(t)}}
\label{eq5:efficiency}    
\end{equation}
Note that we refer to $ \mathcal{F}(T,\hat{\varrho}_\mathsf{P}(t)) $ as $ \mathcal{F}(t) $, for brevity. We thus see that the relevant figure of merit to be maximised is $ \eta_\mathcal{T}(t) \coloneqq \mathcal{F}(t)/t $, and not $ \mathcal{F}(t) $ alone \cite{liuzzo2017energy,Huelga1997}. The largest $ \eta_\mathcal{T}(t) $ is attained at some temperature-dependent optimal interrogation time when the probe is reset to its ground state after every measurement \cite{jevtic2015single,PhysRevLett.114.220405,de2017estimating,cavina2018bridging} (see figure \ref{fig5-1}). This is indeed the case for two-level and harmonic probes, and holds for any spectral density, provided that it is consistent with the approximations invoked in the derivation of equation \eqref{eq5:lindblad} \cite{PhysRevLett.114.220405}. Whenever the probe is a simple two-level system, the optimal measurement at any time throughout the thermalising evolution is again a projection onto the energy basis \cite{cavina2018bridging}.

For simplicity, we focus on this latter qubit case. Let the ground and excited states be $ \ket{g} $ and $ \ket{e} $, respectively, and let us limit ourselves to energy measurements. Since the probe is reset after every interrogation, the corresponding Fisher information can be cast as $ \mathcal{F}_c^{\,\text{res}}(n,t) = n \mathcal{F}_c(\hat{\sigma}_z,t) $, where $ \mathcal{F}_c(\hat{\sigma}_z,t) $ is the FI of a probe prepared in $ \hat{\varrho}_\mathsf{P}(0) $ and relaxing for time $ t $ before being measured. The superindex `res' stands here for `reset'. In turn, $ \mathcal{F}_c(\hat{\sigma}_z,t) $ can be calculated by feeding equation \eqref{eq5:Fisher} with the temperature-dependent probabilities
\begin{equation}
p_{\,\text{res}}(s_i\,\vert T) = \mbox{tr}\,\left\lbrace\Pi_{s_i} \circ e^{\mathcal{L}t} \, \hat{\varrho}_\mathsf{P}(0) \right\rbrace.
\label{eq5:probability_iid}
\end{equation}
Here, $ \Pi_s( \hat{\varrho}_\mathsf{P}) \coloneqq \ketbra{s}{s}\hat{\varrho}_\mathsf{P}\ketbra{s}{s} $ denotes the projective measurement (with $ s\in\{ g, e \} $), $ e^{\mathcal{L}t} $ evolves the state according to equation \eqref{eq5:lindblad}, and $ \circ$  stands for composition. 

It is interesting to compare $ \mathcal{F}_c^{\,\text{res}}(n,t) $ with the FI of a sequential protocol in which the probe is \textit{not} reset after each measurement [see figure \ref{fig5-1}(b)]. That is, when following each interrogation, the corresponding outcome is recorded and the the post-measurement state is left to evolve dissipatively until the next measurement \cite{de2017estimating}. We denote the FI of this sequential setting by $ \mathcal{F}_c^{\,\text{seq}}(n,t) $. After $ n $ steps, a list of outcomes $ \boldsymbol{s} = ( s_1, s_2, \cdots , s_n )^\mathsf{T} $ is registered with probability
\begin{equation}
p_{\,\text{sec}}\left(\boldsymbol{s}\,\vert T\right) = \mbox{tr}\,\left\lbrace \Pi_{s_n}\circ e^{\mathcal{L}t} \circ \cdots \circ \Pi_{s_2} \circ e^{\mathcal{L}t} \circ \Pi_{s_1} \circ e^{\mathcal{L}t}\,\hat{\varrho}_\mathsf{P}(0) \right\rbrace.    
\label{eq5:probability_sequential}
\end{equation}
Not surprisingly, the FI of the `measure-and-reset' scheme is larger than $ \mathcal{F}_c^{\,\text{seq}}(n,t) $ when $ \hat{\varrho}_\mathsf{P}(0) = \ketbra{g}{g} $. However, the sequential setting features a larger FI when the initial state of the probe is unknown. That is, $ \mathcal{F}_c^{\,\text{seq}}(n,t) > \mathcal{F}_c^{\,\text{res}}(n,t) $ when averaging over $ \hat{\varrho}_\mathsf{P}(0) $. This is so because, after the first measurement in the sequential strategy, the information about the initial state is essentially erased. Likewise, the gap between the performance of the best and worst possible preparations is much narrower in the sequential strategy. In any case, the difference between the two setups fades away as the evolution time $ t $ and/or the number of measurements $ n $ grow.

\subsubsection{The role of coherence and entanglement.}\label{sec:coherence-entanglement}

Up to now, we have seen that probes initialised in their ground state are more sensitive thermometers in dynamical scenarios. Still keeping simple thermalising dissipation, we wonder whether genuinely \textit{non-classical} features, such as coherence or entanglement, can be advantageous in some cases.

For instance, the problem of temperature discrimination was studied in reference \cite{jevtic2015single} by looking at the Euclidean distance between the Bloch vectors of two copies of a qubit, each of which being weakly coupled to a different heat bath. The baths have two different (and known) temperatures. The idea is to search for the conditions maximising the `distinguishability' of the two possible trajectories, as this increases the chances of guessing $ T $ correctly when observing the probe relax in either of the baths. In this scenario, the distinguishability is again globally maximised at short times by preparing the probe in its ground state $ \hat{\varrho}_\mathsf{P}(0) = \ketbra{g}{g} $. However, if we were forced to wait until longer times before measuring the probe, we would find that \textit{maximally-coherent} preparations [i.e., $ \hat{\varrho}_\mathsf{P}(0) = \ketbra{+}{+} $, with $ \ket{+} = \frac{1}{\sqrt{2}}(\ket{e}+\ket{g}) $] perform better than any other initial state. Furthermore, supplementing the probe with an ancillary qubit decoupled from the heat bath, and preparing them in an entangled state also leads to enhanced distinguishability, provided that one measures globally \textit{both} qubits \cite{jevtic2015single}. A discussion on an experimental demonstration of this idea can be found in reference \cite{tham2016simulating}.

In fact, as noted in reference \cite{kiilerich2018dynamical}, entanglement with such isolated ancilla is not essential to achieve a thermometric advantage; it is enough to set up probe-ancilla interactions. Indeed, an ancilla prepared in a maximally coherent state (and uncorrelated from the dissipative probe) can acquire almost all the temperature information available in the probe-ancilla composite after a long-enough evolution time.

Importantly, the transient performance of a thermalising probe can be larger than its asymptotic value \cite{PhysRevLett.114.220405,guo2015ring,jevtic2015single}. Let us look, for instance, at the model introduced in reference \cite{guo2015ring}, consisting of $ N $ two-level spins arranged in a ring structure. Recall that the steady-state properties of this setup were briefly discussed in section \ref{sec-GGS}. The dipole moments of the $ N $ spins are assumed to be parallel, and to couple linearly and weakly to the surrounding electromagnetic field. The dissipative evolution can thus be described with a Markovian master equation of the type \eqref{eq5:lindblad} (for a full derivation see, e.g., reference \cite{higgins2014superabsorption}). Specifically, the renormalisation term $ \Delta\hat{H} $ would introduce an effective \textit{coherent radiative coupling} between different spins. Assuming that this coupling is limited to nearest neighbours brings us back to equation \eqref{eq3:ring_interaction}, i.e., $ \Delta\hat{H} = \Omega \sum_{i} (\hat{\sigma}^-_{i}\hat{\sigma}^+_{i+1} + \mbox{h.c.}) $ \cite{higgins2014superabsorption}. In fact the radiative interactions are short-ranged and decrease with the distance as $ 1/r^3 $. 

The sign of $ \Omega $ can be controlled by tuning the angle between the dipole moments of the spins and the plane of the ring, and its magnitude, by adjusting the inter-spin distance. As it turns out, preparing the probe in a maximally entangled Greenberger--Horne--Zeilinger (GHZ) state \cite{greenberger1989going} and choosing $ \Omega > 0 $ leads to a maximum transient quantum Fisher information larger than its stationary value. In turn, this optimal value increases (specially at low temperatures) as the interaction strength $ \Omega $ grows \cite{guo2015ring}. 

\subsection{Interferometric quantum thermometry}\label{sec:interferometric}

\begin{figure*}[t!]
	\centering
	\includegraphics[width=0.60\linewidth]{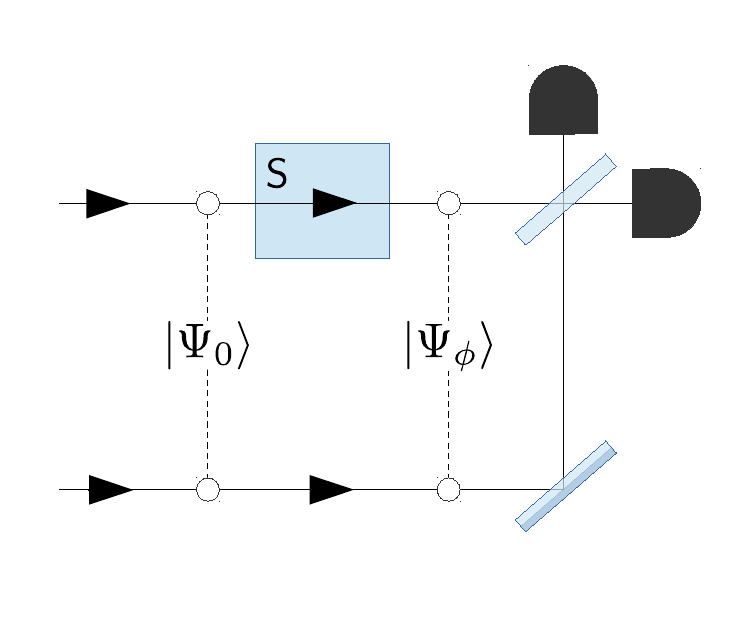}
	\caption{Sketch of an interferometric setup for temperature estimation. A single atom (or photon) in the path-coherent superposition $ \ket{\Psi_0} = \frac{1}{\sqrt{2}}(\ket{1,0} + \ket{0,1}) $ picks up a temperature dependent phase after interacting with the system $ \mathsf{S} $, located in the upper arm of the interferomenter. This results in $ \ket{\Psi_\phi} = \frac{1}{\sqrt{2}}(e^{i\phi}\ket{1,0}+\ket{0,1}) $. A faint beam of $ N $ such particles have a QFI of $ \mathcal{F}(T,\ketbra{\Psi_\phi}{\Psi_\phi}^{\otimes N}) = N \mathcal{F}(T,\ketbra{\Psi_\phi}{\Psi_\phi}) = N (\partial_T\phi)^2 $. On top of that, if the input is an $ N $-particle beam in a path-entangled NOON state $ \ket{\Psi_0} = \frac{1}{\sqrt{2}}(\ket{N,0} + \ket{0,N}) $, the phase accrued is $ N \phi $; that is, $ \ket{\Psi_\phi} = \frac{1}{\sqrt{2}}(e^{iN\phi}\ket{N,0} + \ket{0,N}) $. Accordingly, the QFI evaluates to $ \mathcal{F}(T,\ketbra{\Psi_\phi}{\Psi_\phi}) = \partial_T (N \phi)^2 $, thus scaling \textit{super-extensively} with the probe size.}
	\label{fig5-2}
\end{figure*}

Temperature can be estimated from the global phase picked up by the probe during its (weak) interaction with the system \cite{stace2010quantum,martinez2013berry,sabin2014impurities,jarzyna2015quantum}. In turn, this can be accessed by means of an interferometric setup (see figure \ref{fig5-2}). For instance, following \cite{stace2010quantum}, one can think of a faint $ N $-atom beam entering an interferometer through a beam splitter, so that each probe atom is brought into the `path-coherent' state $ \frac{1}{\sqrt{2}}(\ket{1,0} + \ket{0,1}) $, i.e., into a superposition of travelling through either of the two arms of the interferometer. Upon interacting with the system---placed in one of the arms---information about its temperature is finally mapped into a measurable \textit{relative} phase $ \phi $. It is well known that, in this setting, the estimation of $ \phi $ is shot-noise-limited, i.e., $ \Delta\phi \sim 1/\sqrt{N} $ \cite{PhysRevA.65.052104}. Nonetheless, a better (Heisenberg-limited) scaling of $ \Delta \phi \sim 1/N $ becomes possible when the probe is prepared in a `path-entangled' state of the form $ \frac{1}{\sqrt{2}}(\ket{N,0} + \ket{0,N}) $ \cite{PhysRevA.40.2417,PhysRevLett.85.2733}, also referred-to as `NOON state'. This is so since the relative phase acquired by the probe is amplified to $ N \phi $ (cf. caption of figure \ref{fig5-2}). 

As a simple illustration, let us consider a probe and a system formed of identical non-interacting two-level atoms with Hamiltonian $ \hat{H}_\alpha = \Omega \sum_i {\ketbra{e_i}{e_i}}_\alpha $, where $ \alpha\in\{\mathsf{P},\mathsf{S}\} $ and $ \ket{e_i}_\alpha $ stands for the excited state of the $i$-th atom in $ \alpha $. Under the non-dissipative coupling 
\begin{equation}
\hat{H}_{\mathsf{P}\leftrightarrow\mathsf{S}} = \alpha\sum_{i=1}^N\sum_{j=1}^M \ketbra{e_i}{e_i}_\mathsf{P}\otimes\ketbra{e_j}{e_j}_\mathsf{S}
\label{eq5:interferometer_interaction}
\end{equation}
acting for a time $ t $, a \textit{single} probe atom in its excited state would pick up a phase of $ \phi = \alpha M t p_e $, where $ p_e = (e^{\Omega/T}+1)^{-1} $ is the thermal occupation of the excited state of each of the $ M $ system atoms \cite{stace2010quantum}---temperature may be then directly extracted from $ \phi $. However, $ N $ of such excited probe atoms interacting \textit{simultaneously} with the system would acquire the larger global phase $ \phi = \alpha M N t p_e $. This way, even if $ N \ll M $ so as to minimise the disturbance on the system, very precise thermometry is possible.

Even in the absence of path-entanglement, interferometric thermometry can be a practical alternative to state-of-the-art pyrometers. Interferometric thermometers can indeed detect small temperature fluctuations in, e.g., non-linear crystals, which is relevant for quantum optics \cite{jarzyna2015quantum}. Shining a single-mode Gaussian state of light through a piece of crystal results in a phase shift $ \phi = \omega c \ell $, where $ \ell = n L $ is the corresponding `optical path length', $ n $ stands for the index of refraction, and $ L $, for its thickness \footnote{This process can be modelled with a Markovian quantum master equation (cf. section \ref{sec:markovian}).}. Any infinitesimal variation of temperature changes both the index of refraction, and the length of the crystal. Therefore, the optical path $ \ell $ experiences small variations
\begin{equation}
\delta\ell = L (\partial n/\partial T) \delta T + n (\partial L/\partial T) \delta T.
\label{eq5:optical_length}
\end{equation}
In turn, this translates into measurable phase shifts $ \delta\phi $ relative to a fiducial beam. In reference \cite{jarzyna2015quantum}, the thermal sensitivity of these shifts was computed and benchmarked against the performance of an ideal pyrometer. These measure temperature non-invasively, by fitting the flux of thermal radiation emitted by a system (i.e., its `irradiance') to the Stefan--Boltzmann law \cite{michalski2001pyrometers}. Equivalently, one may think of performing photon counting measurements on every mode of the outcoming equilibrium radiation. Recall from section \ref{sec-best-individual} that such energy measurements are optimal for inferring temperature directly on equilibrium systems. The result of the comparison is that an interferometric quantum thermometer can outperform the best pyrometers, when fed either squeezed or coherent input states with a large number of photons---although not so large so as to cause substantial heating of the system. 

References \cite{martinez2013berry,sabin2014impurities} consider a similar model to that of section \ref{sec:micromechanical}, i.e., a two-level thermometer ($ \hat{H}_\mathsf{P} = \Omega \hat{\sigma}_z $) dissipatively coupled to a single near-resonant thermalised mode ($ \hat{H}_\mathsf{S} = \omega\hat{a}^\dagger\hat{a} $), representing a quantised field in a cavity. As an alternative to population measurements, one could estimate temperature from the relative phase, between the actual thermometer and a reference probe placed in a cavity of known temperature. In the interaction picture with respect to $ \hat{H}_0 = \hat{H}_\mathsf{S} + \hat{H}_\mathsf{P} $ the Hamiltonian of interest becomes
\begin{equation}
\hat{H}_{\mathsf{P}\leftrightarrow\mathsf{S}}(t) = \gamma(\hat{\sigma}^-e^{-i\Omega t} + \hat{\sigma}^+ e^{i\Omega t})(\hat{a}e^{-i\omega t} + \hat{a}^\dagger e^{i\omega t}),
\label{eq5:Jaynes-Cummings-inter}    
\end{equation}
where $ \hat{\sigma}^- $ ($ \hat{a} $) is the lowering (annihilation) operator in the Hilbert space of the probe (system). Provided that, $ \delta = \Omega - \omega $ is sufficiently small, one may perform the rotating-wave approximation, leading to $ \hat{H}_{\mathsf{P}\leftrightarrow\mathsf{S}}(t) = \gamma(\hat{\sigma}^-\hat{a}^\dagger e^{-i \delta t} + \hat{\sigma}^+\hat{a} e^{i\delta t}) $.

Due to the slow time-dependence of $ \hat{H}_{\mathsf{P}\leftrightarrow\mathsf{S}}(t) $ the adiabatic theorem \cite{born1928beweis,kato1950adiabatic} can be invoked to calculate the accumulated phase over a period of time. In particular, this may be split into a \textit{dynamical} and a \textit{geometric} contribution (or Berry phase) \cite{berry1984quantal}. This phases depend on the state of the cavity mode and, ultimately, on the temperature of the cavity. While reference \cite{martinez2013berry} focuses on the thermal sensitivity of the geometric phase, reference \cite{sabin2014impurities} argues that the dynamical phase can perform substantially better as a temperature estimator. Furthermore, a proposal is put forward to realise this thermometric scheme in a Bose--Einstein condensate at sub-$ \si{\nano\kelvin} $ temperatures. 

To conclude, let us mention a recent proposal to measure the temperature of the vibrational modes of linearly trapped ions by coupling them to their internal degrees of freedom \cite{ivanov2019quantum} using a bichromatic laser beam \cite{schneider2012experimental,lee2005phase}. Eventually, temperature information might be optimally extracted from the collective spin of the system, measured via Ramsey spectroscopy. 

\subsection{Thermometry under dynamical control} \label{sec:dynamical_control}

\subsubsection{Temperature estimation from fluctuation relations.} \label{sec:tasaki-crooks}

A somewhat different proposal for dynamical thermometry consists in quenching an already equilibrated system and exploiting non-equilibrium fluctuation theorems to extract the desired temperature information \cite{PhysRevA.93.053619}. The idea is as follows: Let us consider an equilibrium system at temperature $ T $ with Hamiltonian $ \hat{H}(\lambda) = \hat{H}_0 + \lambda \hat{V} $, where $ \lambda $ is a quench parameter. Implementing a protocol in which $ \lambda $ varies from $ \lambda(0) $ to $ \lambda(\tau) $ takes the system to the non-equilibrium state $ \hat{\varrho}(\lambda(\tau)) = \hat{U}(\tau)\,\hat{\varrho}(\lambda(0))\,\hat{U}^{\dagger}(\tau) $, where $ \hat{U}(\tau) = \mathbb{T}\,\exp{\big\lbrace -i\int_0^{\tau}dt~\hat{H}(\lambda(t))\big\rbrace}$, and $ \mathbb{T} $ denotes time ordering. 

During this protocol, the probability that an amount $ w $ of `work' is performed on the system is given by
\begin{align}
P_F(w) = \sum_{l,m} p_m^0\,p_{l \vert m}^{\tau}\delta(w - (E^{\prime}_l - E_m)),
\label{eq5:forward_probability}
\end{align}
where $ p_m^0 $ is the probability to find the equilibrium system in an eigenstate of $ \hat{H}(\lambda(0)) $ with energy $ E_m $ at $ t = 0 $, and $ p^\tau_{l \vert m} $ is the probability of finding it in an eigenstate of $ \hat{H}(\lambda(\tau)) $ with energy $ E'_l $ at $ t = \tau $ \textit{conditioned} on the outcome of the first measurement being $ E_m $. The subindex $ F $ in equation \eqref{eq5:forward_probability} stands for `forward protocol'. Similarly, one can define the probability distribution for the `backwards protocol', which evolves the equilibrium state $ \propto \exp{\{ -\hat{H}(\lambda(\tau))/T \}} $ according to the reversed prescription $ \lambda_B(t) \coloneqq \lambda(\tau - t) $. The Tasaki--Crooks fluctuation theorem states that \cite{tasaki2000jarzynski,PhysRevE.60.2721,talkner2007tasaki,RevModPhys.83.771}
\begin{align}\label{eq:tasaki-crooks}
R = \frac{P_F(w)}{P_B(-w)} = e^{\beta (w - \Delta F)},
\end{align}
where $ \beta = 1/T $ and $ \Delta F = F(\tau) - F(0) $, with $ F(t) \coloneqq T \log \text{tr}\big(\exp{\{-\hat{H}(\lambda(t)/T) \}}\big) $. Therefore, the initial temperature might be extracted from $ R $. Importantly, the work distribution or, more precisely, its characteristic function
\begin{align}
\chi(u) \coloneqq \int_{-\infty}^{\infty} dw~e^{i w u}P(w)
\end{align}
can be measured experimentally by attaching an auxiliary two-level probe to the system \cite{PhysRevLett.110.230601,PhysRevLett.110.230602,PhysRevE.75.050102,PhysRevLett.113.140601}. The information about $ \chi $ encoded in the state of the probe may be then extracted by measuring different Pauli operators on the qubit state. 

In this framework, the temperature of, e.g., ultracold atomic systems might be estimated in the sub-$ \si{\nano\kelvin} $ domain without detailed knowledge of their Hamiltonian. However, a perfect control of the system-probe interactions is necessary. Likewise, multiple sources of error propagate onto the final temperature estimate \cite{PhysRevA.93.053619}.

\subsubsection{Thermometry with periodically driven probes.}\label{sec:periodically_driven}

We now return to the thermalising dissipative dynamics from section \ref{sec:markovian}. However, rather than letting the probe equilibrate passively, we modulate its spectrum periodically. To describe this, we shall remain in the range of validity of the Born--Markov approximation, and introduce a periodic time dependence on the probe Hamiltonian, so that $ \hat{H}_\mathsf{P}(t) = \hat{H}_\mathsf{P}(t+\tau) $ and $ \tau = 2\pi/\nu $. 

If the driving was slow compared to any other time scale in the problem, one could simply use equation \eqref{eq5:lindblad} as discussed in \cite{alicki1979engine}. In general, however, it is necessary to modify it by resorting to Floquet theory \cite{alicki2014quantum,szczygielski2014application,alicki2015non}. The first step to generalise \eqref{eq5:lindblad} is to search for a `time-averaged' Hamiltonian $ \hat{\mathcal{H}} $, such that 
\begin{equation}
\hat{U}(\tau) = \mathbb{T}\exp{\left\lbrace -i \int_0^\tau ds\,\hat{H}_\mathsf{P}(s) \right\rbrace} \coloneqq e^{-i \hat{\mathcal{H}} \tau}.
\label{eq5:time_averaged}    
\end{equation}
Recall from section \ref{sec:markovian} that the standard choice for the probe--system coupling is $ \hat{H}_{\mathsf{P}\leftrightarrow\mathsf{S}} = \hat{P}\otimes\hat{S} $. After moving to the interaction picture with respect to $ \hat{H}_\mathsf{P}(t) + \hat{H}_\mathsf{S} $, the probe coupling operator $ \hat{P} $ can be decomposed as \cite{szczygielski2014application}
\begin{equation}
\hat{P}_I(t) \coloneqq \hat{U}^\dagger(t)\,\hat{P}\,\hat{U}(t) = \sum_{\omega}\sum_{q\in\mathbb{Z}} \hat{A}_{\omega + q \nu}\,e^{-i(\omega + q \nu) t},
\label{eq5:probe-sample-Fourier}    
\end{equation}
where the first summation runs through the Bohr frequencies $ \omega $ of $ \hat{\mathcal{H}} $ and $ \hat{A}_{\omega + q \nu} $ are suitable non-Hermitian operators. 

In general, finding $ \hat{\mathcal{H}} $ and $ \hat{A}_{\omega + q \nu} $ is far from trivial, but it can be done by mere inspection in simple cases \cite{alicki2014quantum}. For a periodically modulated harmonic thermometer with $ \hat{H}_\mathsf{P} = \omega(t)\, \hat{a}^\dagger\hat{a} $ and $ \hat{P} = \hat{a} + \hat{a}^\dagger $, one finds \cite{mukherjee2017high,alicki2014quantum} 
\begin{subequations}
	\begin{align}
	\hat{A}_{\omega + q \nu} &= P(q)\,\hat{a} \\
	P(q) &= \tau^{-1}\int\nolimits_0^\tau dt\,\exp{\left\lbrace i \int\nolimits_0^t ds\,\left( \omega(s) - \omega_0 \right) \right\rbrace} e^{-iq \nu t}.\\
	\omega_0 &\coloneqq \tau^{-1}\int\nolimits_0^\tau dt\,\omega(t).
	\end{align}
	\label{eq5:sideband_weight} 
\end{subequations}
Combining equations \eqref{eq5:probe-sample-Fourier} and \eqref{eq5:sideband_weight} with the standard derivation of \eqref{eq5:lindblad} (cf. reference \cite{BRE02}) yields the generalised master equation 
\begin{subequations}
	\begin{align}
	\frac{d\hat{\varrho}_\mathsf{P}}{dt} = -i[\hat{H}_\mathsf{P}(t),\hat{\varrho}_\mathsf{P}] + \sum_{q\in\mathbb{Z}}\vert P(q) \vert^2 \mathcal{L}_{q} \hat{\varrho}_\mathsf{P}. 
	\end{align}
	\vspace{-15pt}
	\begin{multline}
	\mathcal{L}_q\hat{\varrho}_\mathsf{P}  \coloneqq \sum_{q\in\mathbb{Z}}\vert P(q) \vert^2\,\left[\Gamma_{\omega_0 + q\nu}\left(\hat{a}\hat{\varrho}_\mathsf{P}\hat{a}^\dagger-\frac12\hat{a}^\dagger\hat{a}\hat{\varrho}_\mathsf{P}-\frac12\hat{\varrho}_\mathsf{P}\hat{a}^\dagger\hat{a}\right)\right. \\
	+ \left. \Gamma_{-\omega_0 - q\nu}\left(\hat{a}^\dagger\hat{\varrho}_\mathsf{P}\hat{a}-\frac12\hat{a}\hat{a}^\dagger\hat{\varrho}-\frac12\hat{\varrho}_\mathsf{P}\hat{a}\hat{a}^\dagger\right) \right].    
	\end{multline}
	\label{eq5:floquet}
\end{subequations}

Looking closely at equations \eqref{eq5:floquet}, we notice that the periodically-driven oscillator can seemingly interact with the system at various frequencies $ \{\omega_0 + q \nu\}_{q\in\mathbb{Z}} $ \textit{simultaneously}. In particular, the QFI about $ T $ of its `time-averaged' stationary solution $ \hat{\varrho}(\infty) $ (i.e., $ \sum_q\vert P(q) \vert^2\mathcal{L}_q\,\hat{\varrho}(\infty) = 0 $) can be cast as $ \mathcal{F}(T) = \sum_{q\in\mathbb{Z}} \mathcal{F}_q(T) $ \cite{mukherjee2017high}. 

Interestingly, according to equation \eqref{eq5:sideband_weight}, the number of sidebands to be considered and their share in the total QFI is controlled by $ \omega(t) $, which allows to engineer tailored multi-peaked profiles for the thermal sensitivity. In this way, the \textit{same} dynamically controlled probe could either remain sensitive over a wide range of temperatures, measure very precisely only in the neighbourhood of certain temperature of interest, or estimate arbitrarily low temperatures with a non-diverging relative error \cite{mukherjee2017high,hovhannisyan2017probing}. This should not be confused with the multi-peaked QFI profiles arising from equilibrated probes with degenerate energy spectra discussed in references \cite{2058-9565-3-2-025002,plodzien2018few} and section \ref{sec-best-individual}.

\subsubsection{Quantum heat pumps as nano-thermometers.}\label{sec:quantum_heat_pump}

\begin{figure*}[t!]
	\centering
	\includegraphics[width=0.60\linewidth]{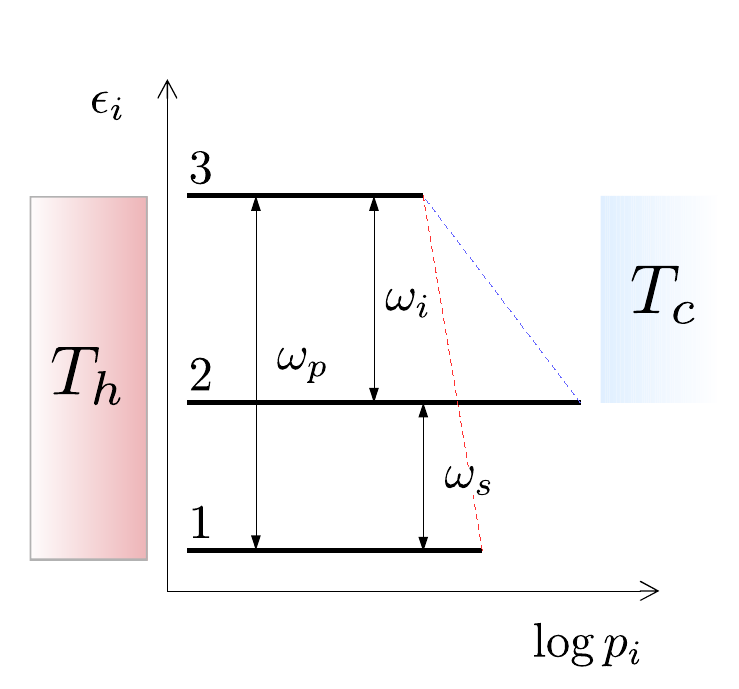}
	\caption{Schematic representation of a three-level maser. The states $ \{\ket{1},\ket{2},\ket{3}\} $, are represented by lines spaced vertically according to their energy. As in reference \cite{PhysRev.156.343}, the length of these lines is proportional to the logarithm of their steady-state population $ p_i $. The `pump transition' $ \ket{1}\leftrightarrow\ket{3} $ is thermally coupled to a high-temperature bath at $ T_h $, while the `idler transition' $ \ket{2}\leftrightarrow\ket{3} $ couples to a cold bath at $ T_c $. The slope of the red and blue dashed lines is thus proportional to $ -1/T_h $ and $ -1/T_c $, respectively. Power extraction would then occur at the population-inverted `signal transition' $ \ket{1}\leftrightarrow\ket{2} $, provided that the pump is coupled to an external field.}
	\label{fig5-3}
\end{figure*}

Another way to bring an open quantum system into a useful non-equilibrium stationary state is to couple it simultaneously to several heat baths at different temperatures. The direction of the stationary energy flows in the resulting `quantum heat pump' can be engineered to realise various continuous energy conversion processes, such as quantum refrigerators or quantum heat engines. These nanoscale heat devices have long been instrumental in \textit{quantum thermodynamics} to unveil the connections between thermodynamics and quantum theory \cite{e15062100,1310.0683v1,gelbwaser2015thermodynamics,vinjanampathy2016quantum,goold2016role}. Furthermore, proof-of-principle implementations of quantum energy conversion cycles have been realised in a variety of experimental platforms, including solid-state electronic spins \cite{5123190}, cold atomic gases \cite{PhysRevLett.119.050602}, trapped ions \cite{rossnagel2015single,maslennikov2017quantum}, or Nitrogen-vacancy centres in diamond \cite{klatzow2017experimental}. Interestingly enough, the earliest proposed practical application for a quantum heat pump was indeed temperature estimation \cite{PhysRev.156.343}.

Following reference \cite{PhysRev.107.1579}, let us consider the three-level `maser' depicted in figure \ref{fig5-3}. This is meant to attain a non-equilibrium stationary regime in which thermal excitations from a hot bath are traded for coherent emission at some lower frequency $ \omega_s $. The waste heat generated in the process is dumped into an auxiliary (and colder) `entropy sink'. Using the terminology of figure \ref{fig5-3}, this operation mode (or `maser action') is realised as long as the signal transition is population-inverted when decoupled from the work repository; i.e., when 
\begin{equation}
\frac{p_2}{p_1} = \frac{p_2}{p_3}\frac{p_3}{p_1} = \exp{\left(\frac{\omega_i}{T_c}\right)} \exp{\left(-\frac{\omega_p}{T_h}\right)} = \exp{\left[\frac{\omega_s}{T_c}\left(\frac{\omega_p}{\omega_s}\left(1-\frac{T_c}{T_h}\right)-1\right)\right]} > 1.
\label{eq5:engine_operation}    
\end{equation}

Due to stationarity, heat absorption, heat rejection, and coherent emission must all occur at the same rate. Hence, the \textit{maser efficiency}---or the ratio of the output emission at the signal transition to the input heat---is simply equal to $ \eta_M = \omega_s/\omega_p $. According to equation \eqref{eq5:engine_operation} this entails
\begin{equation}
\eta_M < 1 - \frac{T_c}{T_h} = \eta_C,   
\label{eq5:carnot_eff}
\end{equation}
where $ \eta_C $ is the efficiency of a Carnot cycle operating between temperatures $ T_h $ and $ T_c $. This interesting analogy between a maser and a heat engine was first noted in the pioneering 1959 paper by Scovil and Shulz-DuBois \cite{PhysRevLett.2.262}---that is, two decades before the formulation of the theory of weakly coupled open quantum systems \cite{davies1974markovian,gorini1976completely,lindblad1976generators} and the subsequent development quantum thermodynamics \cite{alicki1979engine,kosloff1984quantum,binder2018thermodynamics}. 

The temperature $ T_c $ of the cold bath in figure \ref{fig5-3} can be estimated by measuring $ T_h $ and tuning $ \omega_s $ so that the energy conversion rate vanishes, which heralds the saturation of the upper bound in equation \eqref{eq5:carnot_eff}. Alternatively, it may be easier to adjust $ T_h $ while keeping all frequencies fixed \cite{hofer2017machine}. Once in the target configuration, the signal transition must feature complete `transparency', i.e., stimulated emission and absorption cancel exactly, while the pump transition must look thermalised at temperature $ T_h $ \cite{PhysRev.156.343}. Spectroscopic measurements of $ \omega_s $ and $ \omega_p $ would then allow to evaluate $ \eta_M = \omega_s/\omega_p = 1-T_c/T_h $ and thus infer $ T_c $.

Interestingly, nothing prevents one of the reservoirs from exhibiting a \textit{negative} absolute temperature \cite{ramsey1956thermodynamics}. Whether or not negative temperatures are thermodynamically consistent has fuelled a longstanding debate in the literature (see, e.g., recent papers against \cite{dunkel2014consistent} and in favour \cite{frenkel2014gibbs}), and is certainly outside the scope of this review. In any case, as pointed out in reference \cite{PhysRev.156.343}, an unknown negative temperature may be measured in exactly the same way as a positive one with a three-level maser thermometer.

A circuit QED implementation of such thermometric scheme was studied in reference \cite{hofer2017machine}; namely, two coupled harmonic oscillators, each one in contact with a local heat bath. This can be realised as a limiting case of two microwave cavities coupled through a Josephson junction \cite{hofer2016quantum}. Using the language of figure \ref{fig5-3}, the cavities (at frequencies $ \omega_i $ and $ \omega_p $) would correspond to the `idler' and `pump' thermal contacts with the cold and hot baths. The point of vanishing dc electrical current, which signals the maximisation of the efficiency, can be found by varying $ T_h $. Hence, errors in the measurement of the current $ \Delta \langle\hat{I}\rangle $ and the hot temperature $ \Delta T_h $ propagate to the final estimate of $ T_c $. Remarkably, precisions as good as $ \Delta T_c \leq \SI{2}{\milli\kelvin} $ at $ T_c = \SI{15}{\milli\kelvin} $ might be possible for realistic experimental parameters \cite{hofer2017machine}. Adopting the viewpoint of quantum estimation theory, it can also be seen the dc current is not an optimal temperature estimator in this setting (particularly at low $ T_c $). Therefore, alternative estimators might yield even better sensitivities in this setup.

We finally note that exploiting the transport properties of non-equilibrium systems is reminiscent of the way in which `local intensive temperature' is defined in mesoscopic physics---by means of ideal non-invasive scanning probes \cite{PhysRevB.24.1151}. In essence, a temperature can be locally assigned to a small sample, kept out of equilibrium by means of external `thermostats', if a probe lead is weakly coupled to it \cite{PhysRevB.85.125120}. In fact, the probe can be tuned to a unique temperature and chemical potential, so that both electric and heat currents vanish at the interface with the sample \cite{PhysRevB.94.155433}. In turn, this provides an unambiguous operational definition for intensive thermodynamic properties arbitrarily far from equilibrium. Pushing the analogy, in the heat-engine thermometer above the thermostats would be the driving field and the cold bath, while the tunable hot bath would play the role of the probing lead. 

\section{Summary and outlook}\label{sec6}

In this review we have formulated thermometry as a problem of inference and statistical error propagation, where `temperature' has been defined merely as an unknown parameter in the \textit{ad hoc} Gibbs state of the system (cf. section \ref{sec2}). We have classified the growing body of theoretical works on quantum thermometry into three main categories. Namely,
\begin{itemize}
	\item Quantum thermometry from (stationary) \textit{equilibrium} states (section \ref{sec3}).
	\item Quantum thermometry from \textit{stationary non-equilibrium} states (section \ref{sec4}).
	\item Quantum thermometry from \textit{time-evolving} (non-equilibrium) states (section \ref{sec5}).
\end{itemize}

When it comes to the first point above, a neat universal formulation can be made, leading to model-independent results. Namely, we have seen that the signal-to-noise ratio for temperature measurements in any equilibrium system is upper-bounded by its heat capacity. Moreover, energy measurements turn out to saturate such bound \cite{PhysRevE.83.011109} (cf. section \ref{sec-Thermometry-Precision}). 

A number of additional questions may be then answered. For instance, we have shown that the most sensitive $ N $-dimensional system at any given temperature is an effective qubit system with an $ (N-1) $-fold degenerate excited state and a suitable gap \cite{PhysRevLett.114.220405}. We have also seen how precision can be traded for versatility, while discussing the advantages of sub-optimal highly-degenerate energy spectra \cite{plodzien2018few,2058-9565-3-2-025002} (cf. section \ref{sec-best-individual}). In particular, interacting fermionic gases \cite{serwane2011deterministic,wenz2013few,PhysRevLett.108.075303,PhysRevLett.111.175302,PhysRevA.88.033607,PhysRevLett.111.045302,PhysRevA.87.060502} have been shown to be a promising platform to craft highly sensitive quantum thermometers with a large degree of control. Looking closely into the interplay between internal interactions in multipartite probes and thermal sensitivity, by resorting to a simple example, we have seen how internal couplings may be harnessed as a control knob to fine-tune the energy spectrum of a system and substantially enhance its sensitivity, e.g., at low temperatures \cite{1367-2630-19-10-103003} (cf. section \ref{sec-Coupled-ho-Thermal}). 

Taking a step forward, we have scaled up the models of interest from few interacting parts to quantum many-body systems in the thermodynamic limit. This has allowed us to discuss the effects of quantum criticality on low-temperature sensing (cf. section \ref{sec3:criticality}). In particular, we have seen how low-temperature thermometry can be greatly enhanced close to the critical lines of the paradigmatic XY model, owing to its closing gap \cite{1367-2630-17-5-055020,paris2015achieving} (cf. section \ref{sec-XY}). In practice, one can benefit from this enhancement by collecting temperature information from global measurements that are, in principle, feasible via non-demolition methods, such as Faraday polarization spectroscopy \cite{Eckert2007a,RevModPhys.90.035005,RevModPhys.82.1041,roscilde09}. We have also noted that the ultimate thermometric precision scales \textit{polynomially} with temperature in cold atomic lattice gases at criticality \cite{hofer2017fundamental,PhysRevA.94.042121}, and in homogeneous and non-homogeneous quantum gases \cite{PhysRevA.88.063609} (cf. \ref{sec3:size_temp_scaling}). In this respect, one interesting open question is how does thermal sensitivity scale in the neighbourhood of quantum phase crossovers in near-ground-state many-body systems that (contrary to the XY model) \textit{cannot} be mapped into non-interacting quasi-particles.

Another interesting aspect of thermometry with an equilibrium system is that imposing symmetries can drastically change (and potentially, improve) its thermal sensitivity. Indeed, forcing the system to occupy only a sector of its Hilbert state space might very significantly enhance its low-temperature thermal sensitivity (cf. section \ref{sec-GGS}). To illustrate this point, we have looked into the Bose--Hubbard model (with conserved particle number) \cite{1367-2630-19-10-103003}, and into a ring of radiatively coupled two-level emitters (with conserved total angular momentum) \cite{guo2015ring}. This appears as yet another promising line for further investigation.

An important issue is the measurement \textit{back action} on the system; clearly, one would like to extract the maximum amount of information causing the minimum possible disturbance \cite{stace2010quantum,jarzyna2015quantum}. Information-theoretic quantifiers of measurement-induced disturbance \cite{seveso2018trade,PhysRevA.98.012115} have been also briefly discussed and related to thermometric performance in section \ref{sec-inf-dist}. 

Jumping now to the second point in the list above, we may find ourselves needing to estimate temperature from \textit{local} measurements on a large system. Such restriction leaves us with access only to the temperature information encoded in a \textit{non-equilibrium} marginal of the equilibrium state of the global system. An alternative way to phrase the same problem is to think of the stationary state of a thermometer strongly coupled to an initially thermal reservoir. This can be tackled from an open-system perspective. 

When speaking about local thermometry, one usually needs to focus on specific models, thus giving up generality. Nevertheless, we have discussed some fundamental results on the regime of validity of local energy-temperature uncertainty relations \cite{de2016local,de2017universal}, and also on how these can be generalised to hold in any regime of parameters \cite{miller2018mean} (cf. section \ref{sec:local_temperature_fluctuations}). We have also shown very generally that local thermometry on gapped short-range-interacting many-body systems is always \textit{exponentially inefficient} \cite{hovhannisyan2017probing}. As we have already mentioned, enabling sub-exponential scaling calls instead for a \textit{gapless} global spectrum \cite{hovhannisyan2017probing,hofer2017fundamental} (cf. section \ref{sec:exponential_ineff_gapped}). The effects of introducing long-ranged interactions remains, however, an open problem.

For instance, the scaling of the thermal sensitivity of a Brownian thermometer immersed in a bosonic bath has been shown to be \textit{always} quadratic \cite{hovhannisyan2017probing}, provided that the setup is well described by the (critical) Caldeira--Leggett model and that the corresponding dissipative interactions are of the Ohmic type. Elaborating more on this quantum Brownian model for temperature sensing, we have also reported how the dissipation strength might be harnessed for high precision thermometry, so long as that the temperature is sufficiently low \cite{correa2016low} (cf. section \ref{sec:strong_coupling_thermometry}). Interestingly, in a suitable regime of parameters, this setup accurately describes the relaxation of an impurity embedded in a Bose--Einstein condensate, which is an example of sub-nanokelvin thermometry in an technologically relevant situation  \cite{lampo2018open,mehboudi2018using}.

The final point in our list has been \textit{dynamical} quantum thermometry; that is, the study of the precision of temperature estimates drawn from measurements on (non-equilibrium) systems as they undergo some evolution. We have seen that, in such dynamical settings, one needs to find, not only the optimal measurement allowing for the minimum uncertainty in the temperature estimate, but also the best initial state of the probe, and the optimal interrogation time \cite{brunelli2011qubit,higgins2012quantum,brunelli2012qubit,jevtic2015single,PhysRevLett.114.220405,guo2015ring,de2017estimating,ivanov2019quantum} (cf. sections \ref{sec:micromechanical} and \ref{sec:markovian}). This is critical whenever the lifetime of the system is so short that the estimation time becomes a scarce resource \cite{PhysRevLett.114.220405}. We have also argued that all the existing proposals for \textit{interferometric} quantum thermometry \cite{stace2010quantum,jarzyna2015quantum,martinez2013berry,sabin2014impurities} belong to this category, since they rely on the interrogation of a probe beam that interacts for a \textit{finite} time (either dispersively or dissipatively) with the equilibrium system of interest (cf. section \ref{sec:interferometric}). 

It is clear that any genuinely quantum features present in the preparation of a temperature probe become irrelevant once it has fully equilibrated with the system of interest. However, we have pointed out that initial coherences \cite{jevtic2015single,de2017estimating} or entanglement \cite{jevtic2015single,guo2015ring} might indeed play a relevant role in improving thermometric bounds during the \textit{transient} dynamics (cf. section \ref{sec:coherence-entanglement}). Finally, we have illustrated how supplementing passive thermalising evolution with active dynamical control yields very versatile tunable thermometers \cite{PhysRevA.93.053619,mukherjee2017high,PhysRev.156.343,hofer2017machine} (see section \ref{sec:dynamical_control}).

We would not like to finish this review without acknowledging the crucial importance and relevance of the experimental advances in quantum thermometry. Controlling and measuring the temperature in the different devices and platforms that operate in the quantum regime is essential for any application. Clear examples are ultracold atoms or ions acting as quantum simulators of condensed matter physics, quantum chemistry, or high energy physics \cite{lewenstein2012ultracold,kim2010quantum}; quantum processors in cold-ion platforms, where the initialisation near the motional ground state is crucial \cite{Lanyon57,blatt2012quantum,rossnagel2015fast}; or quantum heat engines \cite{rossnagel2015single,klatzow2017experimental}, just to mention some.

Current quantum thermometric methods differ greatly depending on the experimental platform, the required precision, and the temperature range of interest. For instance, standard techniques to measure the ultracold temperatures of laser-cooled atomic clouds have traditionally relied on time-of-fight techniques (see e.g. \cite{Bloch2005}), \textit{in situ} imaging \cite{gemelke2009situ}, or used internal degrees of freedom \cite{olf2015thermometry}, or impurities in fermionionc gases (see \cite{Mckay2010,McKay2011} and references therein). However, very recently nanofibers have been successfully used as thermometers for laser-cooled and trapped atomic samples  \cite{PhysRevA.92.013850}. 
Moreover, nanofibers can act as potential thermometers for hybrid quantum systems, e.g., superconductors that posses quite different cryogenic requirements.  

At the mesoscopic scale there is a wide variety of thermometric schemes operating in the Kelvin or subkelvin regime, usually achieved by steering electron and phonon currents (see e.g. \cite{RevModPhys.78.217,Pekola2013,karimi2018non} and references therein). Typical on-chip thermometers are made by using tunneling junctions between superconductors. A remarkable achievement is the very robust thermometry---operating in the millikelvin regime and with nanometric spatial resolution---that has been successfully developed using fluorescence in NV centres in diamonds \cite{Toyli21052013, kucsko2013nanometre,doi:10.1021/nl401216y}. Yet another spectacular example has been recently put forward using nanophotonic cavities coupled to a nano-optomechanic resonator. In such setup, a chip-integrated Brownian motion thermometer has been demonstrated, which provides a path towards quantum primary thermometry \cite{Purdy1265}.

As the ``toy'' open-system models used in theoretical quantum thermometry continue to become more refined and realistic (e.g., accounting for non-Markovian dissipative effects or non-linearities in the system Hamiltonian), it is to be expected that the thermometric bounds and optimal measurement protocols formulated from the theory side will start to inform novel and practical temperature-sensing techniques. Should these prove capable to beat the current precision standards for ultra-low temperature measurements, they will have a crucial impact on upcoming quantum technologies.

\ack We gratefully acknowledge Antonio Ac\'in, Gerardo Adesso, Robert Alicki, Emili Bagan, Jonathan  Bohr  Brask, Nicolas Brunner, Steve Campbell, John Calsamiglia, Christos Charalambous, Francesco Cosco, Gabriele De Chiara, Antonella De Pasquale, Miguel \'Angel Garc\'ia-March, Senaida Hernandez-Santana, Karen V. Hovhannisyan, Ronnie  Kosloff, Aniello Lampo, Maciej Lewenstein, Harry Miller, Alex Monras, Mariona Moreno-Cardoner, Mauro Paternostro, Jukka Pekola, Mart\'i Perarnau-Llobet, and Patrick P. Potts for useful discussions on quantum thermometry. We acknowledge funding from the Spanish MINECO (QIBEQI FIS2016-80773-P, FIS2016-80681-P (AEI/FEDER, UE), and Severo Ochoa SEV-2015-0522), Generalitat de Catalunya (CIRIT 2017-SGR-1127, CERCA Programme), Fundaci\'{o} Privada Cellex, the European Research Council (StG GQCOP, Grant No. 637352), and the US National Science Foundation (Grant No. NSF PHY1748958). LAC thanks the Kavli Institute for Theoretical Physics for their warm hospitality during the program ``Thermodynamics of quantum systems: Measurement, engines, and control''.

\section*{References}
\providecommand{\newblock}{}

\end{document}